%% file: main.tex
\documentclass[11pt]{article}

\usepackage[utf8]{inputenc}
\usepackage[left=25mm,right=25mm,top=25mm,bottom=25mm]{geometry}

\include{commands}

\include{packages}

\usetikzlibrary{external}
\newtheorem{definition}{Definition}[section]
\newtheorem{prop}[definition]{Proposition}%[section]
\newtheorem{theorem}[definition]{Theorem}%[section]
\newtheorem{corollary}[definition]{Corollary}%[section]
\newtheorem{remark}[definition]{Remark}%[section]
\title{Polyconvex anisotropic hyperelasticity with neural networks\vspace{1ex}}

\author[1,*]{Dominik~K.~Klein}
\author[2]{Mauricio~Fernández}
\author[3]{Robert~J.~Martin}
\author[3]{Patrizio~Neff}
\author[1]{\;and~Oliver~Weeger}

\affil[1]{\footnotesize Cyber-Physical Simulation Group, Department of Mechanical Engineering \& Centre for Computational Engineering, Technical University of Darmstadt, Dolivostr.~15, 64293 Darmstadt, Germany}
\affil[2]{\footnotesize Access e.V., Intzestr.~5, 52072 Aachen, Germany}
\affil[3]{\footnotesize Chair for Nonlinear Analysis and Modeling, Faculty of Mathematics, University of Duisburg-Essen, Thea-Leymann-Str.~9, 45127 Essen, Germany}
\affil[*]{\footnotesize Corresponding author, email: \href{mailto:klein@cps.tu-darmstadt.de}{klein@cps.tu-darmstadt.de}}
\date{October 29, 2021}
\begin{document}

\maketitle
\vspace*{-5mm}

%   include abstract
\par\noindent\rule{\textwidth}{0.4pt}
\input{chapter/abstract}
\par\noindent\rule{\textwidth}{0.4pt}\vspace*{2pt}
\small{Accepted version of manuscript published in the \emph{Journal of the Mechanics and Physics of Solids}. \\ 
Date accepted: October 29, 2021. DOI: \href{https://doi.org/10.1016/j.jmps.2021.104703}{10.1016/j.jmps.2021.104703}. License: \href{https://creativecommons.org/licenses/by-nc-nd/4.0/legalcode}{CC BY-NC-ND 4.0}}\vspace*{-1.4mm}
\par\noindent\rule{\textwidth}{0.4pt}

%   include chapters
\input{chapter/01_introduction}

\input{chapter/02_material_theory}

\input{chapter/03_ANN_based_material_models}

\input{chapter/04_application_to_cubic_lattice_metamaterials}

\input{chapter/05_ti}
\input{chapter/06_critique}
\input{chapter/07_conclusion}

\vspace*{3ex}

\noindent
\textbf{Conflict of interest.} The authors declare that they have no conflict of interest.
\vspace*{1ex}

\noindent
\textbf{Acknowledgment.} The work of Dominik Klein is supported by the Graduate School CE within the Centre for Computational Engineering at Technical University of Darmstadt. 
Patrizio Neff acknowledges support in the framework of the DFG-Priority Programme 2256 ``Variational Methods for Predicting Complex Phenomena in Engineering Structures and Materials'', Neff 902/10-1, Project-No. 440935806.
\vspace*{1ex}

\noindent
\textbf{Data availability.}
The authors provide access to the complete simulation data required to reproduce the results through the public GitHub repository \url{https://github.com/CPShub/sim-data}.

\appendix
\numberwithin{equation}{section}
\input{chapter/A_01_convex_neural_networks}

\renewcommand*{\bibfont}{\footnotesize}
\printbibliography

\end{document}

%% file: commands.tex
\newcommand{\uuA}{{{A}}}
\newcommand{\uuB}{{{B}}}
\newcommand{\uuC}{{{C}}}

\newcommand{\uuF}{{{F}}}

\newcommand{\uuI}{\mathbbm{1}}

\newcommand{\uuQ}{{{Q}}}
\newcommand{\uuX}{{{X}}}
\newcommand{\uA}{{A}}
\newcommand{\ua}{{a}}
\newcommand{\ub}{{b}}
\newcommand{\ue}{{e}}
\newcommand{\uB}{{B}}

\newcommand{\uD}{{D}}

\newcommand{\uX}{{X}}
\newcommand{\up}{{p}}
\newcommand{\uv}{{v}}
\newcommand{\uw}{{w}}
\newcommand{\ux}{{x}}

\newcommand{\uxi}{{\xi}}

\newcommand{\SP}{\mathcal{SP}}
\newcommand{\LSE}{\mathcal{LSE}}
\newcommand{\tr}{\operatorname{tr}}
\newcommand{\cof}{\operatorname{Cof}}
\newcommand{\diag}{\operatorname{diag}}
\newcommand{\SO}{\text{SO}}

\newcommand{\tMF}[1]{{\color{black}#1}}
\newcommand{\tDK}[1]{{\color{black}#1}}
\newcommand{\tOW}[1]{{\color{black}#1}}

\newcommand{\tRM}[1]{{\color{black}#1}}
\newcommand{\ttDK}[1]{{\color{black}#1}}

\newcommand{\cA}{{\mathcal{A}}}

\newcommand{\cG}{{\mathcal{G}}}

\newcommand{\cI}{{\mathcal{I}}}

\newcommand{\cP}{{\mathcal{P}}}

\newcommand{\ucA}{{\cA}}
\newcommand{\ucI}{{\cI}}

%% file: packages.tex
\usepackage{multirow}
\usepackage{booktabs}

\usepackage{amssymb}
\usepackage{amsmath}
\usepackage{amsthm}
\usepackage{nicefrac}
\usepackage{bbm}

\theoremstyle{definition}   %   keine kursive Schrift in der "theorem" - Umgebung
\usepackage{chngcntr}
\usepackage{apptools}
\usepackage[usenames,dvipsnames,svgnames]{xcolor}
\usepackage{tikz}
\usepackage{pgfplots}
\usepgfplotslibrary{groupplots}
\usepackage{pgfplotstable}
\usepackage{subcaption}

\usepackage[font=small]{caption}

\usepackage{wrapfig}

\pgfplotsset{compat=newest}
\usepgfplotslibrary{fillbetween}
\usetikzlibrary{patterns}
\usepgfplotslibrary{colormaps}

\definecolor{color11}{RGB}{236, 188, 0}
\definecolor{color12}{RGB}{50, 67, 121}
\definecolor{color13}{RGB}{222, 222, 222}
\definecolor{color21}{RGB}{92, 134, 196}
\definecolor{color22}{RGB}{249, 156, 0}
\definecolor{color23}{RGB}{222, 222, 222}
\definecolor{color31}{RGB}{222, 222, 222}
\definecolor{color32}{RGB}{222, 222, 222}
\definecolor{color33}{RGB}{191, 60, 60}

\pgfplotscreateplotcyclelist{mycolorlist}{%
only marks,solid, mark options={solid,fill=color11,fill opacity=1,mark size=4pt},mark=square*,color11\\
only marks,solid, mark options={solid,fill=color12,fill opacity=1,mark size=4pt},mark=square*,color12\\
only marks,solid, mark options={solid,fill=color13,fill opacity=1,mark size=4pt},mark=square*,color13\\
only marks,solid, mark options={solid,fill=color21,fill opacity=1,mark size=4pt},mark=square*,color21\\
only marks,solid, mark options={solid,fill=color22,fill opacity=1,mark size=4pt},mark=square*,color22\\
only marks,solid, mark options={solid,fill=color23,fill opacity=1,mark size=4pt},mark=square*,color23\\
only marks,solid, mark options={solid,fill=color31,fill opacity=1,mark size=4pt},mark=square*,color31\\
only marks,solid, mark options={solid,fill=color32,fill opacity=1,mark size=4pt},mark=square*,color32\\
only marks,solid, mark options={solid,fill=color33,fill opacity=1,mark size=4pt},mark=square*,color33\\
color11, dashed, very thick,mark=*, mark options={solid}, mark repeat = 15\\
color12, dashed, very thick ,mark=*, mark options={solid}, mark repeat = 15\\
color13, dashed, very thick ,mark=*, mark options={solid}, mark repeat = 15\\
color21, dashed, very thick ,mark=*, mark options={solid}, mark repeat = 15\\
color22, dashed, very thick ,mark=*, mark options={solid}, mark repeat = 15\\
color23, dashed, very thick ,mark=*, mark options={solid}, mark repeat = 15\\
color31, dashed, very thick,mark=*, mark options={solid} , mark repeat = 15\\
color32, dashed, very thick ,mark=*, mark options={solid}, mark repeat = 15\\
color33, dashed, very thick,mark=*, mark options={solid} , mark repeat = 15\\
color11, very thick \\
color12, very thick \\
color13, very thick \\
color21, very thick \\
color22, very thick \\
color23, very thick \\
color31, very thick \\
color32, very thick \\ 
color33, very thick \\
}

\pgfplotscreateplotcyclelist{mycolorlistWFXcell}{%
only marks,solid, mark options={solid,fill=color11,fill opacity=1,mark size=4pt},mark=square*,color11\\
only marks,solid, mark options={solid,fill=color22,fill opacity=1,mark size=4pt},mark=square*,color22\\
only marks,solid, mark options={solid,fill=color33,fill opacity=1,mark size=4pt},mark=square*,color33\\
color11, dashed, very thick,mark=*, mark options={solid}, mark repeat = 15\\
color22, dashed, very thick,mark=*, mark options={solid}, mark repeat = 15\\
color33, dashed, very thick,mark=*, mark options={solid}, mark repeat = 15\\
color11, very thick \\
color22, very thick \\
color33, very thick \\
}

\pgfplotscreateplotcyclelist{mycolorlistvol}{%
color11, dashed, very thick,mark=*, mark options={solid}, mark repeat = 15\\
color21, very thick \\
color33, very thick \\
}

\pgfplotscreateplotcyclelist{mycolorlist2}{%
only marks,solid, mark options={solid,fill=color11,fill opacity=1,mark size=4pt},mark=square*,color11\\
only marks,solid, mark options={solid,fill=color12,fill opacity=1,mark size=4pt},mark=square*,color12\\
only marks,solid, mark options={solid,fill=color13,fill opacity=1,mark size=4pt},mark=square*,color13\\
only marks,solid, mark options={solid,fill=color21,fill opacity=1,mark size=4pt},mark=square*,color21\\
only marks,solid, mark options={solid,fill=color22,fill opacity=1,mark size=4pt},mark=square*,color22\\
only marks,solid, mark options={solid,fill=color23,fill opacity=1,mark size=4pt},mark=square*,color23\\
only marks,solid, mark options={solid,fill=color31,fill opacity=1,mark size=4pt},mark=square*,color31\\
only marks,solid, mark options={solid,fill=color32,fill opacity=1,mark size=4pt},mark=square*,color32\\
only marks,solid, mark options={solid,fill=color33,fill opacity=1,mark size=4pt},mark=square*,color33\\
color11 \\
color11 \\
color11, dashed, very thick,mark=*, mark options={solid}, mark repeat = 15\\
color12, dashed, very thick ,mark=*, mark options={solid}, mark repeat = 15\\
color13, dashed, very thick ,mark=*, mark options={solid}, mark repeat = 15\\
color21, dashed, very thick ,mark=*, mark options={solid}, mark repeat = 15\\
color22, dashed, very thick ,mark=*, mark options={solid}, mark repeat = 15\\
color23, dashed, very thick ,mark=*, mark options={solid}, mark repeat = 15\\
color31, dashed, very thick,mark=*, mark options={solid} , mark repeat = 15\\
color32, dashed, very thick ,mark=*, mark options={solid}, mark repeat = 15\\
color33, dashed, very thick,mark=*, mark options={solid} , mark repeat = 15\\
color11, very thick \\
color12, very thick \\
color13, very thick \\
color21, very thick \\
color22, very thick \\
color23, very thick \\
color31, very thick \\
color32, very thick \\ 
color33, very thick \\
color11 \\
color11 \\
}

\pgfplotscreateplotcyclelist{mycolorlist123}{%
only marks,solid, mark options={solid,fill=color11,fill opacity=1,mark size=4pt},mark=square*,color11\\
only marks,solid, mark options={solid,fill=color22,fill opacity=1,mark size=4pt},mark=square*,color22\\
only marks,dashed, mark options={solid,fill=color33,fill opacity=1,mark size=4pt},mark=square*,color33\\
color11, very thick,dashed, mark options={solid}, mark repeat = 15\\
color22, very thick , mark options={solid}, mark repeat = 15\\
color33, very thick,dashed , mark options={solid}, mark repeat = 15\\
}

\pgfplotscreateplotcyclelist{mycolorlistSNE}{%
only marks,solid, mark options={solid,fill=color11,fill opacity=1,mark size=4pt},mark=square*,color11\\
only marks,solid, mark options={solid,fill=color22,fill opacity=1,mark size=4pt},mark=square*,color22\\
only marks,dashed, mark options={solid,fill=color33,fill opacity=1,mark size=4pt},mark=square*,color33\\
color11, very thick, dashed,  mark options={solid}, mark repeat = 15,mark=*\\
color22, very thick, dashed , mark options={solid}, mark repeat = 15, mark=*\\
color33, very thick,dashed , mark options={solid}, mark repeat = 15, mark=*\\
color11, very thick \\
color22, very thick \\
color33, very thick \\
}

\usepackage[maxbibnames=50, maxcitenames=3, 
    natbib=true, style=numeric,
    isbn=false, url=false]{biblatex}   % Literaturverzeichnis
\bibliography{bibliography}
\AtBeginBibliography{} % no comma before last author in bibliography

\usepackage[german, main=english]{babel}
\usepackage[autostyle]{csquotes}% Anführungszeichen vereinfacht

\usepackage[T1]{fontenc}

\usepackage{hyperref}

\usepackage[affil-it]{authblk}

%% file: chapter/abstract.tex
\begin{abstract}
\noindent
In the present work, two machine learning based constitutive models for finite deformations are proposed. 
Using input convex neural networks, the models are hyperelastic, anisotropic and fulfill the polyconvexity condition, which implies ellipticity and thus ensures material stability.
The first constitutive model is based on a set of polyconvex, anisotropic and objective invariants.
The second approach is formulated in terms of the deformation gradient, its cofactor and determinant, uses group symmetrization to fulfill the material symmetry condition, and data augmentation to fulfill objectivity approximately. The extension of the dataset for the data augmentation approach is based on mechanical considerations and does not require additional experimental or simulation data. 
The models are calibrated with highly challenging simulation data of cubic lattice metamaterials, including finite deformations and lattice instabilities. A moderate amount of calibration data is used, based on deformations which are commonly applied in experimental investigations. While the invariant-based model shows drawbacks for several deformation modes, the model based on the deformation gradient alone is able to reproduce and predict the effective material behavior very well and exhibits excellent generalization capabilities. 
\tDK{In addition, the models are calibrated with transversely isotropic data, generated with an analytical polyconvex potential. For this case, both models show excellent results, demonstrating the straightforward applicability of the polyconvex neural network constitutive models to other symmetry groups.}
\end{abstract}
\vspace*{2ex}
{\textbf{Key words:} constitutive modeling, nonlinear elasticity, anisotropic hyperelasticity, polyconvexity, ellipticity, material stability, soft materials, metamaterials, invariants, structural tensors, parameter identification, data-driven modeling, machine learning, input convex neural networks}

%% file: chapter/01_introduction.tex
\section{Introduction}

In the last decades, a vast amount of highly specialised materials has been developed and, with advancing requirements in engineering applications, the trend is growing. 
In particular, with recent advances in additive and advanced manufacturing technologies, flexible and functional mechanical metamaterials and composites are being developed \cite{surjadi19}. As a consequence, numerous constitutive models have been formulated, each specifically designed for the mechanical characteristics of a comparatively small class of (soft) materials \cite{chagnon2015,xiang2020a}. However, while the specific formulations may be different, the theoretical considerations that lead to physically sensible and mathematically well-posed models stay the same, i.e., ellipticity, thermodynamic consistency, objectivity, material symmetry, etc.~\cite{Holzapfel2000}.

Recent progress in the field of machine learning (ML) has sparked the development of data-driven numerical methods that avoid the explicit formulation of constitutive models and purely operate on discrete stress-strain data \cite{carrara2020,kirchdoerfer2016,nguyen2018}, as well as of data-driven constitutive models that employ reduced bases \cite{fritzen2019a,fritzen2018,yvonnet2007}, polynomials \cite{khajehtourian2020,YVONNET2009}, or artificial neural networks (ANNs) \cite{flaschel2021,le2015,liu2020,yang2019a} for the representations of nonlinear energy potentials or stress-strain relationships.
In particular, the latter approach offers a high flexibility and applicability to a wide range of materials due to the universal approximation properties of ANNs \cite{Hornik1991}.
Furthermore, they can also be formulated to fulfill important material theoretical considerations, e.g., the material symmetry condition \cite{Fernandez2020,Linka2020}. 
In \textcite{Gonzalez2019}, an analytical model was extended using ML methods. While preserving favorable properties of the analytical model, the ML extension can improve the models performance and, thus, the applicability to different materials, which is shown in \textcite{Gonzalez2020} for vascular soft tissues.
In \textcite{Fernandez2020}, two anisotropic material models for finite deformations are proposed. The first model is based on an analytical model \cite{Itskov2001}, whose function space is extended using ANNs. While preserving favorable analytical properties of the model, the flexibility is significantly improved. For the second approach, the method of group symmetrization is introduced to fulfill the material symmetry. With the six independent components of the right Cauchy-Green tensor $\uuC=\uuF
^T\uuF$ as input quantity for an ANN, the model offers a highly flexible, objective hyperelastic potential, which is able to reproduce the challenging effective behavior of cubic beam-lattice metamaterials, including lattice instabilities. 
The model proposed by \textcite{Linka2020} uses structural tensors, c.f.~\cite{Zheng1993}, processed through ANNs, to formulate a set of invariants reflecting the intrinsic material anisotropy and, using these invariants as input for another ANN, predicts the potential of the hyperelastic structure. 
\tDK{By using a Lagrangian multiplier to enforce incompressibility, c.f. \cite[Section 6.3]{Holzapfel2000}, the model is applicable to elastomeric material behavior, and shows excellent results for Treloar's experimental data on vulcanised rubber.}
Both the models introduced in \cite{Fernandez2020} and \cite{Linka2020} include the objectivity and material symmetry condition in their formulation.
In \textcite{heider2020}, a neural network based constitutive model for anisotropic elastoplastic materials is proposed, and different methods are examined to fulfill the objectivity condition, e.g., the representation of stress and strain tensors in their spectral decomposition for the input and output data of the neural network.

\medskip

In the context of finite elasticity theory, existence and stability of solutions for boundary-value problems are strongly linked to the notion of polyconvexity introduced by \textcite{Ball1977,Ball1976}. From a material theory point of view, polyconvexity is advantageous since it implies ellipticity of the corresponding constitutive model, while being more straightforward to include into the model formulation than the ellipticity condition itself. Ellipticity ensures material stability \cite{neff2015,Zee1983}, and is thus important for numerical applications such as the finite element method.
As will be shown for the model of \textcite{Fernandez2020}, the polyconvexity condition is easily violated when using unrestricted ANNs without caution. As this may lead to a loss of ellipticity and thus to material instability, the polyconvexity condition should be treated with special care in the model formulation. 

The theoretical aspects of polyconvexity are still subject to current research \cite{Ghiba2018,Martin2020,Martin2018b}, and the formulation of polyconvex models remains a challenging task \cite{Martin2018a,Martin2017}. For a long time after its initial conception, the polyconvexity \tOW{condition} was practically restricted to isotropic material response, as no anisotropic formulation was available that ensured at the same time: polyconvexity, objecticity, material symmetry and a stress-free reference configuration. In fact, the fulfillment of multiple constitutive requirements at the same time can be seen  as \enquote{the main open problem of the theory of material behavior} (\textcite[Section 20]{TruesdellNoll}). In a landmark paper, \textcite{Schroeder2003} derived a formulation which fulfills all of the former mentioned requirements, quickly followed by \cite{Ehret2007,Hartmann2003,Itskov2004,Kambouchev2007,schroder2010,Schroeder2005,Schroeder2008}. It was shown that questions of material stability could neatly be avoided while being able to match experimental data \cite{Balzani2006}. The approaches are based on second and fourth order structural tensors, which can reproduce a wide range of anisotropies. Combining the structural tensors with the right Cauchy-Green deformation tensor, sets of invariants can be derived. By a suitable construction of polynomials of these invariants, constitutive models can be created which fulfill all of the former mentioned restrictions. By using polyconvex invariants instead of non-polyconvex invariants, constitutive models can be improved \cite{Cai2021}.
Subsequently, also finite strain finite element methods tailored to the discretization of polyconvex material models were developed \cite{bonet2015,pfefferkorn2019}.

It must be emphasized, however, that the polyconvexity \tOW{condition} is a purely mathematical framework \tOW{that ensures ellipticity, but unlike ellipticity,} to the best knowledge of the authors, is not a fundamental physical requirement. For example, polyconvexity imposes restrictions upon arbitrarily large strains that will never occur in actual experiments. Here, we adopt polyconvexity as a suitable means to ensure overall material stability, i.e., to avoid loss of ellipticity. Otherwise, this would be cumbersome to be checked a posteriori. Note that the only real physical requirement is satisfaction of ellipticity in a compact set including the identity. But even this can be very challenging without polyconvexity \cite{neff2016,neff2015}.

Coming back to the field of ML, the difficulty of formulating physically sensible models, e.g., material models which fulfill multiple constitutive requirements, is still an open problem \cite{e2020,karniadakis2021,wang2020,willard2020}. Several constitutive modeling approaches found in the literature consider convexity properties, however, none of them known to the authors fulfills the polyconvexity condition. 
In \textcite{Ghaderi2020}, an attempt is made to formulate polyconvex potentials based on ANNs. However, for the neural network core, non-convex activation functions are used such as the hyperbolic tangent, and consequently, the potential proposed in \cite{Ghaderi2020} violates the polyconvexity condition, which follows from corollary \ref{corollary:convex_ANN_compositions}. 
In \textcite{vlassis2020a}, a \tDK{feed-forward neural network (FFNN)} based hyperelastic model for anisotropic material behavior is proposed, and the convexity of the resulting potential w.r.t. the right Cauchy-Green tensor $C$ is examined for a special material. However, while the potentials are convex in $C$ for some examined deformation modes, the model only approximates the convexity condition and, consequently, it may be violated for other deformations. Furthermore, convexity in~$C$ does not ensure polyconvexity, as not all components of $C$ are convex in the deformation gradient~$F$. Therefore, convexity in $C$ does not imply ellipticity, and exhibits no physical significance at all.
The constitutive model proposed in \textcite{Xu2020} identifies the Cholesky factor of a tangent stiffness matrix to describe the behavior of the material. By using symmetric positive definite neural networks, the formulation is closely linked to convex potentials, and numerical robustness of the model is ensured. While the model was able to predict both time-dependent and plastic problems, it requires a large amount of data for the calibration.
\tDK{A FFNN based constitutive model for nearly incompressible anisotropic hyperelasticity is proposed in \textcite{tac2021}. By using additional terms in the objective function which enforce semi positive-definiteness of the FFNN's Hessian, the neural network is approximately convex. However, the polyconvexity of the invariants which are used as input quantities for the neural network is not examined. For example, the invariant $\Bar{I}_2=\frac{1}{2}\left(\tr \Bar{C}\right)^2-\frac{1}{2}\tr \Bar{C}^2$ with $\Bar{C}=J^{\nicefrac{-2}{3}}C$ is not polyconvex \cite[Lemma 2.4]{Hartmann2003}. Therefore, in \cite{tac2021}, the polyconvexity condition is violated by the choice of input arguments.}

%%%%%%%%%%%%%%%%%%%%%%%%%%%%%%%%%
%   present work
%%%%%%%%%%%%%%%%%%%%%%%%%%%%%%%%%

\medskip

In the present work, two polyconvex constitutive models are introduced, which are both based on input convex neural networks (ICNNs), see \cite{Amos2016}, and formulated for hyperelastic, anisotropic material behavior and finite deformations. 
\tDK{ICNNs are a special class of FFNNs which are, through a suitable network architecture, constructed as convex functions.} 
\tDK{As already discussed, there is a variety of data-driven or machine learning methods which are used for constitutive modeling. However, for the construction of polyconvex potentials, convex FFNNs ($\widehat{=}$ ICNNs) are a very natural choice, as the simple mathematical structure of FFNNs offers a straightforward application of analytically received convexity conditions, c.f.~Appendix~\ref{app:aux_proofs}.}
While there are several applications for ICNNs in, e.g., convex optimization \cite{Amos2016,Calafiore2020a,Calafiore2020}, the application towards polyconvex constitutive models has, to the best of the authors' knowledge, not been studied yet.
The first model developed in this work is based on a set of invariants already introduced in \textcite{Schroeder2010}, which fulfills the objectivity and material symmetry condition by construction. Using ICNNs, substantially more complex functions for the potential compared to polynomial approaches can be created, while preserving the polyconvexity of the model. 
While this model can be seen as a straightforward extension of invariant-polynomials to the more flexible function space ANNs offer, the second constitutive model is specifically designed for machine learning. Formulated directly in the deformation gradient, its cofactor and determinant, the objectivity condition is not fulfilled by construction. That objectivity, i.e., invariance of the energy $W(Q\,F)=W(F)$ under rotations $Q\in\SO(3)$, is not automatically satisfied, may be surprising at first glance. Usually, hyperelastic formulations come with objectivity built in \textit{a priori}. Indeed, objectivity is not derived from an experimental observation, but is a physical law, c.f.\ sections 17 and 19 in \textcite{TruesdellNoll}.
However, in consideration of a useful model for a compact set of deformation gradients $F=D\varphi$, the possible error in not exactly satisfying objectivity remains controllable. The benefit in giving up exact objectivity is the increased flexibility in combining all the needed constitutive requirements. We remind that the major obstacle in originally extending polyconvexity to the anisotropic setting was coming from \textit{exact} objectivity combined with the stress-free reference configuration and material symmetry. Here, the objectivity condition is approximated using data augmentation, following \cite{Ling2016}. The approach is based on mechanical considerations, and requires no further simulation or experimental data. The anisotropy is taken into account with the group symmetrization already introduced in \textcite{Fernandez2020}. 

Both models can be calibrated with a moderate amount of data, which is shown for the highly challenging behavior of cubic beam-lattice metamaterials, using synthetic homogenization data. Modern additive and advanced manufacturing methods allow for a variety of tailored materials with specifically designed microstructure, consisting of, e.g., beams and shells, which leads to mechanical characteristics not encountered in classical materials, e.g., due to lattice instabilites within the microstructures   \cite{bertoldi2017,glaesener2020b,Jiang2016,khajehtourian2021,Lee2012,Liu2016}. 
Nonlinear multiscale simulations for this class of  metamaterials can be executed by homogenizing the microstructure \cite{Matous2016} either in a current multiscale setting using the FE\textsuperscript{2} method  \cite{glaesener2020b}, or sequentially based on the formulation of an effective constitutive model \cite{Fernandez2020,Jamshidian2020,khajehtourian2020}, as applied here.

The outline of the manuscript is as follows. In Sect.~\ref{sec:basics_of_material_theory}, the present work starts with a short introduction to the basics of constitutive modeling relevant to this work. In Sect.~\ref{sec:polyconvex_constitutive_models_based_on_FFNNs}, the lack of polyconvexity for models with unrestricted ANN architecture is discussed, and the two polyconvex ML based models are introduced. Finally, in Sect.~\ref{sec:application_to_cubic_metamaterials}, the models are calibrated to the highly challenging, homogenized behavior of cubic lattice metamaterials and compared with each other and to a conventional polyconvex model. \tDK{The application to another material symmetry, i.e., transverse isotropy, is demonstrated in Sect.~\ref{sec:application_to_ti}.} 
In Sect.~\ref{sec:critique}, some issues raised by the use of ML techniques in nonlinear elasticity theory are discussed. \tDK{After the conclusion in Sect.~\ref{sec:conclusion}, some general properties of ICNNs are introduced and discussed in Appendix~\ref{app:aux_proofs}.}

\paragraph{Notation}
Throughout this work, tensor compositions and contractions are denoted by $\left(\uuA\,\uuB\right)_{ij}=A_{ik}B_{kj}$, $\ua\cdot\ub=a_ib_i=\langle\ua,\,\ub\rangle$, $\uuA:\uuB=A_{ij}B_{ij}$ and $\left(\mathbb{A}:\uuA\right)_{ij}=\mathbb{A}_{ijkl}A_{kl}$, respectively, with vectors $a$ and $b$, second order tensors $A$ and $B$ and fourth order tensor $\mathbb{A}$. The tensor product is denoted by~$\otimes$, the second order identity tensor by $\uuI$.
The first Fr\'echet derivative of a function $f$ w.r.t. $X$ is denoted by $D_Xf$, the second Fr\'echet derivative (or Hessian) is denoted by $D_X^2f$. For the function composition $f(g(x))$ the compact notation $f\circ g\circ x$ is applied.
The set of invertible second order tensors with positive determinant is denoted by $\text{GL}^+(3):=\big\{\uuX \in\allowbreak \;\mathbb{R}^{3\times 3}\,\rvert\,\allowbreak \det \uuX > 0\big\}$, the special orthogonal group in $\mathbb{R}^3$ by $\SO(3):=\big\{\uuX \in\allowbreak \mathbb{R}^{3\times 3}\;\rvert\allowbreak \;\uuX^T\uuX=\uuI,\;\det \uuX =1\big\}$.

%% file: chapter/02_material_theory.tex
\section{Basics of material theory}
\label{sec:basics_of_material_theory}

The hyperelastic potential
\begin{equation}\label{eq:potential_W}
W\colon\text{GL}^+(3)\rightarrow\mathbb{R}\,,\qquad\uuF\mapsto W(\uuF) 
\end{equation}
corresponds to the strain energy density stored in the body $\Omega\subset\mathbb{R}^3$ due to the deformation 
${\varphi:\Omega\rightarrow\mathbb{R}^3}$, and depends on the deformation gradient $\uuF=D\varphi$ \cite{Holzapfel2000}. To ensure a physically sensible and mathematically well-posed formulation, several restrictions on the potential must be considered. The restrictions relevant to the present work are shortly introduced in the following: 

The reference configuration of the body must be stress-free, i.e., for the first Piola-Kirchhoff~stress
\begin{equation}\label{eq:PK1}
S=D_{\uuF}W(\uuF)
\end{equation}
it holds that $S(\uuI)=0$ \cite{Holzapfel2000}. \tDK{In fact, this is implied by the condition that the potential \eqref{eq:potential_W} attains its unique, global minimum at the identity, i.e., $W(\uuI)=0$ and $W(F)\geq 0$ \cite{Holzapfel2000}.}
The principle of objectivity \cite{TruesdellNoll} states that the material behavior has to be independent of the observer, which means that the potential is invariant under the transformation
\begin{equation}\label{eq:material_objectivity}
\begin{aligned}
W(\uuQ\,\uuF)=W(\uuF) \quad
\forall\uuF\in\text{GL}^+(3)\,,\;\uuQ\in \SO(3)\,.
\end{aligned}
\end{equation}
Following \eqref{eq:PK1}, this implies the invariance of the stress tensor under transformations according to
\begin{equation} \label{eq:stress_objectivity}
\begin{aligned}
S(\uuQ\,\uuF)=\uuQ\,S(\uuF)\quad
\forall\uuF\in\text{GL}^+(3)\,,\;\uuQ\in \SO(3)\,.
\end{aligned}
\end{equation}
For models formulated in terms of the right Cauchy-Green tensor $\uuC=\uuF^T\uuF$,
i.e., as $W=W(\uuC)$, 
the material objectivity condition is automatically fulfilled, which is an advantage compared to models depending directly on the deformation gradient. 
For anisotropic materials with symmetry group $\mathcal{G}\subset\SO(3)$, the strain energy must be invariant under the symmetry transformation \cite{Haupt2002} 
\begin{equation} \label{eq:material_symmetry}
\begin{aligned}
W(\uuF\,\uuQ)=W(\uuF)\quad
\forall\uuF\in\text{GL}^+(3)\,,\;\uuQ\in \mathcal{G}\subset\SO(3)\, ,
\end{aligned}
\end{equation} 
which implies the stress invariance condition
\begin{equation}
\begin{aligned}
S(\uuF\,\uuQ)=S(\uuF)\,\uuQ\quad
\forall\uuF\in\text{GL}^+(3)\,,\;\uuQ\in \mathcal{G}\subset\SO(3)\,.
\end{aligned}
\end{equation} 

The growth condition
\begin{equation}\label{eq:growth}
    W(\uuF)\rightarrow\infty\quad\text{as}\quad\det\uuF\rightarrow 0^+\quad
    \left(\Leftrightarrow\quad\frac{1}{\det\uuF}\rightarrow\infty\right)
\end{equation}
captures the physical consideration that for infinitely large volumetric compression, an infinite amount of energy is required. There are several other growth conditions known in constitutive modeling, which regard the material behavior for very large deformations and are often referred to as coercivity conditions.\footnote{For a coercive function, $f\left(\ux\right)\rightarrow\infty$ as $\left\lVert\ux\right\rVert\rightarrow\infty$ holds.} However, while the case of large volumetric compression \eqref{eq:growth} is important to consider especially for highly compressible materials, coercivity conditions regard material behavior which usually lies outside the considered deformation modes. Therefore, they are of rather theoretical interest, and will not be considered throughout this work. 

\medskip

In finite elasticity theory, the existence of minimizers for the underlying variational functionals is guaranteed if the energy potential \eqref{eq:potential_W} fulfills the polyconvexity condition introduced by \textcite{Ball1977,Ball1976} and an additional coercivity condition \cite{kruzik2019}. 
The potential $W\big(\uuF\big)$ is polyconvex if and only if there exists a function $\mathcal{P}:\mathbb{R}^{3\times 3}\times\mathbb{R}^{3\times 3}\times\mathbb{R}\rightarrow\mathbb{R}$ such that
\begin{equation}{\label{eq:polyconvexity}}
    W(\uuF)=\mathcal{P}(\uuF,\,\cof\uuF,\,\det\uuF)\, ,
\end{equation}
so that $\mathcal{P}$ is convex in its arguments $(\uuF,\,\cof\uuF,\,\det\uuF)$. The function $\mathcal{P}$ is in general non-unique \cite{Schroeder2003}. 
Due to the reason mentioned above, the coercivity condition is not considered in this work. For convex functions which are sufficiently smooth, the Hessian matrix is positive semi-definite \cite{Silhavy2014}, and the polyconvexity condition can be formulated as 
\begin{equation}
\delta \uxi \cdot D_\xi^2\mathcal{P}(\uxi)
\cdot \delta \uxi \geq 0 \quad\forall \;\uxi, \delta\uxi\, ,
\end{equation}
with the rearranged arguments $\uxi=(\uuF,\,\cof\uuF,\,\det\uuF)\in\mathbb{R}^{19}$ and $\delta\uxi\in\mathbb{R}^{19}$ \cite{Ebbing2010}. From this point on, the notation $\uxi=(\uuF,\,\cof\uuF,\,\det\uuF)\in\mathbb{R}^{19}$ will be used to adress the argument in the polyconvexity contexts more compactly.
Polyconvexity implies ellipticity, while being more straightforward to include into the model formulation than the ellipticity condition itself. The ellipticity (or rank-one convexity) condition \cite{neff2015,Zee1983}
\begin{equation} \label{eq:ellipticity}
    \left(\ua\otimes\ub\right):D^2_FW(F): \left(\ua\otimes\ub\right)\geq 0 \quad\forall\;\ua,\ub\in\mathbb{R}^3
\end{equation}
ensures material stability of the constitutive model.

%% file: chapter/03_ANN_based_material_models.tex
\section{Polyconvex constitutive models based on FFNNs}
\label{sec:polyconvex_constitutive_models_based_on_FFNNs}

Feed-forward neural networks (FFNNs) are universal approximators \cite{Hornik1991}, meaning that they can represent continuous functions of arbitrary complexity. FFNN based constitutive models exploit this important property by using them as highly flexible functions within the model formulations, e.g., to represent the energy potential $W$ defined in \eqref{eq:potential_W}.
This is in contrast to conventional approaches, which rely on a human choice for the representation of $W$, thus potentially reducing the possible function space.
Furthermore, through the right choice of network architecture, FFNNs can be constructed as convex functions, which is an often required property in the context of constitutive modeling.

In finite elasticity theory, the construction of potentials which are convex in the deformation gradient, its cofactor and determinant is of special interest, which is referred to as polyconvexity~\cite{Ball1976}. By using convex FFNNs, i.e., ICNNs, hyperelastic potentials can be generated which satisfy the polyconvexity condition, see \eqref{eq:polyconvexity}. Polyconvexity implies ellipticity, and thus ensures material stability of the constitutive models. 
The convexity condition is not trivially fulfilled by arbitrary FFNNs. 
Consequently, for FFNN based models found in the literature a loss of polyconvexity can often be detected, which may lead to a loss of ellipticity and thus material instability. This is now discussed for the approach of \textcite{Fernandez2020}.

\textcite{Fernandez2020} introduced several ML based constitutive models, from which we will examine the potential model.
The model uses a FFNN core for the hyperelastic potential, with the right Cauchy-Green tensor $\uuC=F^TF$ as input, and fulfills the material symmetry condition \eqref{eq:material_symmetry} with a group symmetrization of the potential, see eq. \eqref{eq:group_symmetrization}. With the six independent components of $\uuC$ as input, the model fulfills the objectivity condition \eqref{eq:material_objectivity} per construction. The internal FFNN used in this work is built upon multiple compositions of the Softplus function, cf. \eqref{eq:SP}, with unrestricted weights.
This model violates the polyconvexity condition in two aspects. First of all, only the principal diagonal elements of the right Cauchy-Green tensor $\uuC$ are convex in the deformation gradient~$\uuF$, while the remaining elements of $\uuC$ are neither convex nor concave.\footnote{\tDK{This does not mean that objectivity and polyconvexity contradict each other, it rather shows how challenging it is to combine both requirements.}} Additionally, compositions of Softplus functions with arbitrary parameters are not convex, as is discussed in corollaries \ref{corollary:convex_ANN_compositions} and \ref{cor:SP} in App.~\ref{app:aux_proofs}. Furthermore, while the metamaterials under consideration in~\cite{Fernandez2020} are highly compressible, the volume compression condition \eqref{eq:growth} is not considered in the model formulation. Therefore the model does not ensure a physical sensible behavior for the limiting case of infinitely large volumetric compression.   

\medskip

A brief introduction to convex FFNNs, i.e., ICNNs, and the notation used throughout this work can be found in App.~\ref{app:aux_proofs}. In corollaries~\ref{corollary:convex_ANN_compositions} and \ref{corollary:convex_ANN_bias} the standard conditions for the fulfillment of convexity are provided. Activation functions fulfilling these conditions are presented in theorem~\ref{theorem:LSE} and corollary~\ref{cor:SP}. These activation functions lead to the ICNN cores discussed in proposition~\ref{prop:LSE_SP_cores}. In a nutshell, an ICNN is easily constructed based on standard FFNNs with input vector $\uX$ based on the following restrictions:
\begin{itemize}
    \item the first hidden layer $\uA_1$ is neuron-wise convex with respect to $\uX$ (e.g., by employing convex activation functions as \emph{Softplus}, $s(x) = \log(1+\exp(x))$, in each neuron)
    \item from the second $\uA_2$ to the last hidden layer $\uA_H$, each hidden layer is neuron-wise convex and non-decreasing with respect to the previous hidden layer (e.g., by employing $s(x)$ with non-negative weights in each neuron)
    \item the scalar-valued output layer $a$ is convex and non-decreasing with respect to the last hidden layer.
\end{itemize}
These restrictions imply that the composition $a \circ \uA_H \circ \dotsc \circ \uA_2 \circ \uA_1 \circ \uX = \Bar{a}(\uX)$ is convex in~$\uX$. These architectures are shortly denoted as ICNNs from now on. For several passages, the abbreviation ${\mathcal{A}} = \uA_H \circ \dotsc \circ \uA_1$ will be used for the core (i.e., the hidden layers) of the ICNN. The collection of all weights and biases of the ICNN will be abbreviated by the internal parameter vector $\up$.

We now present two polyconvex ICNN-based approaches for anisotropic hyperelastic constitutive modeling, extending the works of \textcite{Schroeder2010} and \textcite{Fernandez2020}.

%%%%%%%%%%%%%%%%%%%%%%%%%%%%%%%%%
%   "Polyconvex ML model based on Invariants"
%%%%%%%%%%%%%%%%%%%%%%%%%%%%%%%%%

\subsection{Model based on invariants} \label{sec:polyconvex_model_with_invariants}

\paragraph{Polyconvex model based on invariants}
In \textcite{Schroeder2010}, a polyconvex potential for trigonal, tetragonal and cubic symmetry groups is proposed, see also \cite{Ebbing2010}. In the following, we will shortly describe the cubic potential.
It is formulated in polynomials of invariants, which are based on a combination of the right Cauchy-Green tensor and an anisotropic structural tensor. The fourth order tensor $\mathbb{G}$ is called a structural tensor of the material symmetry group $\mathcal{G}\subset \SO(3)$ if
\begin{equation}
\mathbb{G}=\uuQ\ast\mathbb{G}\quad\forall\;\uuQ\in \mathcal{G}\subset \SO(3)\,,
\end{equation}
where $\ast$ denotes the Rayleigh product \cite{bertram2021}.
For some material symmetry groups, second order structural tensors are sufficient to represent the symmetry, while for other symmetries, the introduction of fourth or sixth order tensors is required \cite{Zheng1993}. For cubic symmetry, a fourth-order structural tensor is required.
In \cite{Schroeder2010}, the fourth order structural tensor
\begin{equation}\label{eq:structural_tensor}
\mathbb{G}_{\text{cub}}=\sum_{i=1}^3\ue_i\otimes\ue_i\otimes\ue_i\otimes\ue_i
\end{equation}
is used for the cubic group $\mathcal{G}_7$, with $\ue_i$ being the basis vectors of a Cartesian coordinate system. A description of the cubic group $\mathcal{G}_7$ can be found in \cite[Section 3.7]{Ebbing2010}. Using the structural tensor \eqref{eq:structural_tensor}, a set of five polyconvex invariants can be derived as
\begin{equation}\label{eq:invariants}
\begin{aligned}
&I_1=\tr\uuC\,,\quad\quad\quad\quad\quad I_2=\tr\left(\cof\uuC\right)\,,\quad\quad\quad\quad\quad I_3=\det\uuC\,,
\\
&\quad J_7=\uuC:\mathbb{G}_{\text{cub}}:\uuC\,,\quad\quad\quad\quad\quad
 J_{11}=\cof\uuC:\mathbb{G}_{\text{cub}}:\cof\uuC\,.
\end{aligned}
\end{equation}
The first three invariants are the well-known isotropic invariants, while the remaining two invariants possess the cubic symmetry.  The invariants $I_1$ and $J_7$ are convex in $\uuF$, the invariants $I_2$ and $J_{11}$ are convex in $\cof \uuF$, and the invariant $I_3$ is convex in $\det\uuF$. Since the invariants are formulated in the right Cauchy-Green tensor, they fulfill the principle of objectivity. 
Based on these invariants, \textcite{Schroeder2010} proposed the potential
\begin{equation}\label{eq:SNE}
\begin{aligned}
\tDK{W^\text{SNE}_\text{cub}}=\kappa\,\sum_{r=1}^n\left(\frac{1}{(\alpha_{r}+1)3^{\alpha_{r}}}\,I_1^{(\alpha_{r}+1)}+\frac{3}{(\beta_{r}+1)3^{\beta_{r}}}\,I_2^{(\beta_{r}+1)}
+\frac{1}{(\eta_{r}+1)3^{\eta_{r}}}\,J_{7}^{(\eta_{r}+1)}+\frac{9}{\gamma_{r}}\,I_3^{-\gamma_{r}}\right)\,,
\end{aligned}
\end{equation}
where we introduced the additional parameter $\kappa$. With $\kappa>0,\;\alpha_{r},\,\beta_{r},\,
\eta_{r}\geq0$ and $\gamma_{r}\geq-\nicefrac{1}{2},\;\gamma_r \neq 0$, the potential is polyconvex, coercive and has a stress-free reference configuration.

\paragraph{Extension with neural networks}
A vector of group specific objective invariants $\ucI(\uxi)$ is defined, where each component of $\ucI(\uxi)$ is convex in $\uxi$, i.e., $\ucI(\uxi)$ is a polyconvex vector-valued function. For instance, $\ucI = (I_1, \,I_2,\, I_3,\, J_7,\, J_{11})$ can be considered for cubic materials. This motivates the model
\begin{equation}
    \cP_0(\uxi;\,\up) = a \circ \ucA \circ \ucI \circ \uxi \, ,
    \label{pcnn_inv_based}
\end{equation}
where the layers $a \circ \ucA$ in \eqref{pcnn_inv_based} are restricted to convex and non-decreasing activation functions, since~$\ucI$ represents the first convex layer, see remark \ref{rem:invariant_layer} for a discussion. The model \eqref{pcnn_inv_based} is polyconvex due to the usage of ICNNs und fulfills the objectivity and material symmetry conditions due to the considered invariants. 

\paragraph{Growth condition}
The volumetric growth condition \eqref{eq:growth} can be fulfilled with coercive functions. However, since ICNNs are not necessarily coercive, they are not suited to fulfill this condition, and therefore, an analytical term should be added to the potential \eqref{pcnn_inv_based} according to
\begin{equation} \label{pcnn}
    W=\cP_0(\uxi;\,\up)+W_\text{vol}\left(\det\uuF\right)\,,
\end{equation}
where we choose the polyconvex term \cite{Hartmann2003}
\begin{equation}\label{eq:coercive}
    W_\text{vol}(\det\uuF)=\left(\det F + \frac{1}{\det F}-2\right)^2.
\end{equation}

\tDK{
\paragraph{Computation of stress}
The scalar-valued function $W$ given by \eqref{pcnn}, which consists of the neural network $\cP_0$ from \eqref{pcnn_inv_based} and the volumetric growth term $W_\text{vol}$ from \eqref{eq:coercive}, is then interpreted as a hyperelastic potential, see \eqref{eq:potential_W}, and the first Piola-Kirchhoff stress $S=D_{\uuF}W(\uuF)$ is calculated as its gradient, see \eqref{eq:PK1}. In the present work, automatic differentiation \cite{Baydin2018} is used, which is widely available in modern machine learning libraries. This approach is, of course, not only applicable to the present model but to any differentiable model built in a ML library providing automatic differentiation.

\paragraph{Reference state}
In the reference configuration $F=\uuI$, the model \eqref{pcnn} is not stress-free by construction. However, elastic stress-strain data received from numerical or experimental investigations will always have the property $S(\uuI)=0$. When the model is calibrated with this data, it also approximates the stress behavior for $F=\uuI$, thus fulfilling $S(\uuI)=0$ in an approximate fashion.}
\ttDK{In \textcite{Fernandez2020} a projection approach is proposed which fulfills the stress-free reference configuration in an exact way. While this approach preserves polyconvexity when formulated in terms of $F$ instead of $C$, it is not compatible with the methods of incorporating objectivity and material symmetry which are used in the present work, and therefore cannot be applied.}

%%%%%%%%%%%%%%%%%%%%%%%%%%%%%%%%%
%   "Polyconvex ML model based on the Deformation Gradient"
%%%%%%%%%%%%%%%%%%%%%%%%%%%%%%%%%

\subsection{Model based on the deformation gradient}
\label{sec:polyconvex_model_F}

\paragraph{Model formulation}
Based on the definition of polyconvexity \eqref{eq:polyconvexity}, the most straightforward inputs for a polyconvex FFNN-based model are the deformation gradient, its cofactor and determinant itself. Thus, when $\uxi = \left(\uuF,\,\cof \uuF,\,\det \uuF\right)\in\mathbb{R}^{19}$  is used as input for an ICNN $\Bar{a}(\uxi) = a\circ\ucA\circ\uxi$, the output $\Bar{a}(\uxi)$ is convex in $\uxi$ and the resulting potential is polyconvex, but does not fulfill the material symmetry condition, in general.

Here, the group symmetrization of a function $\phi(\uuF)$ with respect to a given finite group $\cG \subset \SO(3)$ with $\#(\cG)$ elements, as introduced in \textcite{Fernandez2020} for ML-based models, is of interest
\begin{equation}
    \phi^\cG(\uuF) 
    = 
    \frac{1}{\#(\cG)}
    \sum_{\uuQ \in \cG} \phi(\uuF \, \uuQ)\, ,
\end{equation}
since the group symmetrized function $\phi^\cG(\uuF)$ fulfills the material symmetry conditions of the considered group. Application of group elements $\uuQ \in \cG$ on $\uuF$, i.e, $\uuF \rightarrow \uuF \, \uuQ$, results in the linear transformation of $\uxi = \left(\uuF,\,\cof \uuF,\,\det \uuF\right)$ according to
\begin{equation}
    \uxi\!\ast_s\!Q=(\uuF\,\uuQ,\,(\cof \uuF)\,\uuQ,\,\det\uuF),
\end{equation}
where we introduced the operator $\ast_s$ that specifies the transformation of $\uxi$ by $Q$, and exploited that $\cof(F\,Q)=(\cof F)\,Q$ \cite{Schroeder2003}.
Such a linear transformation does not affect the convexity of the ICNN $\Bar{a}(\uxi)$, see corollary \ref{corollary:convex_ANN_bias} for an explicit proof. This implies that the following potential model
\begin{equation}\label{eq:group_symmetrization}
\begin{aligned}
\cP_0(\uxi;\,\up)=a\circ{\mathcal{A}}\circ\uxi\,,
\qquad
W=\frac{1}{\#\left(\mathcal{G}\right)}\sum_{\uuQ\in \mathcal{G}}\cP_0( \uxi\!\ast_s\!Q;\,\up)+W_\text{vol}
\end{aligned}
\end{equation}
is polyconvex and fulfills the material symmetry and volume compression conditions. It should be noted that compared to the previous invariant based model, the first hidden layer of the core $\ucA$ in \eqref{eq:group_symmetrization} is not restricted to convex and non-decreasing activation functions, but only to convex functions. This allows more flexibility in the first hidden layer.

\tDK{
\paragraph{Infinite groups and finite subgroups}
The anisotropy of many materials can be described by finite groups, e.g., cubic, orthotropic or monoclinic materials, such that the group symmetrization \eqref{eq:group_symmetrization} can be carried out for the exact fulfillment of the material symmetry condition \eqref{eq:material_symmetry}. For some materials, the symmetry group is infinite, e.g., for transversely isotropic materials. For such cases, as also remarked in \textcite{Fernandez2020}, a pragmatic solution for the usage of the group symmetrization \eqref{eq:group_symmetrization} can be obtained by consideration of a finite subgroup of the infinite group. For the case of transverse isotropy, a finite subgroup $\cG_{\text{ti}}^*\subset\cG_{\text{ti}}$ for chosen $N$ can be simply constructed by 
\begin{equation}\label{eq:symm_group_ti}
    \cG_{\text{ti}}^*:=\big\{Q_x^{\alpha}\;\rvert\; \alpha=\nicefrac{2\pi\,n}{N},\,n=1,\,2,\,\dotsc,\,N\big\}
    \quad\subset\quad
    \cG_{\text{ti}}:=\big\{Q_x^{\alpha}\;\rvert\; \alpha\in[0,\,2\,\pi)\big\}
\end{equation}
where $Q_x^\alpha$ denotes a rotation around the preferred axis $x$ by the angle $\alpha$. The finite subgroup $\cG_{\text{ti}}^*$ contains $N$ equidistant elements of $\cG_{\text{ti}}$. As will be shown in Sect.~\ref{sec:application_to_ti}, already $N=6$ elements can be sufficient for an excellent approximation of transverse isotropy.
}

\tMF{
\paragraph{Computation of stress}
As in the invariant-based, differentiable FFNN-based model, the present model uses automatic differentiation for the computation of the stresses.
}

\paragraph{Objectivity}
While the determinant of the deformation gradient is an invariant quantity with respect to change of observers, the deformation gradient itself and its cofactor depend on the choice of observer. Thus, the model $W$ as defined in \eqref{eq:group_symmetrization} is generally not objective. While the objectivity condition could be fulfilled trivially by using the components of the right Cauchy-Green tensor $C$, this would violate the polyconvexity condition, as only the main diagonal elements of $C$ are convex in the deformation gradient. Consequently, in order to meet both polyconvexity and objectivity, we choose a quantity which is suitable to fulfill the polyconvexity condition, and take further steps to approximate the objectivity condition.  
Then, the objectivity of the model is approximated with a data augmentation approach, following \textcite{Ling2016}. Based on the existing calibration dataset for a single observer $D=\left\{\uuF,\,S,\,W\right\}$, the dataset is extended by a finite amount of additional observers according to
\begin{equation}\label{eq:data_augmentation}
\widetilde{D}=
\bigcup\limits_{\uuQ\in \cG}\left\{\uuQ\, \uuF,\,\uuQ\,S,\,W\right\},\quad \cG\subset \SO(3)\,.
\end{equation}
For pragmatic reasons, a finite amount of randomly distributed rotation matrices is chosen. 
\tDK{The data augmentation approach can be visualized as follows: 
Without data augmentation, the constitutive model is calibrated with a single observer. As \textit{ab initio}, the model does not know how to extrapolate the learned material behavior to other observers, it is only applicable to the observer chosen in the calibration dataset. When the data augmentation approach \eqref{eq:data_augmentation} is applied, the model is calibrated with multiple observers. If the model is then evaluated with an arbitrary observer, the model will yield reasonable results, as the material behavior for any observer can be seen as the interpolation between the observers with which the model was calibrated.}
With a sufficient number of observers, the model can be trained to behave nearly independent of the observer, as will be demonstrated in the upcoming examples. 

\medskip

In fact, for hyperelastic potentials, it is sufficient to apply the objectivity condition \eqref{eq:material_objectivity} to the potential only, which directly implies the transformation rule for the stress tensor. In \textcite{Ling2016} only the potential values were extended and no visualization of evaluation of the stress values was provided. However, the model quality can benefit from the additional information the stress tensor provides, and therefore, both quantities are used for the data augmentation in this work.
While this approach increases the size of the training data and thus the required calibration time, it is important to note that the time required for the model evaluation is not affected, and that no additional simulation or experimental data is required.

\medskip

\tDK{
Considering \eqref{eq:material_objectivity} and \eqref{eq:stress_objectivity}, objectivity could be further enforced by adding terms of the form $\vert W(Q\,F)-W(F)\vert^2$ and $\left\lVert S(Q\,F)-Q\,S(F)\right\rVert^2$ to the objective function used for calibration of the model. As, in this approach, both $W$ and $S$ are received by the constitutive model, the choice of $F$ is not restricted to deformation states used in the calibration dataset. Consequently, $F$ may be sampled in a sensible range of deformations in which the model is to be applied, e.g., following the sampling strategy as proposed by \textcite{kunc2019}, together with a finite amount of random rotation matrices $Q$. 
This approach takes into account that objectivity is a physical requirement, which must be fulfilled independent of stress-strain data available for a specific material.
However, in the present work, no additional term is added to the objective function, as the objectivity can already be approximated very well with the data augmentation approach \eqref{eq:data_augmentation}, as will be shown in Section \ref{sec:bcc}.
}

%% file: chapter/04_application_to_cubic_lattice_metamaterials.tex
\section{Application to cubic metamaterials}
\label{sec:application_to_cubic_metamaterials}

%%%%%%%%%%%%%%%%%%%%%%%%%%%%%%%%%
%   "Homogenized behavior of soft beam-lattice metamaterials"
%%%%%%%%%%%%%%%%%%%%%%%%%%%%%%%%%

\subsection{Homogenized behavior of soft beam-lattice metamaterials}

The performance of the models proposed in the previous section is now examined in application to the homogenized behavior of  beam-lattice structures with cubic anisotropy. In \textcite{Fernandez2020}, the homogenized behaviors of the cubic BCC cell and the cubic X cell are numerically investigated for several calibration and test scenarios, with full data availability on the public GitHub repository \tDK{\url{https://github.com/CPShub/sim-data}.} The mechanical behavior of these metamaterials is highly nonlinear and exhibits several challenging characteristics like lattice instabilities, which makes it a good benchmark case for the models. 

\begin{figure}[tp]
\centering
\begin{subfigure}[b]{0.22\textwidth}
\includegraphics[width=\textwidth]{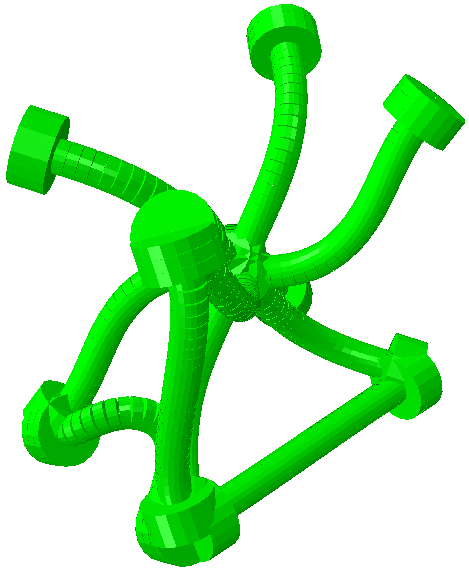}
\caption{$F_{11}=0.63$}
\label{fig:BCC_comp_buckling}
\end{subfigure}
\quad\quad
\begin{subfigure}[b]{0.24\textwidth}
\includegraphics[width=\textwidth]{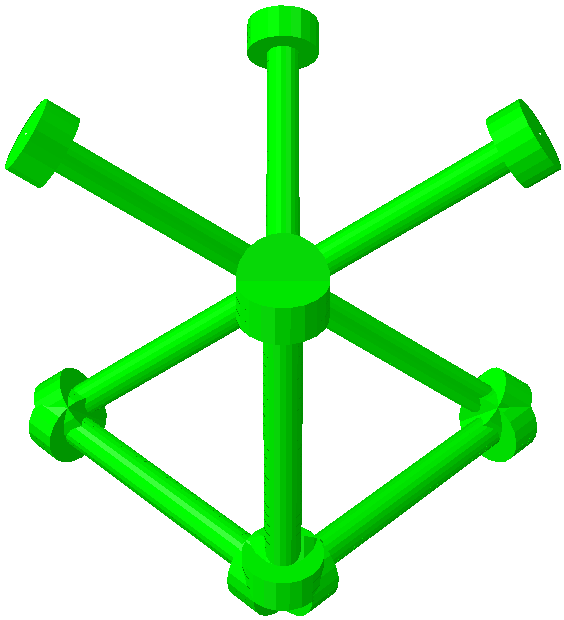}\caption{$F_{11}=1.00$}
\label{fig:BCC_undeformed}
\end{subfigure}
\quad\quad
\begin{subfigure}[b]{0.26\textwidth}
\includegraphics[width=\textwidth]{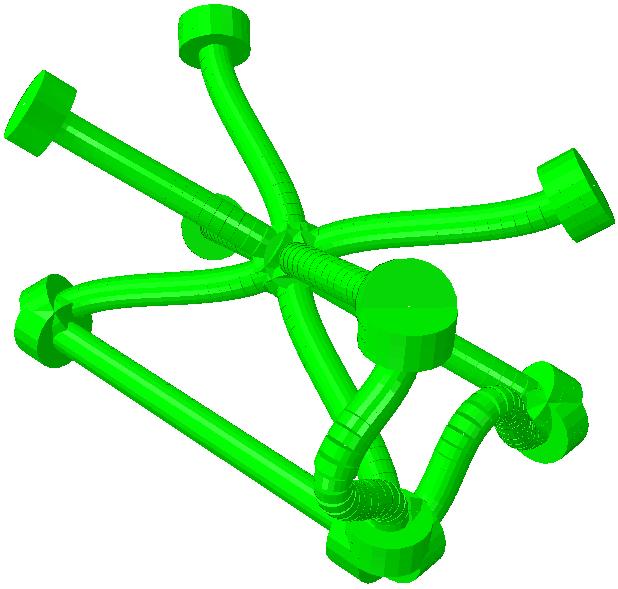}
\caption{$F_{11}=1.50$}
\label{fig:BCC_tension_buckling}
\end{subfigure}
\begin{subfigure}[b]{0.1\textwidth}
\includegraphics[width=\textwidth]{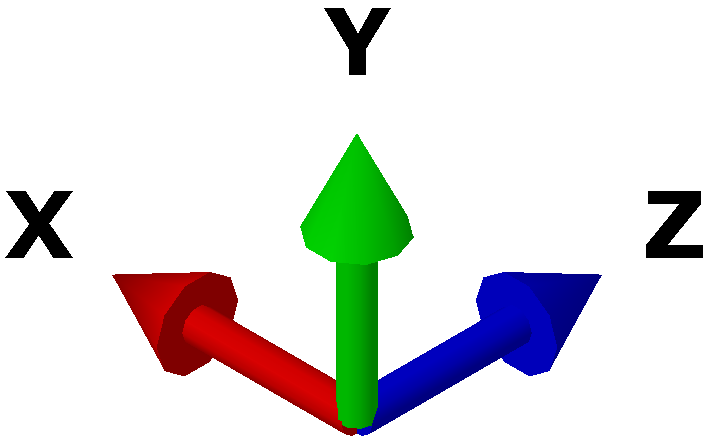}
\end{subfigure}
  \caption{Uniaxial deformation in $x$-direction for the BCC unit cell.\protect\footnotemark}
  \label{fig:BCC_deformed}
\end{figure}

The cubic BCC structure consists of body-centered beams with additional beams along all edges. Identifying the smallest unit from which the structure is built, the BCC unit cell is obtained which, regarding the periodicity, contains beams only at three edges, see Fig.~\ref{fig:BCC_undeformed}. For large structures built from lattice metamaterials, it is convenient to formulate a constitutive model for the homogenized behavior of the cells, instead of simulating every single beam of the structure. 

\footnotetext{Figure from \textcite{Fernandez2020}.}

To determine the homogenized behavior of the BCC unit cell, finite element simulations are carried out, where effective deformation gradients $\uuF$ are applied on the cell using displacements on the outer nodes of the beams. With periodic boundary conditions, the behavior of the unit cell within a larger structure can be simulated. The averaged, or effective, strain energy density $W$ can directly be calculated with the strain energy stored in the beams and the size of the cell. For the structures under consideration, the effective stress response $S$ of the cell can be computed with the reaction forces on the outer nodes and the size of the cell \cite{Fernandez2020}. In contrast to physical experiments, this numerical characterization yields not only the stress response of the material, but also the strain energy density, and the resulting dataset
\begin{equation}
D=\left\{(\uuF_1,\,{S}_1,\,{W}_1\right),\,\dotsc\}
\end{equation}
consists of triplets for corresponding deformation gradient, effective stress and effective strain energy density for each simulation step and applied deformation.

For the calibration data used here, uniaxial, equibiaxial, planar, shear, and volumetric deformations were applied on the unit cell.
These deformations can also be applied in physical experiments. The calibration dataset $D_C$ consists of $905$ triplets. Beside the calibration data, three test cases were examined. For the first two test cases, biaxial deformations with different stretch ratios are applied on the unit cell, while the third test case is a combination of tension and shear deformation. All three test cases exhibit lattice instabilities and complex deformations, which are not included in the calibration data. The test dataset $D_T$ consists of $605$ datapoints. Further technical details on the simulations can be found in \cite{Fernandez2020}. In the present work, the first test case of \cite{Fernandez2020} is referred to as \enquote{biaxial test}, while the third test case is referred to as \enquote{mixed test}.
\tDK{Especially the \enquote{mixed test} is a good benchmark case, as the combination of tension and shear is a very general deformation.}

Due to the soft materials and high slenderness of the beams, lattice instabilities occur for several deformation modes. In \textcite{Fernandez2020}, the microstructure simulation were carried out with an experimentally validated nonlinear post-buckling analysis approach \cite{Jamshidian2020}. Taking a closer look at the uniaxial deformation of the BCC cell, it exhibits instabilities in both compression and tension, which are shown in Fig.~\ref{fig:BCC_deformed}. Even for small uniaxial compression, the load bearing beams show instabilities, while for tension the less stressed beams on the edges show instabilities, which results in a highly differing behavior of the cell for compression and tension, leading to distinctive changes in the slope of the stress components, see Fig.~\ref{fig:BCC_uniaxial_deformation}. This highly challenging behavior is observed for all deformation modes, in both the calibration and test cases. 

\input{chapter/figures/BCC_uniaxial_deformation.tex}

The properties of the constitutive models are examined with the BCC cell, the behavior for the cubic X cell is only shortly discussed. The X cell consists of body-centered beams, without additional beams at the edges like the BCC cell. Therefore, less lattice instabilities occur compared to the BCC cell. While uniaxial and equibiaxial deformation gradients are quite similiar, their stress response for the X cell differ by a factor of ten, which is a challenging behavior for constitutive models.
\tDK{Apart from this case, the stress response of all load scenarios has the same order of magnitude. Furthermore, the amount of data points is roughly equal for all load scenarios. Therefore, for the following investigations, no data normalization is applied.} 
\tOW{If necessary, strategies such as the $L^2$ normalization of deformation cases proposed in \cite{fernandez2021} could be applied.}

%%%%%%%%%%%%%%%%%%%%%%%%%%%%%%%%%
%   "Preparation of the models"
%%%%%%%%%%%%%%%%%%%%%%%%%%%%%%%%%

\subsection{Preparation of the models} \label{subsec:prep}
\paragraph{Analytical model ($W^{\text{SNE}}_{\text{cub}}$)}
The model proposed by \textcite{Schroeder2010} is used as an example for conventional polyconvex models formulated in terms of invariants. For the potential of the model \eqref{eq:SNE}, $n=2$ summands are used.

\paragraph{FFNN model by \textcite{Fernandez2020} ($W^{\text{C}}$)}
The potential model proposed by \cite{Fernandez2020} is used as an example for FFNN based models which are not polyconvex by construction. The model is similar to the one discussed in section \ref{sec:polyconvex_model_F}, with the difference that the six independent components of the right Cauchy-Green tensor are used as input, therefore the model is objective by construction. 
Following \cite{Fernandez2020}, the models core is built from three layers with 16 nodes using Softplus functions in each layer, and unrestricted parameters. 

\paragraph{Polyconvex ICNN model based on invariants ($W^{\text{I}}$)} The set of cubic invariants introduced in \cite{Schroeder2010}, see eq.~\eqref{eq:invariants}, is used, with the additional invariant $I_3^*=-2\sqrt{I_3}$. This additional polyconvex invariant can be received from the last summand in eq.~\eqref{eq:SNE}, and is important for the model to represent negative stress values. Using four isotropic invariants and two cubic invariants, the input of the neural network is the vector of invariants
$
{\mathcal{I}}=\left(I_1,\,I_2,\,I_3,\,I_3^*,\,J_7,\,J_{11}\right)\in\mathbb{R}^6
$.
An ICNN core based on Softplus ($\SP$) functions, see \eqref{eq:SP}, is used, with the number of layers in $\{2,\,3\}$ and the number of nodes in $\{8, \,16,\, 32\}$. The weights in all layers are non-negative.
The restrictions on the networks parameters, which are caused by the convexity condition, are discussed in proposition \ref{prop:LSE_SP_cores}. For a compact notation, the core is referred to as e.g. $\SP\left[8,\,8\right]$ for a core containing two layers with eight nodes with $\SP$ functions in each layer.

\paragraph{Polyconvex ICNN model based on the deformation gradient ($W^{\text{F}}$)} For the second model, the input $\left(\uuF,\,\det \uuF\right)\in\mathbb{R}^{10}$ is chosen. Several simulations showed that, for the metamaterials under consideration, the deformation gradient alone is not enough to represent the material behavior, while there is no improvement when using all arguments of $\left(\uuF,\,\cof\uuF,\,\det\uuF\right)$. The group symmetrization is carried out with the cubic group $\mathcal{G}_7\subset\SO(3)$, which contains 24 orthogonal transformations \cite{Ebbing2010}. The network architectures are chosen as described above for the invariant based model $W^{\text{I}}$, but with arbitrary weights in the first layer, and non-negative weights in all subsequent layers, c.f.~proposition~\ref{prop:LSE_SP_cores}.

\paragraph{Performance measures}
While, from experimental investigations of materials, only stress values are available, the numerical evaluation of the unit cells yields both effective energy density and the stress tensor. For a given dataset $D$, the mean squared error (MSE) of the constitutive model~$W^\square,\,\square\in\left\{ \text{SNE},\, \text{C}, \,\text{I},\,\text{F}\right\}$, is defined as
\begin{equation}\label{eq:MSE}
\begin{aligned}
\text{MSE}^\square\left(\up\right)=\frac{1}{\#\left(D\right)}\sum_{\uuF\in D}\Bigg[  \frac{1}{(\mathrm{J/m^3})^2}\left({W}\left(\uuF\right)-W^\square\left(\uuF;\,\up\right)\right)^2
+\frac{1}{9\text{Pa}^2}\left\lVert{S}\left(\uuF\right)-S^\square\left(\uuF;\,\up\right)\right\rVert^2\Bigg]
\end{aligned}
\end{equation}
where $\#\left(D\right)$ denotes the number of datapoints in $D$. The parameters $\up$ of the models are found as the minimizers of the corresponding MSE.
However, since the minimization of the MSE is a non-convex optimization problem, a local minimum $\up$ may strongly depend on the initial guess and the optimization algorithm. Thus, we introduce the mean deviation of two model instances ($\widetilde{\text{MD}}$) with parameters $\up_i$ and $\up_j$ on a set of deformation gradients $D^F$ as
\begin{equation}\label{eq:MD_tilde}
\begin{aligned}
\widetilde{\text{MD}}^\square\left(\up_i,\,\up_j\right)=\frac{1}{\#\left(D^F\right)}\sum_{\uuF\in D^F}\Bigg[ &  \frac{1}{(\mathrm{J/m^3})^2}\left(W^\square\left(\uuF;\,\up_i\right)-W^\square\left(\uuF;\,\up_j\right)\right)^2
\\
&+\frac{1}{9\text{Pa}^2}\left\lVert S^\square\left(\uuF;\,\up_i\right)-S^\square\left(\uuF;\,\up_j\right)\right\rVert^2 \Bigg]\,.
\end{aligned}
\end{equation}
Then, the overall mean deviation of a model (MD) with $n$ instances is obtained as the averaged $\widetilde{\text{MD}}$ of all possible combinations of model instances:
\begin{equation}\label{eq:MD}
   \text{MD}^\square\left(\up_1,\,\ldots,\,\up_n\right)=
\begin{pmatrix}n\\2\end{pmatrix}^{-1}\;
\sum_{i=1}^{n-1}\sum_{j=i+1}^n \widetilde{\text{MD}}^\square\left(\up_i,\,\up_j\right)\,.
\end{equation}
For the dataset used to calibrate the parameters, different sets of experiments may be applied on a material. As long as each dataset contains all the required information, different calibration datasets should yield the same model behavior. However, for models whose parameters have no physical interpretation (which is the case for ML based models), different calibration datasets may lead to a different model behavior. Thus, we introduce the mean deviation ($\overline{\text{MD}}$) between a single model instance calibrated on the dataset $D_1$ with parameters $p_0$ and multiple model instances trained on another dataset $D_2$ with parameters $\up_1,\,\ldots,\,\up_n$:
\begin{equation}\label{eq:MD_bar}
    \overline{\text{MD}}^\square\left(p_0,\,\up_1,\,\ldots,\,\up_n\right)=\frac{1}{n}\sum_{i=1}^n\widetilde{\text{MD}}^\square\left(\up_0,\,\up_i\right)\,.
\end{equation}

\paragraph{Implementation}
The model $W^\text{SNE}$ has been implemented in MATLAB R2021a, the machine learning models were implemented in TensorFlow 2.3.0 with Python 3.7. 
Each machine learning model has been trained with the full batch of training data and the ADAM optimizer using default settings, every architecture was initialized three times. The models $W^\text{C}$ and $W^\text{I}$ were trained for $20,000$ epochs using the calibration dataset $D_C$. The model based on the deformation gradient $W^\text{F}$ was trained for $15,000$ epochs with the calibration dataset $D_C$, and retrained for $2,000$ epochs with the extended calibration dataset $\widetilde D_C$ according to eq.~\eqref{eq:data_augmentation}.  From our experience, this strategy provides a good balance between accuracy, speed of convergence and computation time required. Since the material objectivity of the model $W^{\text{F}}$ is only approximated, both MSE and MD of the model are evaluated with $1,024$ random observers. \tDK{For the additional observers, uniformly distributed rotation matrices are generated using the \enquote{Spatial Transformation} package provided by SciPy \cite{scipy}.} For the BCC cell, the models are also trained one time with the adapted calibration dataset $D_C^*$ which contains the shear and volumetric deformation cases (from $D_C$), as well as the biaxial and mixed test cases (from $D_T$), following the training strategies given above.

\paragraph{Reproducibility of the models}
The potentials can be reproduced with the information of the network's architecture and parameters, i.e., the number of layers and neurons, the connection of the neurons, choice of input and output, activation functions, as well as the weights and biases for each neuron. 
However, the \emph{training} of the model, i.e., the process of parameter calibration of the weights and biases, is of utmost importance for its performance.
As mentioned above, this calibration corresponds to solving a non-convex optimization problem, for which minima are usually local and may strongly depend on the initial guess and the optimization algorithm. Thus, for the same network architecture, typically several parameter sets are determined based on different random initializations, which renders the calibration process non-reproducible. 
To meet this difficulty, we provide not only the calibration data, but also a compiled version of our TensorFlow model instances as well as their sets of parameters in the public GitHub repository \tDK{\url{https://github.com/CPShub/sim-data}}. With the information about the networks architecture and the parameters used for the potential, it can be reproduced in any program, c.f.~\eqref{eq:FFNN}.

%%%%%%%%%%%%%%%%%%%%%%%%%%%%%%%%%
%   "Model evaluation for BCC cell"
%%%%%%%%%%%%%%%%%%%%%%%%%%%%%%%%%

\subsection{Model evaluation for BCC cell}
\label{sec:bcc}

The performance of the four models calibrated with the training data for the BCC cell is now discussed.
First of all, the evaluation of the analytical model $W^{\text{SNE}}$ only in the uniaxial deformation case is shown in Fig.~\ref{fig:BCC_SNE}. As can be seen, even for this simple case the model fails to represent the material behavior. Thus, this model is not considered for further, more detailed performance evaluations.

\input{chapter/figures/BCC_SNE.tex}

Before the machine learning models can be compared, the required amount of observers for the data augmentation of the model $W^{\text{F}}$ must be examined. The model $W^{\text{F}}$ is calibrated with the full calibration dataset $\widetilde D_C$, using $8$, $16$, $32$ and $64$ random observers, see eq.~\eqref{eq:data_augmentation}, and evaluated after calibration with $1,024$ random observers. The corresponding MSEs \tDK{and relative calibration times}
are shown in Table \ref{tb:loss_variation_of_observers}, and the shear response for each calibration is illustrated in Fig.~\ref{fig:BCC_variation_of_observers}. \tDK{While, for a calibration with $64$ observers, the dataset is eight times as big as the dataset for eight observers, its calibration time is just under six times as long. Although data augmentation increases the calibration time, for increasing datasets, this approach profits from the fast evaluation of ANNs for large batch sizes.} If the model response depends on the choice of observer, the evaluation yields an area containing the stress response for all different observers, whereas for sufficiently well approximated objectivity, minimum and maximum model response coincide, and the area practically reduces to a single line. For a calibration with eight observers, the individual stress components for the shear deformation are within a wide range, and the MSEs are very high. Therefore, the material objectivity is not fulfilled. 
Using more observers for the data augmentation, the approximation quality of the objectivity increases. For $64$ observers in the calibration dataset, the shear responses for different observers are indistinguishable from each other, and the MSEs are sufficiently small. This shows that objectivity can be successfully learned using the data augmentation approach. In the following, the model $W^{\text{F}}$ is calibrated using $64$ observers, and all evaluations for the model are carried out with $1,024$ random observers.

\input{chapter/figures/BCC_variation_of_observers.tex}

\medskip

\begin{table}[tp]
\centering
\begin{tabular}{cllc}
\toprule
Observers for the& \multicolumn{2}{l}{MSE} &Relative cali-\\ 
calibration dataset&$D_T$&$D_C$ &bration time\\ \midrule
$8$ & $1.25\cdot10^4$ & $8.64\cdot10^3$ & $1$ \\
$16$ & $3.97\cdot10^3$ & $1.65\cdot10^3$ & $1.98$\\
$32$ & $3.42\cdot10^3$ & $6.84\cdot10^2$ & $2.95$\\
$64$ & $7.61\cdot10^2$ & $2.57\cdot10^2$ & $5.70$\\
\bottomrule
\end{tabular}
\caption{MSEs of calibrated $W^\text{F}$ models for the BCC cell with augmented data for learning objectivity. Evaluation with $1,024$ random observers. \tDK{Calibration times related to calibration time for $8$ observers.}}
\label{tb:loss_variation_of_observers}
\end{table}

In Table \ref{tb:loss_BCC}, the three best MSEs for the three machine learning models trained on $D_C$ and $\widetilde{D}_C$ are depicted and sorted for decreasing error on the test dataset $D_T$. The model $W^{\text{F}}$ yields far better results than $W^{\text{I}}$ for both calibration and test datasets, which may be caused by a smaller function space of the model $W^{\text{I}}$ due to the human choice of invariants. While the MSEs of $W^\text{F}$ are slightly higher than the ones of $W^\text{C}$ on $D_C$, which is reasonable since the polyconvex model is more restricted in fitting the calibration data than the non-polyconex one, it actually performs slightly better on the test data. This could be rooted in the additional mathematical structure that polyconvexity incorporates into the model, making it more generalizable.

As already encountered in \cite{Fernandez2020}, different neural network architectures using a vastly differing amount of parameters may lead to models of equal quality, with different results for multiple initializations using the same architecture. Also, the stress error in eq.~\eqref{eq:MSE} dominates the overall MSE for all evaluations. 
For the convex neural networks used in this work, the models need a sufficiently high amount of nodes and layers to counteract the loss of flexibility due to the restrictions on the parameters. For the model $W^{\text{F}}$, $\SP\left[16,\,16,\,16\right]$ is the smallest architecture yielding good results, while there were no significant benefits for architectures using more layers or nodes. 

\begin{table}[t]
\centering
\begin{tabular}{llrcllll}
\toprule
\multicolumn{2}{l}{Model} &  Param.& Calibr. &\multicolumn{2}{l}{Deviations}& \multicolumn{2}{l}{MSE}  \\ 
&&&data&$D_T^F$&$D_C^F$&$D_T$&$D_C$\\ \midrule
$W^{\text{C}}$ with&$\SP\left[16,\,16,\,16\right]$  & $673$ &$D_C$ & && $1.02\cdot10^3$ & $1.59\cdot10^2$  \\
&\multicolumn{3}{c}{------\textquotedbl------}& && $1.40\cdot10^3$ & $1.58\cdot10^2$\\
&\multicolumn{3}{c}{------\textquotedbl------}& && $1.41\cdot10^3$ & $1.63\cdot10^2$\\
 &$\SP\left[16,\,16,\,16\right]$  & $673$ &$D_C^*$ & && $2.40\cdot10^2$ & $4.18\cdot10^3$  \\
\cmidrule{2-6}
&&&${\text{MD}}$&$2.05\cdot10^3$&$5.91\cdot 10^0$&&\\
&&&$\overline{\text{MD}}$&$1.47\cdot10^3$&$4.21\cdot 10^3$&&\\
\midrule
$W^{\text{I}}$ with &$\SP\left[32,\,32,\,32\right]$   & $2,369$&$D_C$ & && $1.33\cdot10^4$ & $1.11\cdot10^4$  \\
&$\SP\left[32,\,32\right]$ &$1,313$  &---\textquotedbl---&& & $1.41\cdot10^4$ & $1.10\cdot10^4$  \\
&$\SP\left[16,\,16\right]$ &$401$  &---\textquotedbl---& && $1.44\cdot10^4$ & $1.12\cdot10^4$  \\
&$\SP\left[16,\,16\right]$ &$401$  &$D_C^*$& && $1.77\cdot 10^5$ & $1.30\cdot 10^5$  \\
\cmidrule{2-6}
&&&${\text{MD}}$&$2.37\cdot 10^2$&$2.93\cdot10^2$&&\\
&&&$\overline{\text{MD}}$&$3.47\cdot 10^3$&$1.84\cdot 10^3$&& \\
\midrule
$W^{\text{F}}$ with&$\SP\left[16,\,16,\,16\right]$&$737$ &$\widetilde{D}_C$&  && $7.47\cdot10^2$ & $2.45\cdot10^2$ \\
&\multicolumn{3}{c}{------\textquotedbl------}& && $7.74\cdot10^2$ & $3.43\cdot10^2$ \\
&\multicolumn{3}{c}{------\textquotedbl------}& && $8.05\cdot10^2$ & $2.62\cdot10^2$\\
&$\SP\left[16,\,16,\,16\right]$&$737$ &$\widetilde{D}_C^*$&  && $6.81\cdot 10^2$ & $5.42\cdot 10^2$ \\
\cmidrule{2-6}
&&&${\text{MD}}$&$1.60\cdot 10^2$&$1.11\cdot 10^2$&&\\
&&&$\overline{\text{MD}}$&$4.36\cdot 10^2$&$2.66 \cdot 10^2$&& \\
\bottomrule
\end{tabular}
\caption{Deviations of model instances and MSEs of calibrated ML models for the BCC cell. MD evaluates the deviation for the instances trained on $D_C$, while $\overline{\text{MD}}$ compares the deviation of the instance trained on $D_C^*$ to the instances trained on $D_C$.}
\label{tb:loss_BCC}
\end{table}

Furthermore, in Table \ref{tb:loss_BCC} the mean deviations for the three machine learning models are shown, evaluating the deviations of the different instances of each individual model. The deviations are evaluated for the datasets $D_C^F$ and $D_T^F$, which contain all deformation gradients applied in the calibration dataset $D_C$ and test dataset $D_T$, respectively. MD evaluates the deviation for the instances trained on $D_C$, while $\overline{\text{MD}}$ compares the deviation of the single instance trained on the dataset $D_C^*$ to the instances trained on $D_C$.
For the model $W^\text{C}$ the MD on $D^F_C$ is very low, which may be attributed to the high flexibility of the unrestricted network core, allowing the model to be very similar on $D_C$ for every initialization. However, for $D^F_T$, the MD of the model $W^\text{C}$ is worse than its MSE on $D_T$. 
Furthermore, for both $D_T^F$ and $D_C^F$, the model behavior depends on the dataset used for the calibration, which leads to a high $\overline{\text{MD}}$. Actually, $\overline{\text{MD}}$ on $D_C^F$ is even three magnitudes higher than MD.
For the model $W^{\text{I}}$, the MD on both datasets is about two magnitudes smaller than its MSEs on the datasets. This may be caused by the additional mathematical structure that polyconvexity and invariants incorporate into the model. While the deviation of multiple instances trained on the same dataset is very low, the training on the adapted dataset $D_{{C}}^*$ leads to a different model behavior, which results in a high $\overline{\text{MD}}$. The reason for this is that the model is not flexible enough to capture the material behavior.
The model $W^{\text{F}}$ has a flexible input and, due to the polyconvexity condition, a pronounced mathematical structure. This leads to excellent results for both the MSE and the MD values, making it the only model of this comparison which has both excellent approximation qualities and a consistent behavior within multiple model instances. Furthermore, even for a calibration with the adapted dataset $\widetilde{D}_C^*$ the model behavior stays consistent. The deviation $\overline{\text{MD}}$ has the same magnitude as MD, and also the MSEs for the instances trained with different datasets have the same magnitude.
For the following evaluations, the core  $\SP\left[16,\,16\right]$ is used for  $W^{\text{I}}$, while for $W^{\text{C}}$ and $W^{\text{F}}$ the core with the smallest MSE is chosen. 

\medskip

In Fig.~\ref{fig:BCC_C_model}, the uniaxial deformation case and mixed test case for $W^{\text{C}}$ are shown to examine the ellipticity of the model. As already encountered in \textcite{Fernandez2020}, the model shows excellent approximation properties. However, in both examined cases in Fig.~\ref{fig:BCC_C_model}, the model loses its ellipticity even for small deformations, which causes material instability and would lead to major drawbacks in numerical applications such as the finite element method. Consequently, for the metamaterials under consideration, it is important to include the polyconvexity condition into the model formulation, as it implies ellipticity, and thus ensures material stability. 
Typically, for soft materials, one does not expect loss of material stability for a deformation gradient $F$ in a bounded set including the identity deformation $F=\uuI$ (even if large elastic strains may occur). This is, e.g., the case for isotropic elastic energies defined in the logarithmic strain tensor \cite{neff2016,neff2015,schroeder2018}. The loss of ellipticity in these models occurs only for extremely large strains that cannot be observed in experiments.

\input{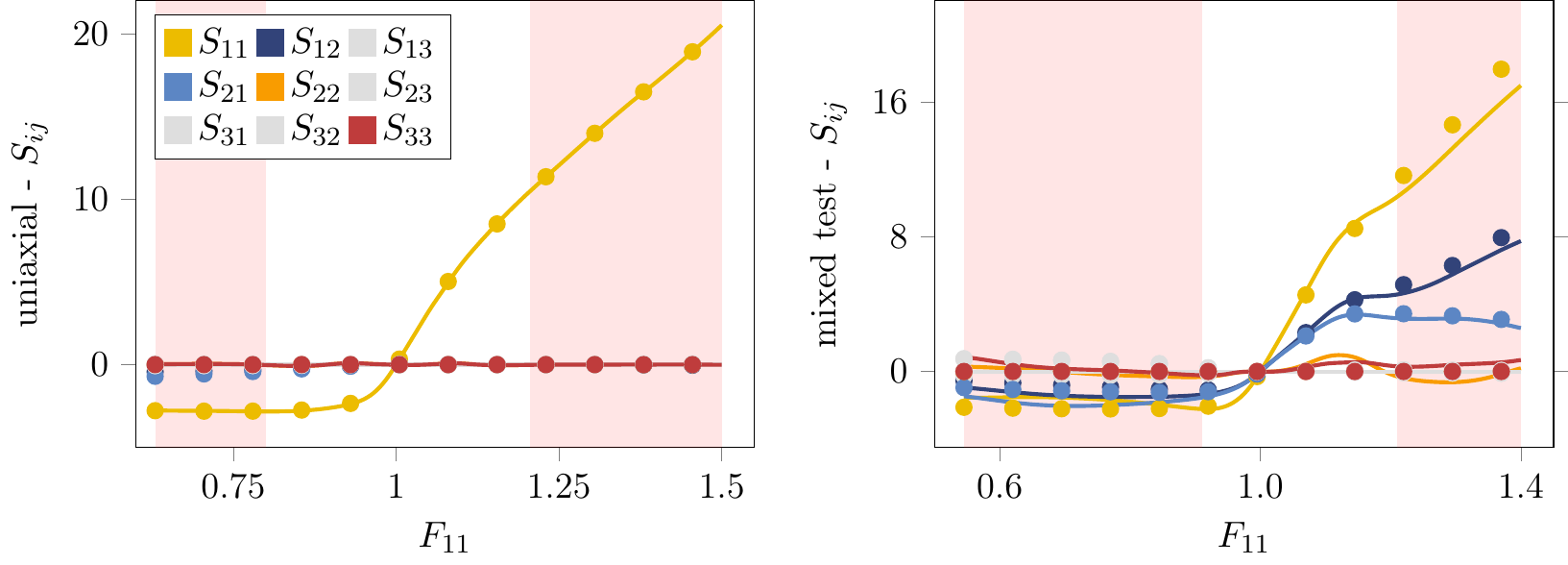}

\input{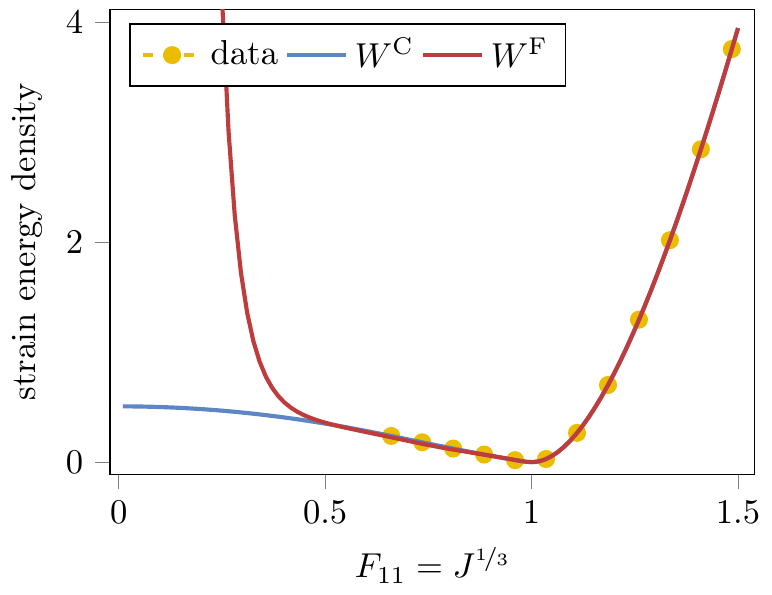}

\medskip

In Fig.~\ref{fig:BCC_volumetric}, the behavior of $W^{\text{C}}$ and $W^{\text{F}}$ is examined for volumetric tension and compression. 
Regarding the volumetric growth condition \eqref{eq:growth}, the strain energy density should rapidly grow for $J=\det\uuF\rightarrow 0^+$. 
The model $W^{\text{F}}$ contains the term $W_\text{vol}$ from \eqref{eq:coercive} and thus fulfills the growth condition. In the formulation of $W^{\text{C}}$, the growth condition is not considered, and consequently it is violated for the extrapolation $J\rightarrow 0^+$. \tDK{For the metamaterials under consideration, lattice instabilities may lead to high volumetric compression, therefore the growth condition should be included in the model formulation.
It must be emphasized that both the loss of ellipticity and the nonphysical behavior for $J\rightarrow 0^+$ are no specific drawbacks of the model $W^{\text{C}}$ as proposed by \textcite{Fernandez2020}. It is very likely that other ML-based constitutive models would show the same behavior when calibrated to the examined data, as long as ellipticity and the volumetric growth condition are not explicitly considered in the model formulation.}

\input{chapter/figures/BCC_invariant_model.tex}
\input{chapter/figures/BCC_deformation_gradient_model.tex}

\medskip

In Fig.~\ref{fig:BCC_invariant_model}, a subset of the calibration and test cases for the model $W^{\text{I}}$ is shown to examine some model characteristics. The model yields acceptable results for deformation gradients with dominating main diagonal elements. For the uniaxial and equibiaxial calibration case, the component $S_{33}$ shows a large deviation from the simulation data, which also transfers to the biaxial test case. For shear deformation, the model completely fails to represent the simulation data, consequently, it also fails to represent the mixed test case.
While, basically, the set of invariants for the model could be extended, it is unlikely that the model behavior for the cubic metamaterials under considerations can be improved with this approach. While the model shows drawbacks especially for shear deformations, the additional invariants can only use main diagonal elements of $\uuC$ in order to be polyconvex, which makes it hard to gain flexibility for shear deformations. \tDK{However, for a wide range of materials with less challenging behavior, the invariant-based model may still be a good choice, which is demonstrated in Section~\ref{sec:application_to_ti} for transverse isotropy.}

\medskip

In Fig.~\ref{fig:BCC_deformation_gradient_model}, a subset of the calibration and test cases for the model $W^{\text{F}}$ is shown to examine the model's characteristics. The model shows excellent results for every deformation mode of the calibration dataset and the test dataset, with only small deviations of the $S_{12}$ component for the mixed test case. After recalibration of the model with the concatenation of calibration and test dataset for $1,000$ epochs, the model can perfectly represent the simulation data for both calibration and test data, which is not shown in Fig.~\ref{fig:BCC_deformation_gradient_model}. The data augmentation approximates the material objectivity so well, that 
no dependence on the observer can be seen at all.

%%%%%%%%%%%%%%%%%%%%%%%%%%%%%%%%%
%   "Model evaluation for X cell"
%%%%%%%%%%%%%%%%%%%%%%%%%%%%%%%%%

%\newpage

\subsection{Model evaluation for X cell}
\label{sec:xcell}

\begin{table}[t]
\centering
\begin{tabular}{llrcllll}
\toprule
\multicolumn{2}{l}{Model} &  Param.& Calibr. &\multicolumn{2}{l}{Deviations}& \multicolumn{2}{l}{MSE}  \\ 
&&&data&$D_T^F$&${D}_C^F$&$D_T$&$D_C$\\ \midrule
$W^{\text{F}}$ with&$\SP\left[16,\,16,\,16\right]$& $737$ &$\widetilde{D}_C$& && $5.41\cdot10^2$ & $2.38\cdot10^2$ \\
&\multicolumn{3}{c}{------\textquotedbl------}& && $5.62\cdot10^2$ & $2.61\cdot10^2$ \\
&\multicolumn{3}{c}{------\textquotedbl------}& && $8.82\cdot10^2$ & $2.54\cdot10^2$\\
\cmidrule{2-6}
&&&${\text{MD}}$&$1.47\cdot 10^2$&$5.64\cdot 10^1$&&\\
\bottomrule
\end{tabular}
\caption{Deviation of model instances and MSEs of calibrated $W^\text{F}$ model for the X cell.}
\label{tb:loss_X}
\end{table}

\input{chapter/figures/X_deformation_gradient_model.tex}
\vspace*{-2mm}

In the following, for the evaluation of the performance of the models calibrated with the training data for the X cell, only the results for the model $W^{\text{F}}$ are shortly discussed; the results for the model~$W^{\text{I}}$ confirmed the observations made for the BCC cell without yielding further insights. 
The model~$W^{\text{F}}$ was initialized three times using three layers with $16$ nodes in each layer, leading to the MDs and MSEs shown in Table~\ref{tb:loss_X}. In Fig.~\ref{fig:F_model_X_cell}, the uniaxial and equibiaxial deformation mode for the model with the best $D_T$ are shown. While the model shows excellent agreement with the simulation data for almost all calibration cases, in the uniaxial tension regime, the model shows a slight dependence on the choice of observer. The reason for this is that the uniaxial and equibiaxial deformations are similar, while their stress response differs by a factor of ten. Therefore it is challenging for the model to capture the material's behavior for both uniaxial and equibiaxial deformations. This may be resolved by an adapted training strategy as proposed in \textcite{fernandez2021}, for which the stress responses of the different deformation cases are scaled to a comparable magnitude for the training. We should also remark that approximate satisfaction of objectivity does not conflict with an existence theorem based on polyconvexity and growth or coercivity conditions.

%% file: chapter/figures/BCC_uniaxial_deformation.tex
\tikzsetnextfilename{BCC_uniaxial_deformation}

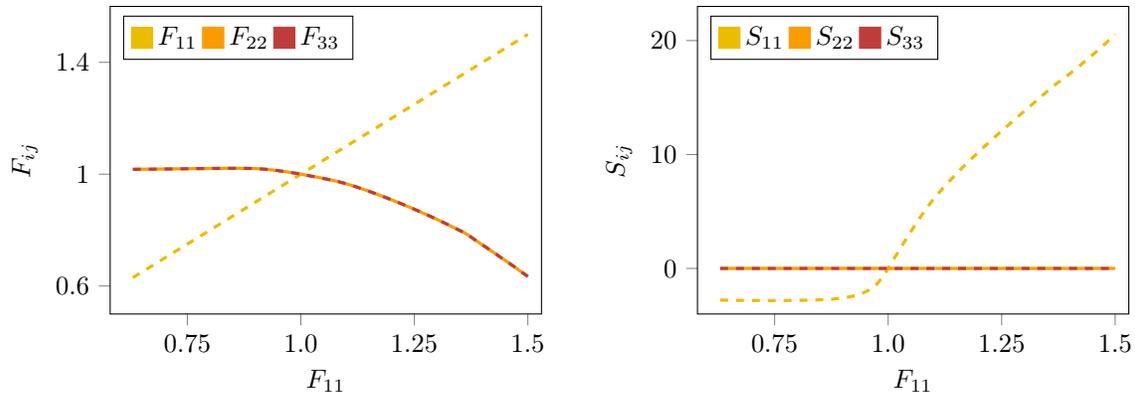
\begin{figure}[tp]
\centering
\pgfplotstableset{
create on use/uniax_F11/.style={create col/copy column from table={plot_data/BCC/calibration_data/uniaxial.txt}{0}}
}

\begin{tikzpicture}

\begin{groupplot}[
	group style = {group size = 2 by 1	, vertical sep = 0.04*\textwidth,
								horizontal sep = 0.125*\textwidth},
	cycle list name=mycolorlist123,
	xtick align = outside,
	ytick align = outside,
	xtick pos = left,
	%ytick pos = left
	]

\nextgroupplot[legend columns=3,legend pos = north west, ytick pos = left, ylabel = {$F_{ij}$}, ytick={0.6,1,1.4},yticklabels={0.6,1,1.4},
width=0.441*\textwidth, height=0.342*\textwidth,
xtick={0.75,1.0,1.25,1.5},xticklabels={0.75,1.0,1.25,1.5},
ymin = 0.5, ymax = 1.6,
xmin = 0.58, xmax = 1.53,
xlabel={$F_{11}$}] % uniax F

\foreach \num in {11, 22, 33}{
 \addplot coordinates { (-2,-2) (-3,-3) };
 \addlegendentryexpanded{$F_{\num}$}
 };
 
    \foreach \num in {0, 4, 8}{
    \addplot table [x =uniax_F11,y index=\num] 
    {plot_data/BCC/calibration_data/uniaxial.txt};
    };
	
	    \nextgroupplot[ylabel = {$S_{ij}$},ytick pos = left, legend pos = north west,legend columns=3, ytick={0,1000,2000},yticklabels={0,10,20},
width=0.441*\textwidth, height=0.342*\textwidth,
xtick={0.75,1.0,1.25,1.5},xticklabels={0.75,1.0,1.25,1.5},
ymin = -400, ymax = 2300,
xmin = 0.58, xmax = 1.53,
xlabel={$F_{11}$}] % unuiax P
\foreach \num in {11, 22, 33}{
\addplot coordinates { (-2,-2) (-3,-3) };
\addlegendentryexpanded{$S_{\num}$}
 };

    \foreach \num in {9, 13, 17}{
    \addplot table [x =uniax_F11,y index=\num] 
    {plot_data/BCC/calibration_data/uniaxial.txt};
    };

\end{groupplot}
\end{tikzpicture}

\caption{Deformation (left side) and first Piola-Kirchhoff stress $S$ (right side) for uniaxial deformation of the cubic BCC cell. Lattice instabilities express themselves by nearly horizontal stress values of $S_{11}$ (compression) and decreasing slope of $S_{11}$ (tension). Stress in $[\text{hPa}]$.}
\label{fig:BCC_uniaxial_deformation}
\end{figure}

%% file: chapter/figures/BCC_SNE.tex
\tikzsetnextfilename{BCC_SNE}

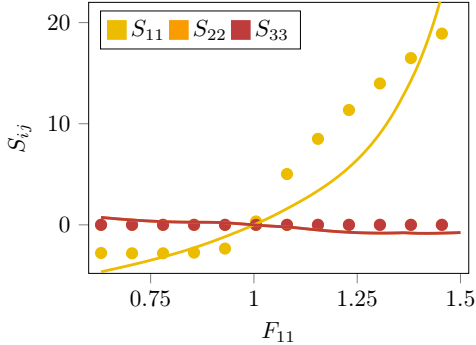
\begin{figure}[tp]
\centering
\pgfplotstableset{
create on use/uniax_F11/.style={create col/copy column from table={plot_data/BCC/calibration_data/uniaxial.txt}{0}}
}

\resizebox{0.4\textwidth}{!}{
\begin{tikzpicture}
\pgfplotsset{
  set layers,% using layers
  mark layer=axis tick labels% defines the layer of the marks
}
\begin{groupplot}[
	group style = {group size = 1 by 1, vertical sep = 0.09*\textwidth,
								horizontal sep = 0.115*\textwidth},
	cycle list name=mycolorlistSNE,
	xtick align = outside,
	ytick align = outside,
	xtick pos = left,
	%ytick pos = left
	]
\nextgroupplot[legend pos = north west,legend columns=3, ylabel = {$S_{ij}$}, ytick={0,1000,2000},yticklabels={0,10,20},
width=0.441*\textwidth, height=0.342*\textwidth,
xtick={0.75,1,1.25,1.5},xticklabels={0.75,1,1.25,1.5},
ymin = -480, ymax = 2200,
xlabel={$F_{11}$},
xmin = 0.6, xmax = 1.52,
ytick pos = left] % uniaxial invariant model
\foreach \num in {11, 22, 33}{
 \addplot coordinates { (-2,-2) (-3,-3) };
\addlegendentryexpanded{$S_{\num}$}
 };
    \foreach \num in {9, 13, 17}{
    \addplot table [x =uniax_F11,y index=\num, only marks] 
    {plot_data/BCC/calibration_data/uniaxial.txt};
    };
        \foreach \num in {0,4,8}{
    \addplot table [dashed, x =uniax_F11,y index=\num] {plot_data/SNE_BCC/P_uniaxial.txt};
    };

\end{groupplot}
\end{tikzpicture}
}

\caption{Evaluation of $W^{\text{SNE}}_{\text{cub}}$ for the BCC cell, calibrated only for the uniaxial deformation case. Points depict the simulation data, while lines depict the calibrated model; stress in $\left[\text{hPa}\right]$. The $S_{11}$ component is not well-fitted.}
\label{fig:BCC_SNE}
\end{figure}

%% file: chapter/figures/BCC_variation_of_observers.tex
\tikzsetnextfilename{BCC_variation_of_observers}

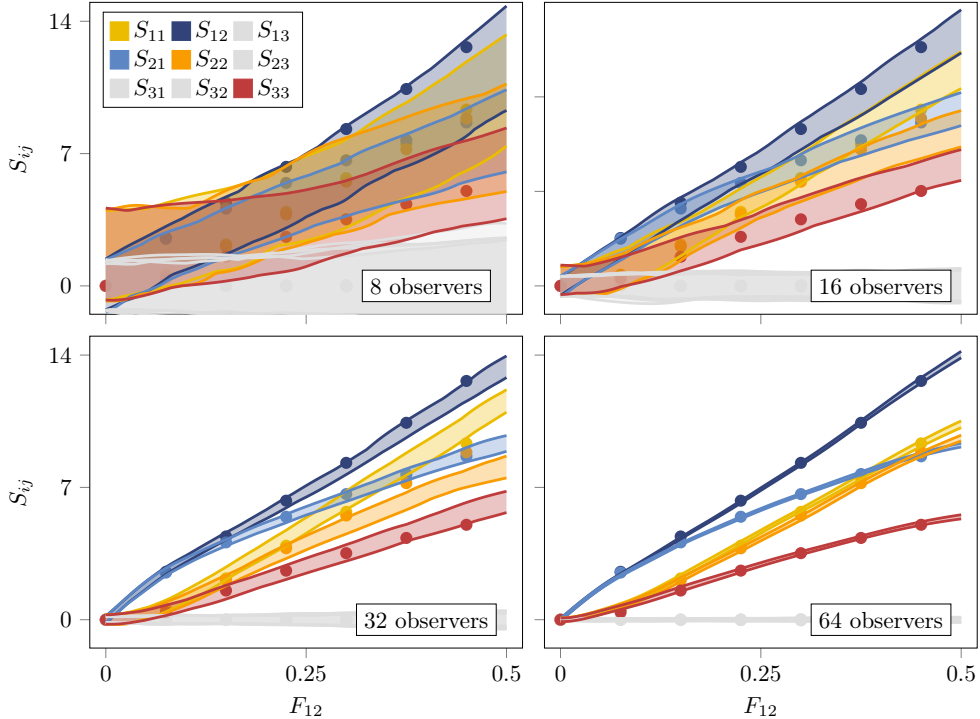
\begin{figure}[t!]
\centering
\pgfplotstableset{
create on use/shear_F12/.style={create col/copy column from table={plot_data/BCC/calibration_data/shear.txt}{1}}
}
\resizebox{0.8\textwidth}{!}{
\begin{tikzpicture}
\pgfplotsset{
  set layers,% using layers
  mark layer=axis tick labels% defines the layer of the marks
}
\begin{groupplot}[
	group style = {group size = 2 by 2, vertical sep = 0.02*\textwidth,
								horizontal sep = 0.02*\textwidth},
	cycle list name=mycolorlist,
	xtick align = outside,
	ytick align = outside,
	xtick pos = left,
	ytick pos = left
	]
  
  \nextgroupplot[align=center,legend pos = north west,legend columns=3,ylabel = {$S_{ij}$}, ytick={0,700,1400},yticklabels={0,7,14},
width=0.49*\textwidth, height=0.38*\textwidth, xlabel={},
xtick={0,0.25,0.5},xticklabels={,,},
ymin = -150, ymax = 1500,
xmin = -0.02, xmax = 0.52] % shear for 8 observers
	\foreach \num in {11, 12, 13, 21, 22, 23, 31, 32, 33}{
 \addplot coordinates { (-2,-2) (-3,-3) };
  \addlegendentryexpanded{$S_{\num}$}
 };
  \foreach \num in {9, 10, 11, 12, 13, 14, 15, 16, 17}{
   \addplot table [x =shear_F12,y index=\num, only marks] 
    {plot_data/BCC/calibration_data/shear.txt};
    };

        \def\observer{8}
\pgfplotstableread{plot_data/variation_of_observers/\observer/P_(1, 1)_shear.txt}\data
\addplot [fill=color11,mark=none, draw=none, opacity=0.3] table {\data};
\pgfplotstableread{plot_data/variation_of_observers/\observer/P_(1, 2)_shear.txt}\data
\addplot [fill=color12,mark=none, draw=none, opacity=0.3] table {\data};
\pgfplotstableread{plot_data/variation_of_observers/\observer/P_(1, 3)_shear.txt}\data
\addplot [fill=color13,mark=none, draw=none, opacity=0.3] table {\data};
\pgfplotstableread{plot_data/variation_of_observers/\observer/P_(2, 1)_shear.txt}\data
\addplot [fill=color21,mark=none, draw=none, opacity=0.3] table {\data};
\pgfplotstableread{plot_data/variation_of_observers/\observer/P_(2, 2)_shear.txt}\data
\addplot [fill=color22,mark=none, draw=none, opacity=0.3] table {\data};
\pgfplotstableread{plot_data/variation_of_observers/\observer/P_(2, 3)_shear.txt}\data
\addplot [fill=color23,mark=none, draw=none, opacity=0.3] table {\data};
\pgfplotstableread{plot_data/variation_of_observers/\observer/P_(3, 1)_shear.txt}\data
\addplot [fill=color31,mark=none, draw=none, opacity=0.3] table {\data};
\pgfplotstableread{plot_data/variation_of_observers/\observer/P_(3, 2)_shear.txt}\data
\addplot [fill=color32,mark=none, draw=none, opacity=0.3] table {\data};
\pgfplotstableread{plot_data/variation_of_observers/\observer/P_(3, 3)_shear.txt}\data
\addplot [fill=color33,mark=none, draw=none, opacity=0.3] table {\data};

\pgfplotstableread{plot_data/variation_of_observers/\observer/P_(1, 1)_shear_min.txt}\data
\addplot [color11, fill=none,mark=none, very thick] table {\data};
\pgfplotstableread{plot_data/variation_of_observers/\observer/P_(1, 1)_shear_max.txt}\data
\addplot [color11, fill=none,mark=none, very thick] table {\data};
\pgfplotstableread{plot_data/variation_of_observers/\observer/P_(1, 2)_shear_min.txt}\data
\addplot [color12, fill=none,mark=none, very thick] table {\data};
\pgfplotstableread{plot_data/variation_of_observers/\observer/P_(1, 2)_shear_max.txt}\data
\addplot [color12, fill=none,mark=none, very thick] table {\data};
\pgfplotstableread{plot_data/variation_of_observers/\observer/P_(1, 3)_shear_min.txt}\data
\addplot [color13, fill=none,mark=none, very thick] table {\data};
\pgfplotstableread{plot_data/variation_of_observers/\observer/P_(1, 3)_shear_max.txt}\data
\addplot [color13, fill=none,mark=none, very thick] table {\data};
\pgfplotstableread{plot_data/variation_of_observers/\observer/P_(2, 1)_shear_min.txt}\data
\addplot [color21, fill=none,mark=none, very thick] table {\data};
\pgfplotstableread{plot_data/variation_of_observers/\observer/P_(2, 1)_shear_max.txt}\data
\addplot [color21, fill=none,mark=none, very thick] table {\data};
\pgfplotstableread{plot_data/variation_of_observers/\observer/P_(2, 2)_shear_min.txt}\data
\addplot [color22, fill=none,mark=none, very thick] table {\data};
\pgfplotstableread{plot_data/variation_of_observers/\observer/P_(2, 2)_shear_max.txt}\data
\addplot [color22, fill=none,mark=none, very thick] table {\data};
\pgfplotstableread{plot_data/variation_of_observers/\observer/P_(2, 3)_shear_min.txt}\data
\addplot [color23, fill=none,mark=none, very thick] table {\data};
\pgfplotstableread{plot_data/variation_of_observers/\observer/P_(2, 3)_shear_max.txt}\data
\addplot [color23, fill=none,mark=none, very thick] table {\data};
\pgfplotstableread{plot_data/variation_of_observers/\observer/P_(3, 1)_shear_min.txt}\data
\addplot [color31, fill=none,mark=none, very thick] table {\data};
\pgfplotstableread{plot_data/variation_of_observers/\observer/P_(3, 1)_shear_max.txt}\data
\addplot [color31, fill=none,mark=none, very thick] table {\data};
\pgfplotstableread{plot_data/variation_of_observers/\observer/P_(3, 2)_shear_min.txt}\data
\addplot [color32, fill=none,mark=none, very thick] table {\data};
\pgfplotstableread{plot_data/variation_of_observers/\observer/P_(3, 2)_shear_max.txt}\data
\addplot [color32, fill=none,mark=none, very thick] table {\data};
\pgfplotstableread{plot_data/variation_of_observers/\observer/P_(3, 3)_shear_min.txt}\data
\addplot [color33, fill=none,mark=none, very thick] table {\data};
\pgfplotstableread{plot_data/variation_of_observers/\observer/P_(3, 3)_shear_max.txt}\data
\addplot [color33, fill=none,mark=none, very thick] table {\data};
      
                    \node[fill=white, draw=black] at (0.4, 0) {{8 observers}};

    \nextgroupplot[legend pos = outer north east,ylabel = {}, ytick={0,500,1000},yticklabels={},
width=0.49*\textwidth, height=0.38*\textwidth, xlabel={},
xtick={0,0.25,0.5},xticklabels={,,},
ymin = -150, ymax = 1500,
xmin = -0.02, xmax = 0.52] % shear for 16 observers
\foreach \num in {11, 12, 13, 21, 22, 23, 31, 32, 33}{
 \addplot coordinates { (-2,-2) (-3,-3) };
% \addlegendentryexpanded{$P_{\num}$}
 };
    \foreach \num in {9, 10, 11, 12, 13, 14, 15, 16, 17}{
    \addplot table [x =shear_F12,y index=\num, only marks]
     {plot_data/BCC/calibration_data/shear.txt};
    };

    \def\observer{16}
\pgfplotstableread{plot_data/variation_of_observers/\observer/P_(1, 1)_shear.txt}\data
\addplot [fill=color11,mark=none, draw=none, opacity=0.3] table {\data};
\pgfplotstableread{plot_data/variation_of_observers/\observer/P_(1, 2)_shear.txt}\data
\addplot [fill=color12,mark=none, draw=none, opacity=0.3] table {\data};
\pgfplotstableread{plot_data/variation_of_observers/\observer/P_(1, 3)_shear.txt}\data
\addplot [fill=color13,mark=none, draw=none, opacity=0.3] table {\data};
\pgfplotstableread{plot_data/variation_of_observers/\observer/P_(2, 1)_shear.txt}\data
\addplot [fill=color21,mark=none, draw=none, opacity=0.3] table {\data};
\pgfplotstableread{plot_data/variation_of_observers/\observer/P_(2, 2)_shear.txt}\data
\addplot [fill=color22,mark=none, draw=none, opacity=0.3] table {\data};
\pgfplotstableread{plot_data/variation_of_observers/\observer/P_(2, 3)_shear.txt}\data
\addplot [fill=color23,mark=none, draw=none, opacity=0.3] table {\data};
\pgfplotstableread{plot_data/variation_of_observers/\observer/P_(3, 1)_shear.txt}\data
\addplot [fill=color31,mark=none, draw=none, opacity=0.3] table {\data};
\pgfplotstableread{plot_data/variation_of_observers/\observer/P_(3, 2)_shear.txt}\data
\addplot [fill=color32,mark=none, draw=none, opacity=0.3] table {\data};
\pgfplotstableread{plot_data/variation_of_observers/\observer/P_(3, 3)_shear.txt}\data
\addplot [fill=color33,mark=none, draw=none, opacity=0.3] table {\data};

\pgfplotstableread{plot_data/variation_of_observers/\observer/P_(1, 1)_shear_min.txt}\data
\addplot [color11, fill=none,mark=none, very thick] table {\data};
\pgfplotstableread{plot_data/variation_of_observers/\observer/P_(1, 1)_shear_max.txt}\data
\addplot [color11, fill=none,mark=none, very thick] table {\data};
\pgfplotstableread{plot_data/variation_of_observers/\observer/P_(1, 2)_shear_min.txt}\data
\addplot [color12, fill=none,mark=none, very thick] table {\data};
\pgfplotstableread{plot_data/variation_of_observers/\observer/P_(1, 2)_shear_max.txt}\data
\addplot [color12, fill=none,mark=none, very thick] table {\data};
\pgfplotstableread{plot_data/variation_of_observers/\observer/P_(1, 3)_shear_min.txt}\data
\addplot [color13, fill=none,mark=none, very thick] table {\data};
\pgfplotstableread{plot_data/variation_of_observers/\observer/P_(1, 3)_shear_max.txt}\data
\addplot [color13, fill=none,mark=none, very thick] table {\data};
\pgfplotstableread{plot_data/variation_of_observers/\observer/P_(2, 1)_shear_min.txt}\data
\addplot [color21, fill=none,mark=none, very thick] table {\data};
\pgfplotstableread{plot_data/variation_of_observers/\observer/P_(2, 1)_shear_max.txt}\data
\addplot [color21, fill=none,mark=none, very thick] table {\data};
\pgfplotstableread{plot_data/variation_of_observers/\observer/P_(2, 2)_shear_min.txt}\data
\addplot [color22, fill=none,mark=none, very thick] table {\data};
\pgfplotstableread{plot_data/variation_of_observers/\observer/P_(2, 2)_shear_max.txt}\data
\addplot [color22, fill=none,mark=none, very thick] table {\data};
\pgfplotstableread{plot_data/variation_of_observers/\observer/P_(2, 3)_shear_min.txt}\data
\addplot [color23, fill=none,mark=none, very thick] table {\data};
\pgfplotstableread{plot_data/variation_of_observers/\observer/P_(2, 3)_shear_max.txt}\data
\addplot [color23, fill=none,mark=none, very thick] table {\data};
\pgfplotstableread{plot_data/variation_of_observers/\observer/P_(3, 1)_shear_min.txt}\data
\addplot [color31, fill=none,mark=none, very thick] table {\data};
\pgfplotstableread{plot_data/variation_of_observers/\observer/P_(3, 1)_shear_max.txt}\data
\addplot [color31, fill=none,mark=none, very thick] table {\data};
\pgfplotstableread{plot_data/variation_of_observers/\observer/P_(3, 2)_shear_min.txt}\data
\addplot [color32, fill=none,mark=none, very thick] table {\data};
\pgfplotstableread{plot_data/variation_of_observers/\observer/P_(3, 2)_shear_max.txt}\data
\addplot [color32, fill=none,mark=none, very thick] table {\data};
\pgfplotstableread{plot_data/variation_of_observers/\observer/P_(3, 3)_shear_min.txt}\data
\addplot [color33, fill=none,mark=none, very thick] table {\data};
\pgfplotstableread{plot_data/variation_of_observers/\observer/P_(3, 3)_shear_max.txt}\data
\addplot [color33, fill=none,mark=none, very thick] table {\data};
        
                    \node[fill=white, draw=black] at  (0.4, 0) {{16 observers}};
    
        \nextgroupplot[ylabel = {$S_{ij}$}, ytick={0,700,1400},yticklabels={0,7,14},
width=0.49*\textwidth, height=0.38*\textwidth, xlabel={$F_{12}$},
xtick={0,0.25,0.5},xticklabels={0,0.25,0.5},
ymin = -150, ymax = 1500,
xmin = -0.02, xmax = 0.52] % shear for 32 observers
\foreach \num in {11, 12, 13, 21, 22, 23, 31, 32, 33}{
 \addplot coordinates { (-2,-2) (-3,-3) };
 };
  \foreach \num in {9, 10, 11, 12, 13, 14, 15, 16, 17}{
   \addplot table [x =shear_F12,y index=\num, only marks] 
    {plot_data/BCC/calibration_data/shear.txt};
    };

        \def\observer{32}
\pgfplotstableread{plot_data/variation_of_observers/\observer/P_(1, 1)_shear.txt}\data
\addplot [fill=color11,mark=none, draw=none, opacity=0.3] table {\data};
\pgfplotstableread{plot_data/variation_of_observers/\observer/P_(1, 2)_shear.txt}\data
\addplot [fill=color12,mark=none, draw=none, opacity=0.3] table {\data};
\pgfplotstableread{plot_data/variation_of_observers/\observer/P_(1, 3)_shear.txt}\data
\addplot [fill=color13,mark=none, draw=none, opacity=0.3] table {\data};
\pgfplotstableread{plot_data/variation_of_observers/\observer/P_(2, 1)_shear.txt}\data
\addplot [fill=color21,mark=none, draw=none, opacity=0.3] table {\data};
\pgfplotstableread{plot_data/variation_of_observers/\observer/P_(2, 2)_shear.txt}\data
\addplot [fill=color22,mark=none, draw=none, opacity=0.3] table {\data};
\pgfplotstableread{plot_data/variation_of_observers/\observer/P_(2, 3)_shear.txt}\data
\addplot [fill=color23,mark=none, draw=none, opacity=0.3] table {\data};
\pgfplotstableread{plot_data/variation_of_observers/\observer/P_(3, 1)_shear.txt}\data
\addplot [fill=color31,mark=none, draw=none, opacity=0.3] table {\data};
\pgfplotstableread{plot_data/variation_of_observers/\observer/P_(3, 2)_shear.txt}\data
\addplot [fill=color32,mark=none, draw=none, opacity=0.3] table {\data};
\pgfplotstableread{plot_data/variation_of_observers/\observer/P_(3, 3)_shear.txt}\data
\addplot [fill=color33,mark=none, draw=none, opacity=0.3] table {\data};

\pgfplotstableread{plot_data/variation_of_observers/\observer/P_(1, 1)_shear_min.txt}\data
\addplot [color11, fill=none,mark=none, very thick] table {\data};
\pgfplotstableread{plot_data/variation_of_observers/\observer/P_(1, 1)_shear_max.txt}\data
\addplot [color11, fill=none,mark=none, very thick] table {\data};
\pgfplotstableread{plot_data/variation_of_observers/\observer/P_(1, 2)_shear_min.txt}\data
\addplot [color12, fill=none,mark=none, very thick] table {\data};
\pgfplotstableread{plot_data/variation_of_observers/\observer/P_(1, 2)_shear_max.txt}\data
\addplot [color12, fill=none,mark=none, very thick] table {\data};
\pgfplotstableread{plot_data/variation_of_observers/\observer/P_(1, 3)_shear_min.txt}\data
\addplot [color13, fill=none,mark=none, very thick] table {\data};
\pgfplotstableread{plot_data/variation_of_observers/\observer/P_(1, 3)_shear_max.txt}\data
\addplot [color13, fill=none,mark=none, very thick] table {\data};
\pgfplotstableread{plot_data/variation_of_observers/\observer/P_(2, 1)_shear_min.txt}\data
\addplot [color21, fill=none,mark=none, very thick] table {\data};
\pgfplotstableread{plot_data/variation_of_observers/\observer/P_(2, 1)_shear_max.txt}\data
\addplot [color21, fill=none,mark=none, very thick] table {\data};
\pgfplotstableread{plot_data/variation_of_observers/\observer/P_(2, 2)_shear_min.txt}\data
\addplot [color22, fill=none,mark=none, very thick] table {\data};
\pgfplotstableread{plot_data/variation_of_observers/\observer/P_(2, 2)_shear_max.txt}\data
\addplot [color22, fill=none,mark=none, very thick] table {\data};
\pgfplotstableread{plot_data/variation_of_observers/\observer/P_(2, 3)_shear_min.txt}\data
\addplot [color23, fill=none,mark=none, very thick] table {\data};
\pgfplotstableread{plot_data/variation_of_observers/\observer/P_(2, 3)_shear_max.txt}\data
\addplot [color23, fill=none,mark=none, very thick] table {\data};
\pgfplotstableread{plot_data/variation_of_observers/\observer/P_(3, 1)_shear_min.txt}\data
\addplot [color31, fill=none,mark=none, very thick] table {\data};
\pgfplotstableread{plot_data/variation_of_observers/\observer/P_(3, 1)_shear_max.txt}\data
\addplot [color31, fill=none,mark=none, very thick] table {\data};
\pgfplotstableread{plot_data/variation_of_observers/\observer/P_(3, 2)_shear_min.txt}\data
\addplot [color32, fill=none,mark=none, very thick] table {\data};
\pgfplotstableread{plot_data/variation_of_observers/\observer/P_(3, 2)_shear_max.txt}\data
\addplot [color32, fill=none,mark=none, very thick] table {\data};
\pgfplotstableread{plot_data/variation_of_observers/\observer/P_(3, 3)_shear_min.txt}\data
\addplot [color33, fill=none,mark=none, very thick] table {\data};
\pgfplotstableread{plot_data/variation_of_observers/\observer/P_(3, 3)_shear_max.txt}\data
\addplot [color33, fill=none,mark=none, very thick] table {\data};
                    \node[fill=white, draw=black] at  (0.4, 0) {{32 observers}};
    
    \nextgroupplot[ legend pos=outer north east,ytick={0,700,1400},yticklabels={,,},
width=0.49*\textwidth, height=0.38*\textwidth, xlabel={$F_{12}$},
xtick={0,0.25,0.5},xticklabels={0,0.25,0.5},
ymin = -150, ymax = 1500,
xmin = -0.02, xmax = 0.52] % shear fpr 64 observers
\foreach \num in {11, 12, 13, 21, 22, 23, 31, 32, 33}{
 \addplot coordinates { (-2,-2) (-3,-3) };
 };
    \foreach \num in {9, 10, 11, 12, 13, 14, 15, 16, 17}{
    \addplot table [x =shear_F12,y index=\num, only marks]
     {plot_data/BCC/calibration_data/shear.txt};
    };
     \def\observer{64}
\pgfplotstableread{plot_data/variation_of_observers/\observer/P_(1, 1)_shear.txt}\data
\addplot [fill=color11,mark=none, draw=none, opacity=0.3] table {\data};
\pgfplotstableread{plot_data/variation_of_observers/\observer/P_(1, 2)_shear.txt}\data
\addplot [fill=color12,mark=none, draw=none, opacity=0.3] table {\data};
\pgfplotstableread{plot_data/variation_of_observers/\observer/P_(1, 3)_shear.txt}\data
\addplot [fill=color13,mark=none, draw=none, opacity=0.3] table {\data};
\pgfplotstableread{plot_data/variation_of_observers/\observer/P_(2, 1)_shear.txt}\data
\addplot [fill=color21,mark=none, draw=none, opacity=0.3] table {\data};
\pgfplotstableread{plot_data/variation_of_observers/\observer/P_(2, 2)_shear.txt}\data
\addplot [fill=color22,mark=none, draw=none, opacity=0.3] table {\data};
\pgfplotstableread{plot_data/variation_of_observers/\observer/P_(2, 3)_shear.txt}\data
\addplot [fill=color23,mark=none, draw=none, opacity=0.3] table {\data};
\pgfplotstableread{plot_data/variation_of_observers/\observer/P_(3, 1)_shear.txt}\data
\addplot [fill=color31,mark=none, draw=none, opacity=0.3] table {\data};
\pgfplotstableread{plot_data/variation_of_observers/\observer/P_(3, 2)_shear.txt}\data
\addplot [fill=color32,mark=none, draw=none, opacity=0.3] table {\data};
\pgfplotstableread{plot_data/variation_of_observers/\observer/P_(3, 3)_shear.txt}\data
\addplot [fill=color33,mark=none, draw=none, opacity=0.3] table {\data};

\pgfplotstableread{plot_data/variation_of_observers/\observer/P_(1, 1)_shear_min.txt}\data
\addplot [color11, fill=none,mark=none, very thick] table {\data};
\pgfplotstableread{plot_data/variation_of_observers/\observer/P_(1, 1)_shear_max.txt}\data
\addplot [color11, fill=none,mark=none, very thick] table {\data};
\pgfplotstableread{plot_data/variation_of_observers/\observer/P_(1, 2)_shear_min.txt}\data
\addplot [color12, fill=none,mark=none, very thick] table {\data};
\pgfplotstableread{plot_data/variation_of_observers/\observer/P_(1, 2)_shear_max.txt}\data
\addplot [color12, fill=none,mark=none, very thick] table {\data};
\pgfplotstableread{plot_data/variation_of_observers/\observer/P_(1, 3)_shear_min.txt}\data
\addplot [color13, fill=none,mark=none, very thick] table {\data};
\pgfplotstableread{plot_data/variation_of_observers/\observer/P_(1, 3)_shear_max.txt}\data
\addplot [color13, fill=none,mark=none, very thick] table {\data};
\pgfplotstableread{plot_data/variation_of_observers/\observer/P_(2, 1)_shear_min.txt}\data
\addplot [color21, fill=none,mark=none, very thick] table {\data};
\pgfplotstableread{plot_data/variation_of_observers/\observer/P_(2, 1)_shear_max.txt}\data
\addplot [color21, fill=none,mark=none, very thick] table {\data};
\pgfplotstableread{plot_data/variation_of_observers/\observer/P_(2, 2)_shear_min.txt}\data
\addplot [color22, fill=none,mark=none, very thick] table {\data};
\pgfplotstableread{plot_data/variation_of_observers/\observer/P_(2, 2)_shear_max.txt}\data
\addplot [color22, fill=none,mark=none, very thick] table {\data};
\pgfplotstableread{plot_data/variation_of_observers/\observer/P_(2, 3)_shear_min.txt}\data
\addplot [color23, fill=none,mark=none, very thick] table {\data};
\pgfplotstableread{plot_data/variation_of_observers/\observer/P_(2, 3)_shear_max.txt}\data
\addplot [color23, fill=none,mark=none, very thick] table {\data};
\pgfplotstableread{plot_data/variation_of_observers/\observer/P_(3, 1)_shear_min.txt}\data
\addplot [color31, fill=none,mark=none, very thick] table {\data};
\pgfplotstableread{plot_data/variation_of_observers/\observer/P_(3, 1)_shear_max.txt}\data
\addplot [color31, fill=none,mark=none, very thick] table {\data};
\pgfplotstableread{plot_data/variation_of_observers/\observer/P_(3, 2)_shear_min.txt}\data
\addplot [color32, fill=none,mark=none, very thick] table {\data};
\pgfplotstableread{plot_data/variation_of_observers/\observer/P_(3, 2)_shear_max.txt}\data
\addplot [color32, fill=none,mark=none, very thick] table {\data};
\pgfplotstableread{plot_data/variation_of_observers/\observer/P_(3, 3)_shear_min.txt}\data
\addplot [color33, fill=none,mark=none, very thick] table {\data};
\pgfplotstableread{plot_data/variation_of_observers/\observer/P_(3, 3)_shear_max.txt}\data
\addplot [color33, fill=none,mark=none, very thick] table {\data};
                    \node[fill=white, draw=black] at  (0.4, 0) {{64 observers}};

\end{groupplot}
\end{tikzpicture}
}
\caption{Variation of observers for $W^{\text{F}}$ for the cubic BCC cell. Points depict the simulation data, while lines and shaded areas depict the calibrated model evaluated with $1,024$ observers. Stress tensor $Q^T\,S(\uuQ\,\uuF)$ \eqref{eq:stress_objectivity} for shear deformation is shown, with normal components in red colors, shear components in blue colors and less important components in gray; stress in $\left[\text{hPa}\right]$.}
\label{fig:BCC_variation_of_observers}
\end{figure}

%% file: chapter/figures/BCC_W_C.tex
\tikzsetnextfilename{BCC_W_C}

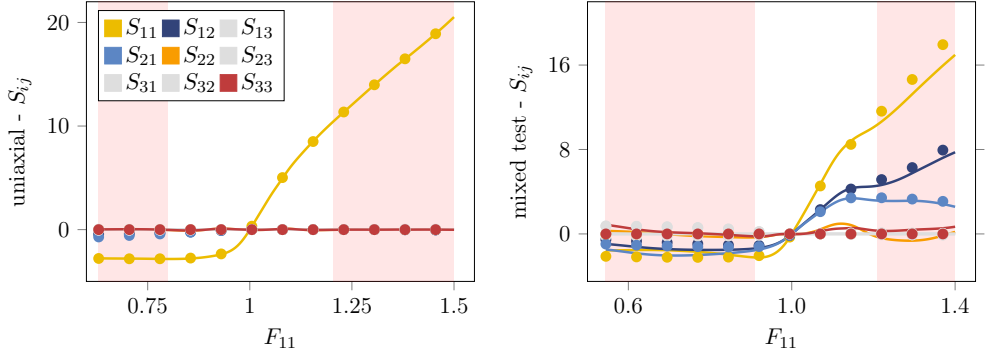
\begin{figure}[tp]
\centering
\pgfplotstableset{
create on use/uniax_F11/.style={create col/copy column from table={plot_data/BCC/calibration_data/uniaxial.txt}{0}}
}
\pgfplotstableset{
create on use/test3_F11/.style={create col/copy column from table={plot_data/BCC/test_data/test3.txt}{0}}
}

\resizebox{0.8\textwidth}{!}{
\begin{tikzpicture}
\begin{groupplot}[
	group style = {group size = 2 by 1, vertical sep = 0.09*\textwidth,
								horizontal sep = 0.115*\textwidth},
	cycle list name=mycolorlist2,
	xtick align = outside,
	ytick align = outside,
	xtick pos = left,
	%ytick pos = left
	]
\nextgroupplot[legend pos = north west,legend columns=3, ylabel = {uniaxial - $S_{ij}$}, ytick={0,1000,2000},yticklabels={0,10,20},
width=0.49*\textwidth, height=0.38*\textwidth,
xtick={0.75,1,1.25,1.5},xticklabels={0.75,1,1.25,1.5},
ymin = -500, ymax = 2200,
xlabel={$F_{11}$},
xmin = 0.6, xmax = 1.55,
ytick pos = left] % uniaxial W_C model

\foreach \num in {11, 12, 13, 21, 22, 23, 31, 32, 33}{
 \addplot coordinates { (-2,-2) (-3,-3) };
  \addlegendentryexpanded{$S_{\num}$}
 };

       \pgfplotstableread[col sep=comma,row sep=crcr]{
    0.63, 2500 \\
    0.8, 2500 \\
    0.8, -500 \\
    0.63, -500 \\
   }\data
    \addplot [fill=red,mark=none, draw=none, opacity=0.1] table {\data};
    
        \pgfplotstableread[col sep=comma,row sep=crcr]{
    1.205, 2500 \\
    1.5, 2500 \\
    1.5, -500 \\
    1.205, -500 \\
   }\data
    \addplot [fill=red,mark=none, draw=none, opacity=0.1] table {\data};

    \foreach \num in {9, 10, 11, 12, 13, 14, 15, 16, 17}{
    \addplot table [x =uniax_F11,y index=\num, only marks]  {plot_data/BCC/calibration_data/uniaxial.txt};
    };
        \foreach \num in {0,1,2,3,4,5,6,7,8}{
    \addplot table [dashed, x =uniax_F11,y index=\num] {plot_data/W_C/P_uniaxial.txt};
    };

   \nextgroupplot[legend pos = outer north east,ylabel = {mixed test - $S_{ij}$}, ytick={0,800,1600},yticklabels={0,8,16},
width=0.49*\textwidth, height=0.38*\textwidth,
xlabel={$F_{11}$},
xtick={0.6,1.0,1.4},xticklabels={0.6,1.0,1.4},
ymin = -450, ymax = 2200,
xmin = 0.5, xmax = 1.45] % test3 

\foreach \num in {11, 12, 13, 21, 22, 23, 31, 32, 33}{
 \addplot coordinates { (-2,-2) (-3,-3) };
 %\addlegendentryexpanded{$P_{\num}$}
 };

        \pgfplotstableread[col sep=comma,row sep=crcr]{
    0.545, 2500 \\
    0.91, 2500 \\
    0.91, -500 \\
    0.545, -500 \\
    }\data
    \addplot [fill=red,mark=none, draw=none, opacity=0.1] table {\data};
    
       \pgfplotstableread[col sep=comma,row sep=crcr]{
   1.21, 2500 \\
    1.4, 2500 \\
    1.4, -2500 \\
    1.21, -2500 \\
    }\data
    \addplot [fill=red,mark=none, draw=none, opacity=0.1] table {\data};

  \foreach \num in {9, 10, 11, 12, 13, 14, 15, 16, 17}{
   \addplot table [x =test3_F11,y index=\num, only marks] {plot_data/BCC/test_data/test3.txt};
    };
    
        \foreach \num in {0,1,2,3,4,5,6,7,8}{
    \addplot table [dashed, x =test3_F11,y index=\num] {plot_data/W_C/P_test3.txt};
    }; 
      
    \end{groupplot}
\end{tikzpicture}
}

\caption{Evaluation of $W^{\text{C}}$ for the BCC cell, shaded areas denote the model's loss of ellipticity. The ellipticity was checked with the Hessian of the potential, using $500$ random \tRM{unit vectors} for each test vector, c.f. \eqref{eq:ellipticity}. Points depict the simulation data, while lines depict the calibrated model. Normal components of the stress tensor in red colors, shear components in blue, less important components in gray; stress in $\left[\text{hPa}\right]$.}
\label{fig:BCC_C_model}
\end{figure}

%% file: chapter/figures/BCC_volumetric.tex
\tikzsetnextfilename{BCC_volumetric}

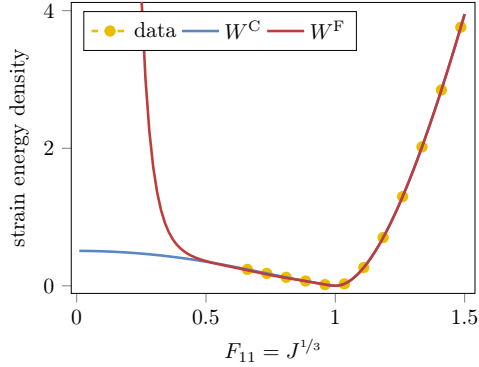
\begin{figure}[t!]
\centering

\pgfplotstableset{
create on use/J_model/.style={create col/copy column from table={plot_data/BCC_volumetric.txt}{0}}
}

\pgfplotstableset{
create on use/J_data/.style={create col/copy column from table={plot_data/BCC/calibration_data/volumetric.txt}{0}}
}

\resizebox{0.4\textwidth}{!}{
\begin{tikzpicture}
\pgfplotsset{
  set layers,% using layers
  mark layer=axis tick labels% defines the layer of the marks
}
\begin{groupplot}[
	group style = {group size = 1 by 1, vertical sep = 0.09*\textwidth,
								horizontal sep = 0.115*\textwidth},
	cycle list name=mycolorlistvol,
	xtick align = outside,
	ytick align = outside,
	xtick pos = left,
	]

   \nextgroupplot[ytick pos = left, legend pos = north west,legend columns=3,ylabel = {strain energy density}, ytick={0,2000,4000},yticklabels={0,2,4},
width=0.49*\textwidth, height=0.38*\textwidth,xlabel={$F_{11}={J}^{\nicefrac{1}{3}}$},
xtick={0,0.5,1,1.5},xticklabels={0,0.5,1,1.5},
ymin = -110, ymax = 4120,
xmin = -0.021, xmax = 1.54] 

   \addplot table [x =J_data,y index=18, only marks] {plot_data/BCC/calibration_data/volumetric.txt};

             \foreach \num in {1,4}{
    \addplot table [dashed, x = J_model, y index=\num,  mark=none] {plot_data/BCC_volumetric.txt};
    }

\addlegendentry{data};
\addlegendentry{$W^{\text{C}}$};
\addlegendentry{$W^{\text{F}}$};

\end{groupplot}
\end{tikzpicture}
}
\caption{Evaluation of $W^{\text{C}}$ and $W^{\text{F}}$ for volumetric deformation of the cubic BCC cell; strain energy density in [kJ/m\textsuperscript{3}].}
\label{fig:BCC_volumetric}
\end{figure}

%% file: chapter/figures/BCC_invariant_model.tex
\tikzsetnextfilename{BCC_invariant_model}

\begin{figure}[tp]
\centering
\pgfplotstableset{
create on use/uniax_F11/.style={create col/copy column from table={plot_data/BCC/calibration_data/uniaxial.txt}{0}}
}
\pgfplotstableset{
create on use/shear_F12/.style={create col/copy column from table={plot_data/BCC/calibration_data/shear.txt}{1}}
}

\pgfplotstableset{
create on use/test1_F11/.style={create col/copy column from table={plot_data/BCC/test_data/test1.txt}{0}}
}
\pgfplotstableset{
create on use/test3_F11/.style={create col/copy column from table={plot_data/BCC/test_data/test3.txt}{0}}
}
\resizebox{0.8\textwidth}{!}{
\begin{tikzpicture}
\pgfplotsset{
  set layers,% using layers
  mark layer=axis tick labels% defines the layer of the marks
}
\begin{groupplot}[
	group style = {group size = 2 by 2, vertical sep = 0.09*\textwidth,
								horizontal sep = 0.115*\textwidth},
	cycle list name=mycolorlist,
	xtick align = outside,
	ytick align = outside,
	xtick pos = left,
	%ytick pos = left
	]
\nextgroupplot[legend pos = north west,legend columns=3, ylabel = {uniaxial - $S_{ij}$}, ytick={0,1000,2000},yticklabels={0,10,20},
width=0.49*\textwidth, height=0.38*\textwidth,
xtick={0.75,1,1.25,1.5},xticklabels={0.75,1,1.25,1.5},
ymin = -500, ymax = 2200,
xlabel={$F_{11}$},
xmin = 0.6, xmax = 1.55,
ytick pos = left] % uniaxial invariant model
\foreach \num in {11, 12, 13, 21, 22, 23, 31, 32, 33}{
 \addplot coordinates { (-2,-2) (-3,-3) };
\addlegendentryexpanded{$S_{\num}$}
 };
    \foreach \num in {9, 10, 11, 12, 13, 14, 15, 16, 17}{
    \addplot table [x =uniax_F11,y index=\num, only marks] 
    {plot_data/BCC/calibration_data/uniaxial.txt};
    };
        \foreach \num in {0,1,2,3,4,5,6,7,8}{
    \addplot table [dashed, x =uniax_F11,y index=\num,  mark=none] {plot_data/BCC/calibrated_models/invariant_model/P_uniaxial.txt};
    };

   \nextgroupplot[ylabel = {shear - $S_{ij}$}, ytick={0,700,1400},yticklabels={0,7,14},
width=0.49*\textwidth, height=0.38*\textwidth, xlabel={$F_{12}$},
xtick={0,0.25,0.5},xticklabels={0,0.25,0.5},
ymin = -150, ymax = 1500,
xmin = -0.02, xmax = 0.53,
ytick pos = left] % shear F det F model
\foreach \num in {11, 12, 13, 21, 22, 23, 31, 32, 33}{
 \addplot coordinates { (-2,-2) (-3,-3) };
 };
  \foreach \num in {9, 10, 11, 12, 13, 14, 15, 16, 17}{
   \addplot table [x =shear_F12,y index=\num, only marks] 
    {plot_data/BCC/calibration_data/shear.txt};
    };
    
        \foreach \num in {0,1,2,3,4,5,6,7,8}{
    \addplot table [dashed, x =shear_F12,y index=\num,  mark=none] {plot_data/BCC/calibrated_models/invariant_model/P_shear.txt};
    }; 
    \nextgroupplot[legend pos = north west,legend columns=3, ylabel = {biaxial test - $S_{ij}$}, ytick={0,1100,2200},yticklabels={0,11,22},
width=0.49*\textwidth, height=0.38*\textwidth,
xtick={0.8,1.0,1.2,1.4},xticklabels={0.8,1.0,1.2,1.4},
ymin = -400, ymax = 2800,
xlabel={$F_{11}$},
xmin = 0.65, xmax = 1.45,
ytick pos = left] % test1 F det F model

\foreach \num in {11, 12, 13, 21, 22, 23, 31, 32, 33}{
 \addplot coordinates { (-2,-2) (-3,-3) };
 };
 
    \foreach \num in {9, 10, 11, 12, 13, 14, 15, 16, 17}{
    \addplot table [x =test1_F11,y index=\num, only marks] 
    {plot_data/BCC/test_data/test1.txt};
    };

    \foreach \num in {0,1,2,3,4,5,6,7,8}{
    \addplot table [dashed, x =test1_F11,y index=\num] {plot_data/BCC/calibrated_models/invariant_model/P_test1.txt};
    };

       \nextgroupplot[ylabel = {mixed test - $S_{ij}$},xlabel={$F_{11}$}, ytick={0,800,1600},yticklabels={0,8,16},
width=0.49*\textwidth, height=0.38*\textwidth,
xtick={0.6,1.0,1.4},xticklabels={0.6,1.0,1.4},
ymin = -450, ymax = 2200,
xmin = 0.5, xmax = 1.45,
ytick pos = left] % test3 F det F model

\foreach \num in {11, 12, 13, 21, 22, 23, 31, 32, 33}{
 \addplot coordinates { (-2,-2) (-3,-3) };
 };
     \foreach \num in {9, 10, 11, 12, 13, 14, 15, 16, 17}{
    \addplot table [x =test3_F11,y index=\num, only marks] 
    {plot_data/BCC/test_data/test3.txt};
    };

    \foreach \num in {0,1,2,3,4,5,6,7,8}{
    \addplot table [dashed, x =test3_F11,y index=\num] {plot_data/BCC/calibrated_models/invariant_model/P_test3.txt};
    };

\end{groupplot}
\end{tikzpicture}
}
\caption{Evaluation of $W^{\text{I}}$ for the cubic BCC cell. Points depict the simulation data, while lines depict the calibrated model. Normal components of the stress tensor in red colors, shear components in blue, less important components in gray; stress in $\left[\text{hPa}\right]$.}
\label{fig:BCC_invariant_model}
\end{figure}
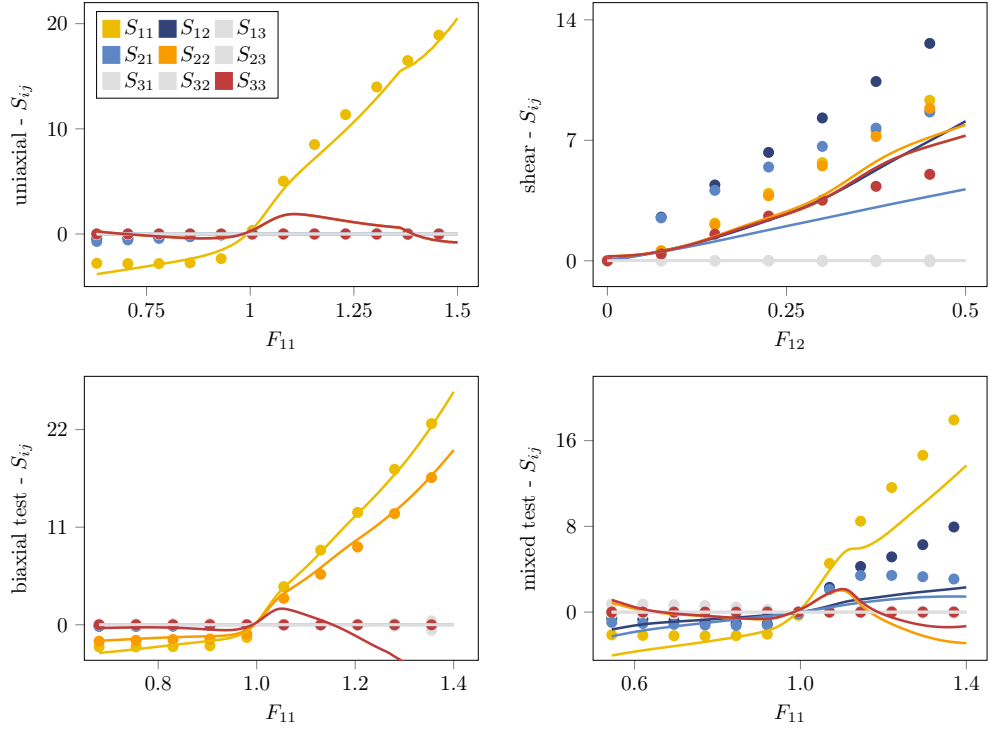

%% file: chapter/figures/BCC_deformation_gradient_model.tex
\tikzsetnextfilename{BCC_deformation_gradient_model}

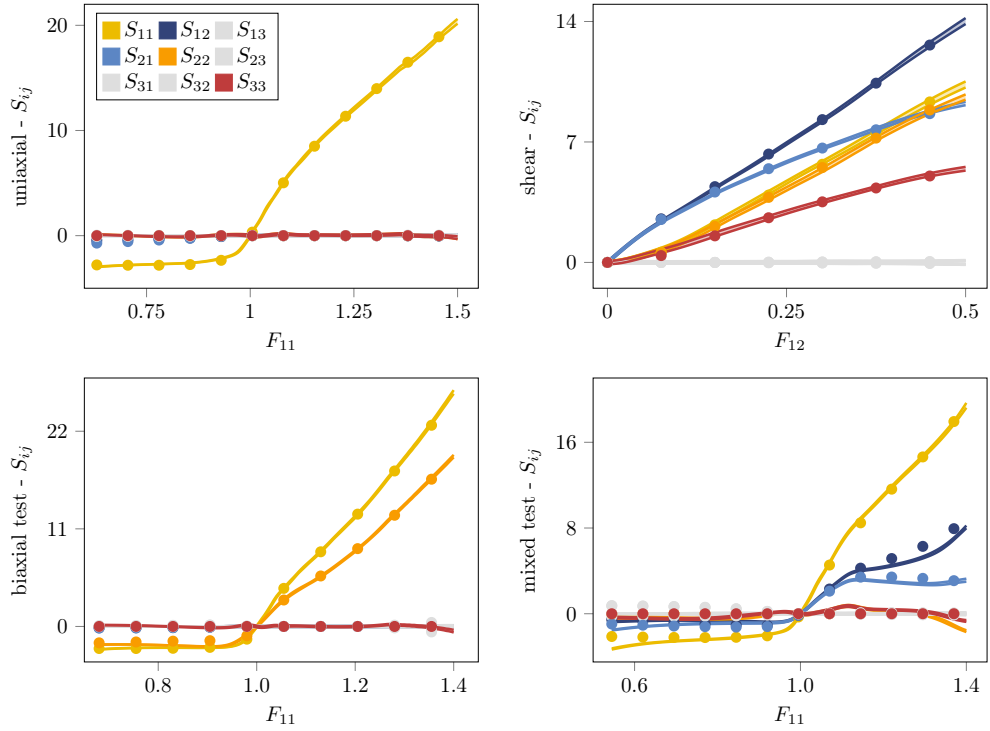
\begin{figure}[tp]
\centering
\pgfplotstableset{
create on use/uniax_F11/.style={create col/copy column from table={plot_data/BCC/calibration_data/uniaxial.txt}{0}}
}

\pgfplotstableset{
create on use/shear_F12/.style={create col/copy column from table={plot_data/BCC/calibration_data/shear.txt}{1}}
}

\pgfplotstableset{
create on use/test1_F11/.style={create col/copy column from table={plot_data/BCC/test_data/test1.txt}{0}}
}
\pgfplotstableset{
create on use/test3_F11/.style={create col/copy column from table={plot_data/BCC/test_data/test3.txt}{0}}
}
\resizebox{0.8\textwidth}{!}{
\begin{tikzpicture}
\begin{groupplot}[
	group style = {group size = 2 by 2, vertical sep = 0.09*\textwidth,
								horizontal sep = 0.115*\textwidth},
	cycle list name=mycolorlist,
	xtick align = outside,
	ytick align = outside,
	xtick pos = left,
	%ytick pos = left
	]
\nextgroupplot[legend pos = north west,legend columns=3, ylabel = {uniaxial - $S_{ij}$}, ytick={0,1000,2000},yticklabels={0,10,20},
width=0.49*\textwidth, height=0.38*\textwidth,
xtick={0.75,1,1.25,1.5},xticklabels={0.75,1,1.25,1.5},
ymin = -500, ymax = 2200,
xlabel={$F_{11}$},
xmin = 0.6, xmax = 1.55,
ytick pos = left] % uniaxial F det F model
\foreach \num in {11, 12, 13, 21, 22, 23, 31, 32, 33}{
 \addplot coordinates { (-2,-2) (-3,-3) };
\addlegendentryexpanded{$S_{\num}$}
 };
    \foreach \num in {9, 10, 11, 12, 13, 14, 15, 16, 17}{
    \addplot table [x =uniax_F11,y index=\num, only marks] 
    {plot_data/BCC/calibration_data/uniaxial.txt};
    };
\pgfplotstableread{plot_data/BCC/calibrated_models/F_det_F_model/P_(1, 1)_uniaxial.txt}\data
\addplot [fill=color11,mark=none, draw=none, opacity=0.3] table {\data};
\pgfplotstableread{plot_data/BCC/calibrated_models/F_det_F_model/P_(1, 2)_uniaxial.txt}\data
\addplot [fill=color12,mark=none, draw=none, opacity=0.3] table {\data};
\pgfplotstableread{plot_data/BCC/calibrated_models/F_det_F_model/P_(1, 3)_uniaxial.txt}\data
\addplot [fill=color13,mark=none, draw=none, opacity=0.3] table {\data};
\pgfplotstableread{plot_data/BCC/calibrated_models/F_det_F_model/P_(2, 1)_uniaxial.txt}\data
\addplot [fill=color21,mark=none, draw=none, opacity=0.3] table {\data};
\pgfplotstableread{plot_data/BCC/calibrated_models/F_det_F_model/P_(2, 2)_uniaxial.txt}\data
\addplot [fill=color22,mark=none, draw=none, opacity=0.3] table {\data};
\pgfplotstableread{plot_data/BCC/calibrated_models/F_det_F_model/P_(2, 3)_uniaxial.txt}\data
\addplot [fill=color23,mark=none, draw=none, opacity=0.3] table {\data};
\pgfplotstableread{plot_data/BCC/calibrated_models/F_det_F_model/P_(3, 1)_uniaxial.txt}\data
\addplot [fill=color31,mark=none, draw=none, opacity=0.3] table {\data};
\pgfplotstableread{plot_data/BCC/calibrated_models/F_det_F_model/P_(3, 2)_uniaxial.txt}\data
\addplot [fill=color32,mark=none, draw=none, opacity=0.3] table {\data};
\pgfplotstableread{plot_data/BCC/calibrated_models/F_det_F_model/P_(3, 3)_uniaxial.txt}\data
\addplot [fill=color33,mark=none, draw=none, opacity=0.3] table {\data};

\pgfplotstableread{plot_data/BCC/calibrated_models/F_det_F_model/P_(1, 1)_uniaxial_min.txt}\data
\addplot [color11, fill=none,mark=none, very thick] table {\data};
\pgfplotstableread{plot_data/BCC/calibrated_models/F_det_F_model/P_(1, 1)_uniaxial_max.txt}\data
\addplot [color11, fill=none,mark=none, very thick] table {\data};
\pgfplotstableread{plot_data/BCC/calibrated_models/F_det_F_model/P_(1, 2)_uniaxial_min.txt}\data
\addplot [color12, fill=none,mark=none, very thick] table {\data};
\pgfplotstableread{plot_data/BCC/calibrated_models/F_det_F_model/P_(1, 2)_uniaxial_max.txt}\data
\addplot [color12, fill=none,mark=none, very thick] table {\data};
\pgfplotstableread{plot_data/BCC/calibrated_models/F_det_F_model/P_(1, 3)_uniaxial_min.txt}\data
\addplot [color13, fill=none,mark=none, very thick] table {\data};
\pgfplotstableread{plot_data/BCC/calibrated_models/F_det_F_model/P_(1, 3)_uniaxial_max.txt}\data
\addplot [color13, fill=none,mark=none, very thick] table {\data};
\pgfplotstableread{plot_data/BCC/calibrated_models/F_det_F_model/P_(2, 1)_uniaxial_min.txt}\data
\addplot [color21, fill=none,mark=none, very thick] table {\data};
\pgfplotstableread{plot_data/BCC/calibrated_models/F_det_F_model/P_(2, 1)_uniaxial_max.txt}\data
\addplot [color21, fill=none,mark=none, very thick] table {\data};
\pgfplotstableread{plot_data/BCC/calibrated_models/F_det_F_model/P_(2, 2)_uniaxial_min.txt}\data
\addplot [color22, fill=none,mark=none, very thick] table {\data};
\pgfplotstableread{plot_data/BCC/calibrated_models/F_det_F_model/P_(2, 2)_uniaxial_max.txt}\data
\addplot [color22, fill=none,mark=none, very thick] table {\data};
\pgfplotstableread{plot_data/BCC/calibrated_models/F_det_F_model/P_(2, 3)_uniaxial_min.txt}\data
\addplot [color23, fill=none,mark=none, very thick] table {\data};
\pgfplotstableread{plot_data/BCC/calibrated_models/F_det_F_model/P_(2, 3)_uniaxial_max.txt}\data
\addplot [color23, fill=none,mark=none, very thick] table {\data};
\pgfplotstableread{plot_data/BCC/calibrated_models/F_det_F_model/P_(3, 1)_uniaxial_min.txt}\data
\addplot [color31, fill=none,mark=none, very thick] table {\data};
\pgfplotstableread{plot_data/BCC/calibrated_models/F_det_F_model/P_(3, 1)_uniaxial_max.txt}\data
\addplot [color31, fill=none,mark=none, very thick] table {\data};
\pgfplotstableread{plot_data/BCC/calibrated_models/F_det_F_model/P_(3, 2)_uniaxial_min.txt}\data
\addplot [color32, fill=none,mark=none, very thick] table {\data};
\pgfplotstableread{plot_data/BCC/calibrated_models/F_det_F_model/P_(3, 2)_uniaxial_max.txt}\data
\addplot [color32, fill=none,mark=none, very thick] table {\data};
\pgfplotstableread{plot_data/BCC/calibrated_models/F_det_F_model/P_(3, 3)_uniaxial_min.txt}\data
\addplot [color33, fill=none,mark=none, very thick] table {\data};
\pgfplotstableread{plot_data/BCC/calibrated_models/F_det_F_model/P_(3, 3)_uniaxial_max.txt}\data
\addplot [color33, fill=none,mark=none, very thick] table {\data};

   \nextgroupplot[ylabel = {shear - $S_{ij}$}, ytick={0,700,1400},yticklabels={0,7,14},
width=0.49*\textwidth, height=0.38*\textwidth, xlabel={$F_{12}$},
xtick={0,0.25,0.5},xticklabels={0,0.25,0.5},
ymin = -150, ymax = 1500,
xmin = -0.02, xmax = 0.53,
ytick pos = left] % shear F det F model
\foreach \num in {11, 12, 13, 21, 22, 23, 31, 32, 33}{
 \addplot coordinates { (-2,-2) (-3,-3) };
 };
  \foreach \num in {9, 10, 11, 12, 13, 14, 15, 16, 17}{
   \addplot table [x =shear_F12,y index=\num, only marks] 
    {plot_data/BCC/calibration_data/shear.txt};
    };
\pgfplotstableread{plot_data/BCC/calibrated_models/F_det_F_model/P_(1, 1)_shear.txt}\shear
\addplot [fill=color11, mark=none, draw=none, opacity=0.3] table {\shear};
\pgfplotstableread{plot_data/BCC/calibrated_models/F_det_F_model/P_(1, 2)_shear.txt}\shear
\addplot [fill=color12, mark=none, draw=none, opacity=0.3] table {\shear};
\pgfplotstableread{plot_data/BCC/calibrated_models/F_det_F_model/P_(1, 3)_shear.txt}\shear
\addplot [fill=color13, mark=none, draw=none, opacity=0.3] table {\shear};
\pgfplotstableread{plot_data/BCC/calibrated_models/F_det_F_model/P_(2, 1)_shear.txt}\shear
\addplot [fill=color21, mark=none, draw=none, opacity=0.3] table {\shear};
\pgfplotstableread{plot_data/BCC/calibrated_models/F_det_F_model/P_(2, 2)_shear.txt}\data
\addplot [fill=color22,mark=none, draw=none, opacity=0.3] table {\data};
\pgfplotstableread{plot_data/BCC/calibrated_models/F_det_F_model/P_(2, 3)_shear.txt}\data
\addplot [fill=color23,mark=none, draw=none, opacity=0.3] table {\data};
\pgfplotstableread{plot_data/BCC/calibrated_models/F_det_F_model/P_(3, 1)_shear.txt}\data
\addplot [fill=color31,mark=none, draw=none, opacity=0.3] table {\data};
\pgfplotstableread{plot_data/BCC/calibrated_models/F_det_F_model/P_(3, 2)_shear.txt}\data
\addplot [fill=color32,mark=none, draw=none, opacity=0.3] table {\data};
\pgfplotstableread{plot_data/BCC/calibrated_models/F_det_F_model/P_(3, 3)_shear.txt}\data
\addplot [fill=color33,mark=none, draw=none, opacity=0.3] table {\data};

\pgfplotstableread{plot_data/BCC/calibrated_models/F_det_F_model/P_(1, 1)_shear_min.txt}\data
\addplot [color11, fill=none,mark=none, very thick] table {\data};
\pgfplotstableread{plot_data/BCC/calibrated_models/F_det_F_model/P_(1, 1)_shear_max.txt}\data
\addplot [color11, fill=none,mark=none, very thick] table {\data};
\pgfplotstableread{plot_data/BCC/calibrated_models/F_det_F_model/P_(1, 2)_shear_min.txt}\data
\addplot [color12, fill=none,mark=none, very thick] table {\data};
\pgfplotstableread{plot_data/BCC/calibrated_models/F_det_F_model/P_(1, 2)_shear_max.txt}\data
\addplot [color12, fill=none,mark=none, very thick] table {\data};
\pgfplotstableread{plot_data/BCC/calibrated_models/F_det_F_model/P_(1, 3)_shear_min.txt}\data
\addplot [color13, fill=none,mark=none, very thick] table {\data};
\pgfplotstableread{plot_data/BCC/calibrated_models/F_det_F_model/P_(1, 3)_shear_max.txt}\data
\addplot [color13, fill=none,mark=none, very thick] table {\data};
\pgfplotstableread{plot_data/BCC/calibrated_models/F_det_F_model/P_(2, 1)_shear_min.txt}\data
\addplot [color21, fill=none,mark=none, very thick] table {\data};
\pgfplotstableread{plot_data/BCC/calibrated_models/F_det_F_model/P_(2, 1)_shear_max.txt}\data
\addplot [color21, fill=none,mark=none, very thick] table {\data};
\pgfplotstableread{plot_data/BCC/calibrated_models/F_det_F_model/P_(2, 2)_shear_min.txt}\data
\addplot [color22, fill=none,mark=none, very thick] table {\data};
\pgfplotstableread{plot_data/BCC/calibrated_models/F_det_F_model/P_(2, 2)_shear_max.txt}\data
\addplot [color22, fill=none,mark=none, very thick] table {\data};
\pgfplotstableread{plot_data/BCC/calibrated_models/F_det_F_model/P_(2, 3)_shear_min.txt}\data
\addplot [color23, fill=none,mark=none, very thick] table {\data};
\pgfplotstableread{plot_data/BCC/calibrated_models/F_det_F_model/P_(2, 3)_shear_max.txt}\data
\addplot [color23, fill=none,mark=none, very thick] table {\data};
\pgfplotstableread{plot_data/BCC/calibrated_models/F_det_F_model/P_(3, 1)_shear_min.txt}\data
\addplot [color31, fill=none,mark=none, very thick] table {\data};
\pgfplotstableread{plot_data/BCC/calibrated_models/F_det_F_model/P_(3, 1)_shear_max.txt}\data
\addplot [color31, fill=none,mark=none, very thick] table {\data};
\pgfplotstableread{plot_data/BCC/calibrated_models/F_det_F_model/P_(3, 2)_shear_min.txt}\data
\addplot [color32, fill=none,mark=none, very thick] table {\data};
\pgfplotstableread{plot_data/BCC/calibrated_models/F_det_F_model/P_(3, 2)_shear_max.txt}\data
\addplot [color32, fill=none,mark=none, very thick] table {\data};
\pgfplotstableread{plot_data/BCC/calibrated_models/F_det_F_model/P_(3, 3)_shear_min.txt}\data
\addplot [color33, fill=none,mark=none, very thick] table {\data};
\pgfplotstableread{plot_data/BCC/calibrated_models/F_det_F_model/P_(3, 3)_shear_max.txt}\data
\addplot [color33, fill=none,mark=none, very thick] table {\data};

    \nextgroupplot[legend pos = north west,legend columns=3, ylabel = {biaxial test - $S_{ij}$}, ytick={0,1100,2200},yticklabels={0,11,22},
width=0.49*\textwidth, height=0.38*\textwidth,
xtick={0.8,1.0,1.2,1.4},xticklabels={0.8,1.0,1.2,1.4},
ymin = -400, ymax = 2800,
xlabel={$F_{11}$},
xmin = 0.65, xmax = 1.45,
ytick pos = left] % test1 F det F model

\foreach \num in {11, 12, 13, 21, 22, 23, 31, 32, 33}{
 \addplot coordinates { (-2,-2) (-3,-3) };
 };
 
    \foreach \num in {9, 10, 11, 12, 13, 14, 15, 16, 17}{
    \addplot table [x =test1_F11,y index=\num, only marks] 
    {plot_data/BCC/test_data/test1.txt};
    };

\pgfplotstableread{plot_data/BCC/calibrated_models/F_det_F_model/P_(1, 1)_test1.txt}\data
\addplot [fill=color11,mark=none, draw=none, opacity=0.3] table {\data};
\pgfplotstableread{plot_data/BCC/calibrated_models/F_det_F_model/P_(1, 2)_test1.txt}\data
\addplot [fill=color12,mark=none, draw=none, opacity=0.3] table {\data};
\pgfplotstableread{plot_data/BCC/calibrated_models/F_det_F_model/P_(1, 3)_test1.txt}\data
\addplot [fill=color13,mark=none, draw=none, opacity=0.3] table {\data};
\pgfplotstableread{plot_data/BCC/calibrated_models/F_det_F_model/P_(2, 1)_test1.txt}\data
\addplot [fill=color21,mark=none, draw=none, opacity=0.3] table {\data};
\pgfplotstableread{plot_data/BCC/calibrated_models/F_det_F_model/P_(2, 2)_test1.txt}\data
\addplot [fill=color22,mark=none, draw=none, opacity=0.3] table {\data};
\pgfplotstableread{plot_data/BCC/calibrated_models/F_det_F_model/P_(2, 3)_test1.txt}\data
\addplot [fill=color23,mark=none, draw=none, opacity=0.3] table {\data};
\pgfplotstableread{plot_data/BCC/calibrated_models/F_det_F_model/P_(3, 1)_test1.txt}\data
\addplot [fill=color31,mark=none, draw=none, opacity=0.3] table {\data};
\pgfplotstableread{plot_data/BCC/calibrated_models/F_det_F_model/P_(3, 2)_test1.txt}\data
\addplot [fill=color32,mark=none, draw=none, opacity=0.3] table {\data};
\pgfplotstableread{plot_data/BCC/calibrated_models/F_det_F_model/P_(3, 3)_test1.txt}\data
\addplot [fill=color33,mark=none, draw=none, opacity=0.3] table {\data};

\pgfplotstableread{plot_data/BCC/calibrated_models/F_det_F_model/P_(1, 1)_test1_min.txt}\data
\addplot [color11, fill=none,mark=none, very thick] table {\data};
\pgfplotstableread{plot_data/BCC/calibrated_models/F_det_F_model/P_(1, 1)_test1_max.txt}\data
\addplot [color11, fill=none,mark=none, very thick] table {\data};
\pgfplotstableread{plot_data/BCC/calibrated_models/F_det_F_model/P_(1, 2)_test1_min.txt}\data
\addplot [color12, fill=none,mark=none, very thick] table {\data};
\pgfplotstableread{plot_data/BCC/calibrated_models/F_det_F_model/P_(1, 2)_test1_max.txt}\data
\addplot [color12, fill=none,mark=none, very thick] table {\data};
\pgfplotstableread{plot_data/BCC/calibrated_models/F_det_F_model/P_(1, 3)_test1_min.txt}\data
\addplot [color13, fill=none,mark=none, very thick] table {\data};
\pgfplotstableread{plot_data/BCC/calibrated_models/F_det_F_model/P_(1, 3)_test1_max.txt}\data
\addplot [color13, fill=none,mark=none, very thick] table {\data};
\pgfplotstableread{plot_data/BCC/calibrated_models/F_det_F_model/P_(2, 1)_test1_min.txt}\data
\addplot [color21, fill=none,mark=none, very thick] table {\data};
\pgfplotstableread{plot_data/BCC/calibrated_models/F_det_F_model/P_(2, 1)_test1_max.txt}\data
\addplot [color21, fill=none,mark=none, very thick] table {\data};
\pgfplotstableread{plot_data/BCC/calibrated_models/F_det_F_model/P_(2, 2)_test1_min.txt}\data
\addplot [color22, fill=none,mark=none, very thick] table {\data};
\pgfplotstableread{plot_data/BCC/calibrated_models/F_det_F_model/P_(2, 2)_test1_max.txt}\data
\addplot [color22, fill=none,mark=none, very thick] table {\data};
\pgfplotstableread{plot_data/BCC/calibrated_models/F_det_F_model/P_(2, 3)_test1_min.txt}\data
\addplot [color23, fill=none,mark=none, very thick] table {\data};
\pgfplotstableread{plot_data/BCC/calibrated_models/F_det_F_model/P_(2, 3)_test1_max.txt}\data
\addplot [color23, fill=none,mark=none, very thick] table {\data};
\pgfplotstableread{plot_data/BCC/calibrated_models/F_det_F_model/P_(3, 1)_test1_min.txt}\data
\addplot [color31, fill=none,mark=none, very thick] table {\data};
\pgfplotstableread{plot_data/BCC/calibrated_models/F_det_F_model/P_(3, 1)_test1_max.txt}\data
\addplot [color31, fill=none,mark=none, very thick] table {\data};
\pgfplotstableread{plot_data/BCC/calibrated_models/F_det_F_model/P_(3, 2)_test1_min.txt}\data
\addplot [color32, fill=none,mark=none, very thick] table {\data};
\pgfplotstableread{plot_data/BCC/calibrated_models/F_det_F_model/P_(3, 2)_test1_max.txt}\data
\addplot [color32, fill=none,mark=none, very thick] table {\data};
\pgfplotstableread{plot_data/BCC/calibrated_models/F_det_F_model/P_(3, 3)_test1_min.txt}\data
\addplot [color33, fill=none,mark=none, very thick] table {\data};
\pgfplotstableread{plot_data/BCC/calibrated_models/F_det_F_model/P_(3, 3)_test1_max.txt}\data
\addplot [color33, fill=none,mark=none, very thick] table {\data};
   
       \nextgroupplot[ytick pos = left, ylabel = {mixed test - $S_{ij}$},xlabel={$F_{11}$}, ytick={0,800,1600},yticklabels={0,8,16},
width=0.49*\textwidth, height=0.38*\textwidth,
xtick={0.6,1.0,1.4},xticklabels={0.6,1.0,1.4},
ymin = -450, ymax = 2200,
xmin = 0.5, xmax = 1.45] % test3 F det F model

\foreach \num in {11, 12, 13, 21, 22, 23, 31, 32, 33}{
 \addplot coordinates { (-2,-2) (-3,-3) };
 };
     \foreach \num in {9, 10, 11, 12, 13, 14, 15, 16, 17}{
    \addplot table [x =test3_F11,y index=\num, only marks] 
    {plot_data/BCC/test_data/test3.txt};
    };
\pgfplotstableread{plot_data/BCC/calibrated_models/F_det_F_model/P_(1, 1)_test3.txt}\data
\addplot [fill=color11,mark=none, draw=none, opacity=0.3] table {\data};
\pgfplotstableread{plot_data/BCC/calibrated_models/F_det_F_model/P_(1, 2)_test3.txt}\data
\addplot [fill=color12,mark=none, draw=none, opacity=0.3] table {\data};
\pgfplotstableread{plot_data/BCC/calibrated_models/F_det_F_model/P_(1, 3)_test3.txt}\data
\addplot [fill=color13,mark=none, draw=none, opacity=0.3] table {\data};
\pgfplotstableread{plot_data/BCC/calibrated_models/F_det_F_model/P_(2, 1)_test3.txt}\data
\addplot [fill=color21,mark=none, draw=none, opacity=0.3] table {\data};
\pgfplotstableread{plot_data/BCC/calibrated_models/F_det_F_model/P_(2, 2)_test3.txt}\data
\addplot [fill=color22,mark=none, draw=none, opacity=0.3] table {\data};
\pgfplotstableread{plot_data/BCC/calibrated_models/F_det_F_model/P_(2, 3)_test3.txt}\data
\addplot [fill=color23,mark=none, draw=none, opacity=0.3] table {\data};
\pgfplotstableread{plot_data/BCC/calibrated_models/F_det_F_model/P_(3, 1)_test3.txt}\data
\addplot [fill=color31,mark=none, draw=none, opacity=0.3] table {\data};
\pgfplotstableread{plot_data/BCC/calibrated_models/F_det_F_model/P_(3, 2)_test3.txt}\data
\addplot [fill=color32,mark=none, draw=none, opacity=0.3] table {\data};
\pgfplotstableread{plot_data/BCC/calibrated_models/F_det_F_model/P_(3, 3)_test3.txt}\data
\addplot [fill=color33,mark=none, draw=none, opacity=0.3] table {\data};

\pgfplotstableread{plot_data/BCC/calibrated_models/F_det_F_model/P_(1, 1)_test3_min.txt}\data
\addplot [color11, fill=none,mark=none, very thick] table {\data};
\pgfplotstableread{plot_data/BCC/calibrated_models/F_det_F_model/P_(1, 1)_test3_max.txt}\data
\addplot [color11, fill=none,mark=none, very thick] table {\data};
\pgfplotstableread{plot_data/BCC/calibrated_models/F_det_F_model/P_(1, 2)_test3_min.txt}\data
\addplot [color12, fill=none,mark=none, very thick] table {\data};
\pgfplotstableread{plot_data/BCC/calibrated_models/F_det_F_model/P_(1, 2)_test3_max.txt}\data
\addplot [color12, fill=none,mark=none, very thick] table {\data};
\pgfplotstableread{plot_data/BCC/calibrated_models/F_det_F_model/P_(1, 3)_test3_min.txt}\data
\addplot [color13, fill=none,mark=none, very thick] table {\data};
\pgfplotstableread{plot_data/BCC/calibrated_models/F_det_F_model/P_(1, 3)_test3_max.txt}\data
\addplot [color13, fill=none,mark=none, very thick] table {\data};
\pgfplotstableread{plot_data/BCC/calibrated_models/F_det_F_model/P_(2, 1)_test3_min.txt}\data
\addplot [color21, fill=none,mark=none, very thick] table {\data};
\pgfplotstableread{plot_data/BCC/calibrated_models/F_det_F_model/P_(2, 1)_test3_max.txt}\data
\addplot [color21, fill=none,mark=none, very thick] table {\data};
\pgfplotstableread{plot_data/BCC/calibrated_models/F_det_F_model/P_(2, 2)_test3_min.txt}\data
\addplot [color22, fill=none,mark=none, very thick] table {\data};
\pgfplotstableread{plot_data/BCC/calibrated_models/F_det_F_model/P_(2, 2)_test3_max.txt}\data
\addplot [color22, fill=none,mark=none, very thick] table {\data};
\pgfplotstableread{plot_data/BCC/calibrated_models/F_det_F_model/P_(2, 3)_test3_min.txt}\data
\addplot [color23, fill=none,mark=none, very thick] table {\data};
\pgfplotstableread{plot_data/BCC/calibrated_models/F_det_F_model/P_(2, 3)_test3_max.txt}\data
\addplot [color23, fill=none,mark=none, very thick] table {\data};
\pgfplotstableread{plot_data/BCC/calibrated_models/F_det_F_model/P_(3, 1)_test3_min.txt}\data
\addplot [color31, fill=none,mark=none, very thick] table {\data};
\pgfplotstableread{plot_data/BCC/calibrated_models/F_det_F_model/P_(3, 1)_test3_max.txt}\data
\addplot [color31, fill=none,mark=none, very thick] table {\data};
\pgfplotstableread{plot_data/BCC/calibrated_models/F_det_F_model/P_(3, 2)_test3_min.txt}\data
\addplot [color32, fill=none,mark=none, very thick] table {\data};
\pgfplotstableread{plot_data/BCC/calibrated_models/F_det_F_model/P_(3, 2)_test3_max.txt}\data
\addplot [color32, fill=none,mark=none, very thick] table {\data};
\pgfplotstableread{plot_data/BCC/calibrated_models/F_det_F_model/P_(3, 3)_test3_min.txt}\data
\addplot [color33, fill=none,mark=none, very thick] table {\data};
\pgfplotstableread{plot_data/BCC/calibrated_models/F_det_F_model/P_(3, 3)_test3_max.txt}\data
\addplot [color33, fill=none,mark=none, very thick] table {\data};
\end{groupplot}
\end{tikzpicture}
}
\caption{Evaluation of $W^{\text{F}}$ for the cubic BCC cell. Points depict the simulation data, lines depict the calibrated model. Normal components of the stress tensor in red colors, shear components in blue, less important components in gray; stress in $\left[\text{hPa}\right]$.}
\label{fig:BCC_deformation_gradient_model}
\end{figure}

%% file: chapter/figures/X_deformation_gradient_model.tex
\tikzsetnextfilename{X_deformation_gradient_model}

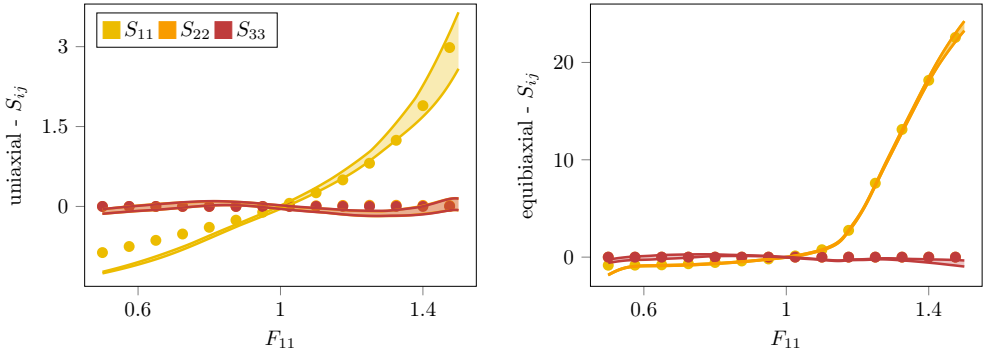
\begin{figure}[tp]
\centering
\pgfplotstableset{
create on use/uniax_F11/.style={create col/copy column from table={plot_data/X/calibration_data/uniaxial.txt}{0}}
}

\pgfplotstableset{
create on use/biaxial_F11/.style={create col/copy column from table={plot_data/X/calibration_data/biaxial.txt}{0}}
}

\resizebox{0.8\textwidth}{!}{
\begin{tikzpicture}
\pgfplotsset{
  set layers,% using layers
  mark layer=axis tick labels% defines the layer of the marks
}
\begin{groupplot}[
	group style = {group size = 2 by 1, vertical sep = 0.09*\textwidth,
								horizontal sep = 0.115*\textwidth},
	cycle list name=mycolorlistWFXcell,
	xtick align = outside,
	ytick align = outside,
	xtick pos = left,
	%ytick pos = left
	]
\nextgroupplot[legend pos = north west,legend columns=3, ylabel = {uniaxial - $S_{ij}$}, ytick={0,150,300},yticklabels={0,1.5,3},
width=0.49*\textwidth, height=0.38*\textwidth,
xtick={0.6,1,1.4},xticklabels={0.6,1,1.4},
ymin = -150, ymax = 380,
xlabel={$F_{11}$},
xmin = 0.45, xmax = 1.55,
ytick pos = left] % uniaxial F det F model
\foreach \num in {11, 22, 33}{
 \addplot coordinates { (-2,-2) (-3,-3) };
\addlegendentryexpanded{$S_{\num}$}
 };
    \foreach \num in {9, 12, 17}{
    \addplot table [x =uniax_F11,y index=\num, only marks] 
    {plot_data/X/calibration_data/uniaxial.txt};
    };
\pgfplotstableread{plot_data/X/calibrated_models/F_det_F_model/P_(1, 1)_uniaxial.txt}\data
\addplot [fill=color11,mark=none, draw=none, opacity=0.3] table {\data};
\pgfplotstableread{plot_data/X/calibrated_models/F_det_F_model/P_(2, 2)_uniaxial.txt}\data
\addplot [fill=color22,mark=none, draw=none, opacity=0.3] table {\data};
\pgfplotstableread{plot_data/X/calibrated_models/F_det_F_model/P_(3, 3)_uniaxial.txt}\data
\addplot [fill=color33,mark=none, draw=none, opacity=0.3] table {\data};

\pgfplotstableread{plot_data/X/calibrated_models/F_det_F_model/P_(1, 1)_uniaxial_min.txt}\data
\addplot [color11, fill=none,mark=none, very thick] table {\data};
\pgfplotstableread{plot_data/X/calibrated_models/F_det_F_model/P_(1, 1)_uniaxial_max.txt}\data
\addplot [color11, fill=none,mark=none, very thick] table {\data};
\pgfplotstableread{plot_data/X/calibrated_models/F_det_F_model/P_(2, 2)_uniaxial_min.txt}\data
\addplot [color22, fill=none,mark=none, very thick] table {\data};
\pgfplotstableread{plot_data/X/calibrated_models/F_det_F_model/P_(2, 2)_uniaxial_max.txt}\data
\addplot [color22, fill=none,mark=none, very thick] table {\data};
\pgfplotstableread{plot_data/X/calibrated_models/F_det_F_model/P_(3, 3)_uniaxial_min.txt}\data
\addplot [color33, fill=none,mark=none, very thick] table {\data};
\pgfplotstableread{plot_data/X/calibrated_models/F_det_F_model/P_(3, 3)_uniaxial_max.txt}\data
\addplot [color33, fill=none,mark=none, very thick] table {\data};

   \nextgroupplot[ytick pos = left, ylabel = {equibiaxial - $S_{ij}$},xlabel={$F_{11}$}, ytick={0,1000,2000},yticklabels={0,10,20},
width=0.49*\textwidth, height=0.38*\textwidth,
xtick={0.6,1,1.4},xticklabels={0.6,1,1.4},
ymin = -300, ymax = 2600,
xmin = 0.45, xmax = 1.55] % biaxial F det F model
\foreach \num in {11, 22, 33}{
 \addplot coordinates { (-2,-2) (-3,-3) };
 };
    \foreach \num in {9,12,17}{
    \addplot table [x =biaxial_F11,y index=\num,  only marks] 
    {plot_data/X/calibration_data/biaxial.txt};
    };
\pgfplotstableread{plot_data/X/calibrated_models/F_det_F_model/P_(1, 1)_biaxial.txt}\data
\addplot [fill=color11,mark=none, draw=none, opacity=0.3] table {\data};
\pgfplotstableread{plot_data/X/calibrated_models/F_det_F_model/P_(2, 2)_biaxial.txt}\data
\addplot [fill=color22,mark=none, draw=none, opacity=0.3] table {\data};
\pgfplotstableread{plot_data/X/calibrated_models/F_det_F_model/P_(3, 3)_biaxial.txt}\data
\addplot [fill=color33,mark=none, draw=none, opacity=0.3] table {\data};

\pgfplotstableread{plot_data/X/calibrated_models/F_det_F_model/P_(1, 1)_biaxial_min.txt}\data
\addplot [color11, fill=none,mark=none, very thick] table {\data};
\pgfplotstableread{plot_data/X/calibrated_models/F_det_F_model/P_(1, 1)_biaxial_max.txt}\data
\addplot [color11, fill=none,mark=none, very thick] table {\data};
\pgfplotstableread{plot_data/X/calibrated_models/F_det_F_model/P_(2, 2)_biaxial_min.txt}\data
\addplot [color22, fill=none,mark=none, very thick] table {\data};
\pgfplotstableread{plot_data/X/calibrated_models/F_det_F_model/P_(2, 2)_biaxial_max.txt}\data
\addplot [color22, fill=none,mark=none, very thick] table {\data};
\pgfplotstableread{plot_data/X/calibrated_models/F_det_F_model/P_(3, 3)_biaxial_min.txt}\data
\addplot [color33, fill=none,mark=none, very thick] table {\data};
\pgfplotstableread{plot_data/X/calibrated_models/F_det_F_model/P_(3, 3)_biaxial_max.txt}\data
\addplot [color33, fill=none,mark=none, very thick] table {\data};

\end{groupplot}
\end{tikzpicture}
}
\caption{Evaluation of $W^{\text{F}}$ for the cubic X cell. Points depict the simulation data, while lines and shaded areas depict the calibrated model; stress in $[\text{hPa}]$.}
\label{fig:F_model_X_cell}
\end{figure}

%% file: chapter/05_ti.tex
\section{Application to transverse isotropy}
\label{sec:application_to_ti}
\tDK{
After the detailed examinations for cubic lattice metamaterials in Section~\ref{sec:application_to_cubic_metamaterials}, the application of the polyconvex ML models to transverse isotropy is now briefly discussed. In doing so, we demonstrate the straightforward applicability of our models to other symmetry groups. Also, for the highly challenging behavior of cubic lattice metamaterials, the invariant-based model $W^\text{I}$ showed poor approximation quality. With the following example we show that, for a wide range of materials, $W^\text{I}$ can still be an appropriate choice.

\subsection{Data generation}

\paragraph{Analytical transversely isotropic potential}
For the following investigations, we generate data with the polyconvex model proposed by \textcite{Schroeder2008}, which is applicable to several symmetry groups, including transverse isotropy. In the following, the preferred axis of the transversely isotropic symmetry group \eqref{eq:symm_group_ti} is chosen as the $x_1$-axis, which motivates the second order structural tensor
\begin{equation}\label{eq:struct_ti}
    G_{\text{ti}}=\diag \left(\beta^2,\,\frac{1}{\beta},\,\frac{1}{\beta}\right)\,.
\end{equation}
Using this structural tensor, the two transversely isotropic invariants
\begin{equation}\label{eq:invariants_ti}
\begin{aligned}
J_4 = \tr \left(C \,G_{\text{ti}}\right)
\,,\quad\quad\quad\quad\quad
J_5 = \tr \left(\cof (C) \,G_{\text{ti}}\right)
\end{aligned}
\end{equation}
can be derived, which are convex in $F$ and $\cof F$, respectively \cite{Schroeder2008}. Together with the isotropic invariants $I_{1-3}$, \textcite{Schroeder2008} proposed the potential
\begin{equation}\label{eq:SNE_ti}
    W^\text{SNE}_\text{ti} = \alpha_1 \,I_1+\alpha_2 \,I_2 + \delta_1 \,I_3 - \delta_2 \,\log \left(\sqrt{I_3}\right)+\frac{\eta_1}{\alpha_4 \left( \tr G_\text{ti}\right)^{\alpha_4}}\,\left(J_4^{\alpha_4}+J_5^{\alpha_4}\right)\, ,
\end{equation}
which is objective by construction, and polyconvex if all parameters are equal to or greater than zero. The parameter $\delta_2$ depends on the other parameters and is chosen such that the model is stress-free in the reference configuration. Note that the parameter $\beta$ of the structural tensor \eqref{eq:struct_ti} needs to be specified as well. Here, we use the parameter values $\left(\beta,\,\alpha_1, \,\alpha_2,\,\delta_1,\,\delta_2,\,\alpha_4,\,\eta_1\right)=(2,\,8,\,0,\,10,\,56,\,2,\,10)$, which were fitted in \cite{Schroeder2008} to referential data of a not further specified real-world material.

Using the potential \eqref{eq:SNE_ti}, transversely isotropic data can be generated. As was already the case for the numerical homogenization of cubic lattice metamaterials, the analytical potential \eqref{eq:SNE_ti} provides both energy and stress values. However, for the following investigations, we will only make use of the stress values, which will demonstrate that energy values, which are typically not available in experiments, are not necessarily required to calibrate the proposed polyconvex ML models. 
Given as an analytical function, the potential \eqref{eq:SNE_ti} can directly be evaluated for a given deformation gradient $F\in\text{GL}^+(3)$. This leaves the question open of how to choose $F$ for the generation of calibration and test datasets.

\paragraph{Calibration dataset}
The calibration dataset $D_C$ consists of uniaxial and equibiaxial tensile tests, as well as a shear deformation. For the special cases of uniaxial and equibiaxial tensile tests, the corresponding boundary-value problem can be directly formulated as systems of non-linear equations. For uniaxial tension in $x_1$-direction, deformation gradient and stress tensor are given by
\begin{equation}
    F=\diag \left(F_{11},\,F_{22},\,F_{33}\right)\,,\quad\quad S=\diag \left(S_{11}(F),\, 0,\, 0\right)\, ,
\end{equation}
where $F_{11}$ is prescribed, $F_{22}=F_{33}$ are unknown and $S_{11}(F)=D_{F_{11}}W^\text{SNE}_\text{ti}(F)$. For equibiaxial tension in $x_1,x_2$-directions, 
\begin{equation}
    F=\diag \left(F_{11},\,F_{22},\,F_{33}\right)\,,\quad\quad S=\diag \left(S_{11}(F),\, S_{22}(F),\, 0\right)
\end{equation}
holds, where $F_{11}=F_{22}$, $F_{33}$ is unknown and $S_{11}(F)=D_{F_{11}}W^\text{SNE}_\text{ti}(F)$, $S_{22}(F)=D_{F_{22}}W^\text{SNE}_\text{ti}(F)$. These non-linear systems of equations are then solved with standard functions provided by MATLAB R2021a, which provides the overall deformation gradients for uniaxial and equibiaxial tensile tests. For this, $200$ equidistant values $F_{11}\in [0.5,\,2]$ are prescribed. The shear deformation $F=\uuI+\gamma(e_1\otimes e_2+e_2\otimes e_1)$ is evaluated for $250$ equidistant values $\gamma\in[0,\,0.5]$. Selected data points are visualized in Fig.~\ref{fig:TI_invariant_model} and Fig.~\ref{fig:TI_deformation_gradient_model}.

\paragraph{Test dataset}
The test dataset $D_T$ consists of a biaxial test and a combined tension-shear test. Note that the \enquote{biaxial test} and \enquote{mixed test} are different from the ones used in Sect.~\ref{sec:application_to_cubic_metamaterials}. For the biaxial test, the system of nonlinear equations
\begin{equation}
    F=\diag \left(F_{11},\,F_{22},\,F_{33}\right)\,,\quad\quad S=\diag \left(S_{11}(F),\, S_{22}(F),\, 0\right)
\end{equation}
with prescribed $F_{11}$, $F_{22}=0.5 \,F_{11}$ is solved for $F_{33}$ for 100 equidistant values $F_{11}\in[0.5,\,2]$. 
The mixed test case
\begin{equation}
    F=
    \begin{pmatrix}
    1+0.2\,\lambda & 0.2\, \lambda & 0 \\
    0 & 1+0.1\,\lambda & 0 \\
    0 & 0 & 1-0.1\,\lambda
    \end{pmatrix}
\end{equation}
is evaluated for $100$ equidistant values $\lambda\in[-1,\,2.5]$.

\medskip

After generation of the deformation gradients $F$ for all calibration and test cases, the first Piola-Kirchhoff stress $S=D_FW^\text{SNE}_\text{ti}(F)$ is evaluated, and the resulting datasets consist of tuples
\begin{equation}
D=\left\{(\uuF_1,\,{S}_1\right),\,\dotsc\}\, .
\end{equation}
Altogether, the calibration dataset $D_C$ consists of $650$ tuples, while the test dataset $D_T$ consists of $200$ tuples.

\subsection{Model preparation}

For the invariant-based ML model $W^{\text{I}}$, the two transversely isotropic invariants from \eqref{eq:invariants_ti} together with the three isotropic invariants from \eqref{eq:invariants} and the additional invariant $I_3^*=-2\sqrt{I_3}$ form the input $\mathcal{I}=\left(I_1,\,I_2,\,I_3,\,I_3^*,\,J_4,\, J_5\right)\in\mathbb{R}^6$ of the neural network. For the deformation gradient based model $W^{\text{F}}$, the input $\left(\uuF,\,\det \uuF\right)\in\mathbb{R}^{10}$ is chosen. 
For the network core of $W^{\text{I}}$, one hidden layer with eight nodes turned out as a sufficiently accurate choice, while for $W^{\text{F}}$ three hidden layers with $32$ nodes in each layer are chosen. 
As before, convexity is ensured by choice of the convex Softplus activation function in each node and restrictions on the network parameters, which are discussed in Section~\ref{subsec:prep} and Proposition~\ref{prop:LSE_SP_cores}, respectively. 
The transversely isotropic symmetry group $\cG_{\text{ti}}$ has an infinite number of elements, see \eqref{eq:symm_group_ti}. In order to apply the group symmetrization approach from \eqref{eq:group_symmetrization} on $W^{\text{F}}$, the group is approximated by six rotations around the $x_1$-direction, c.f. eq. \eqref{eq:symm_group_ti}. 

\medskip

For the model calibration, the MSE
\begin{equation}\label{eq:MSE2}
\begin{aligned}
\text{MSE}^\square\left(\up\right)=\frac{1}{\#\left(D\right)}\sum_{\uuF\in D} \frac{1}{9\text{Pa}^2}\left\lVert{S}\left(\uuF\right)-S^\square\left(\uuF;\,\up\right)\right\rVert^2
\end{aligned}
\end{equation}
with~$W^\square,\,\square\in\left\{\text{I},\,\text{F}\right\}$ is applied, c.f.~Sect.~\ref{subsec:prep}. Here, the MSE for the shear deformation is weighted twice, as its stress response is considerably lower than the stress response of the other calibration cases. For the data augmentation for objectivity of $W^{\text{F}}$, see  \eqref{eq:data_augmentation}, $128$ random rotation matrices are used, while for the group symmetrization for transverse isotropy of $W^{\text{F}}$, see \eqref{eq:group_symmetrization}, six equidistant rotations around the $x_1$-axis are applied.
Since both objectivity and material symmetry are only approximated for $W^{\text{F}}$, the model is evaluated with $1,024$ random observers for the verification of objectivity \eqref{eq:material_objectivity} and with $60$ equidistant rotations around the $x_1$-axis for the material symmetry \eqref{eq:material_symmetry}. Both models are initialized three times and each trained for $5,000$ epochs.

Further technical details are discussed in Sect.~\ref{subsec:prep}.
The MATLAB code used to generate calibration and test data, as well as compiled versions of both ML models and their sets of parameters are provided in the public GitHub repository \url{https://github.com/CPShub/sim-data}.

\subsection{Model evaluation}

\begin{table}[t!]
\tDK{
\centering
\begin{tabular}{lrcll}
\toprule
\multicolumn{2}{l}{Model} &  Param.& \multicolumn{2}{l}{MSE}  \\
&&&$D_T$&$D_C$\\ \midrule
$W^{\text{I}}$ with &$\SP\left[8\right]$   & $65$& $1.52\cdot10^{-1}$ & $2.90\cdot10^{-2}$  \\
&\multicolumn{2}{c}{------\textquotedbl------}& $7.84\cdot10^{-1}$ & $6.41\cdot10^{-2}$  \\
&\multicolumn{2}{c}{------\textquotedbl------}&  $8.47\cdot10^{-1}$ & $3.44\cdot10^{-2}$  \\
\midrule
$W^{\text{F}}$ with&$\SP\left[32,\,32,\,32\right]$&$737$ & $3.20\cdot10^0$ & $2.80\cdot10^{-1}$ \\
&\multicolumn{2}{c}{------\textquotedbl------}& $3.40\cdot10^0$ & $3.44\cdot10^{-1}$ \\
&\multicolumn{2}{c}{------\textquotedbl------}&  $4.12\cdot10^0$ & $2.95\cdot10^{-1}$\\
\bottomrule
\end{tabular}
\caption{MSEs of calibrated ML models for the transversely isotropic data.}
\label{tb:loss_TI}
}
\end{table}

\input{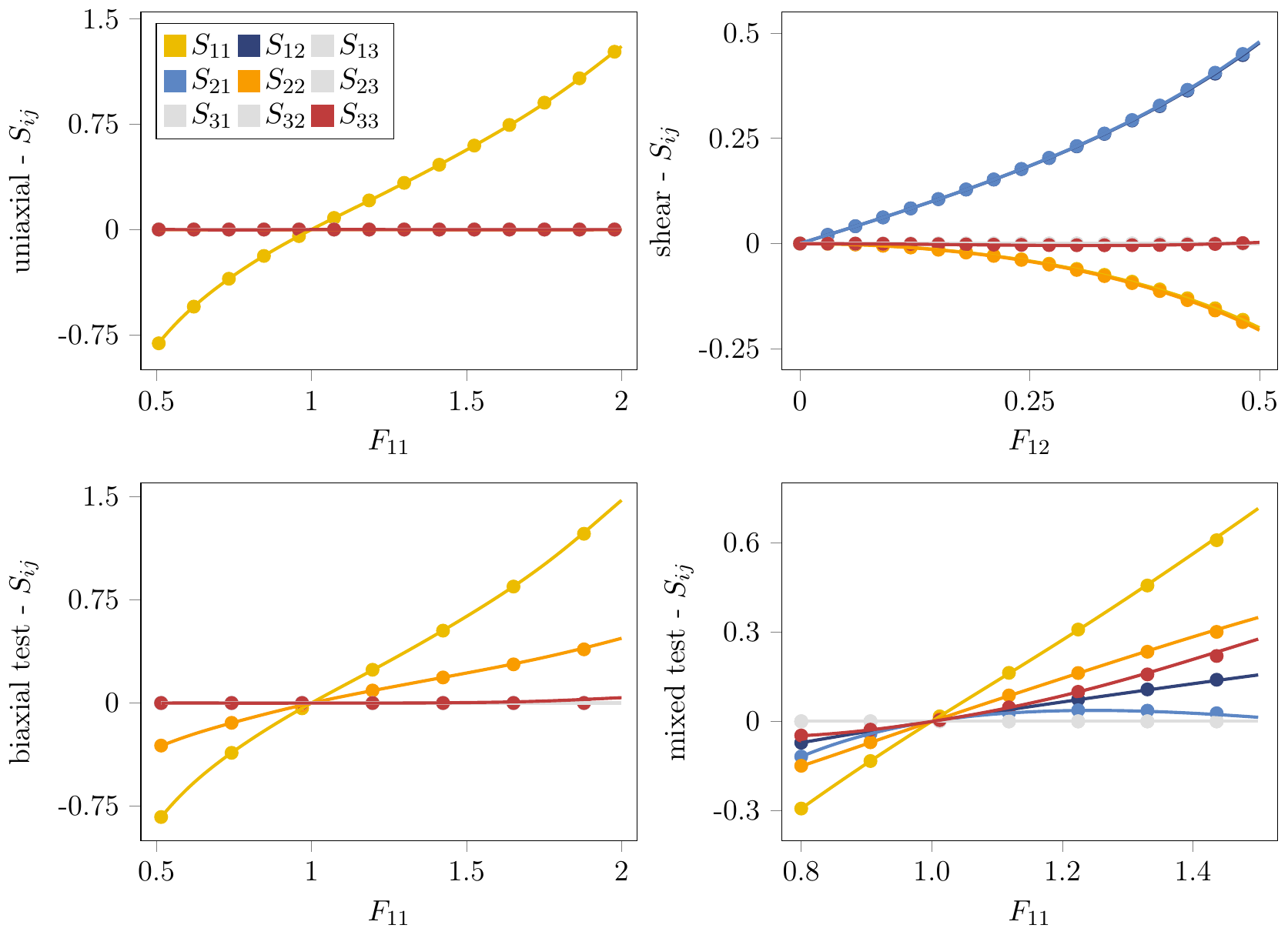}
\input{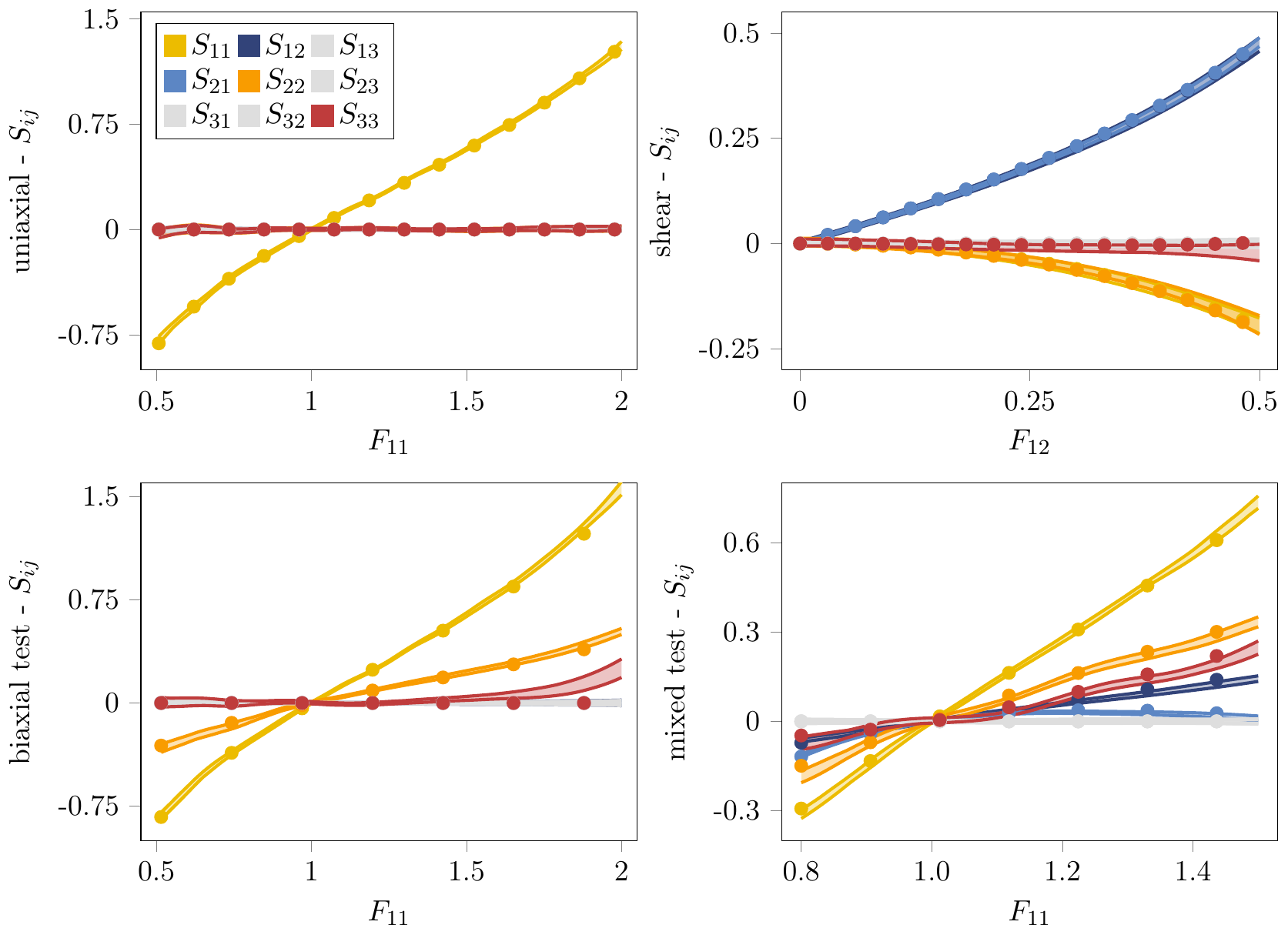}

In Table~\ref{tb:loss_TI}, the MSEs of the different model initializations are shown. The ML models show excellent agreement with both the calibration and test dataset. In particular, even though $W^{\text{I}}$ has only one layer with eight nodes, it can represent the data almost perfectly, which is most likely caused by the fact that the ML model $W^{\text{I}}$ uses the same invariants as the analytical potential \eqref{eq:SNE_ti}. A subset of the training dataset and both test cases are shown for $W^{\text{I}}$ and $W^{\text{F}}$ in Fig.~\ref{fig:TI_invariant_model} and \ref{fig:TI_deformation_gradient_model}, respectively. Again, the model $W^{\text{I}}$ shows excellent agreement for the transversely isotropic data. For the shear calibration case and the test cases, $W^{\text{F}}$ shows a slight dependence on the observer. Overall, both ML models are able to represent the analytical potential \eqref{eq:SNE_ti} very well, and here especially the invariant-based model $W^{\text{I}}$ shows excellent results.
As such analytical potentials are successfully applied in, e.g., modelling of soft biological tissues \cite{Balzani2006}, this implies that the polyconvex ML models are also applicable to a wide range of real-world materials.

}

%% file: chapter/figures/TI_invariant_model.tex
\tikzsetnextfilename{TI_invariant_model}

\begin{figure}[tp]
\centering
\pgfplotstableset{
create on use/F11_training/.style={create col/copy column from table={plot_data/TI/uniaxial.txt}{0}}
}
\pgfplotstableset{
create on use/F12/.style={create col/copy column from table={plot_data/TI/pure_shear.txt}{1}}
}

\pgfplotstableset{
create on use/F11_mixed_test/.style={create col/copy column from table={plot_data/TI/mixed_test.txt}{0}}
}

\pgfplotstableset{
create on use/F11_biaxial_test/.style={create col/copy column from table={plot_data/TI/biax_test.txt}{0}}
}

\resizebox{0.8\textwidth}{!}{
\begin{tikzpicture}
\pgfplotsset{
  set layers,% using layers
  mark layer=axis tick labels% defines the layer of the marks
}
\begin{groupplot}[
	group style = {group size = 2 by 2, vertical sep = 0.09*\textwidth,
								horizontal sep = 0.115*\textwidth},
	cycle list name=mycolorlist,
	xtick align = outside,
	ytick align = outside,
	xtick pos = left,
	%ytick pos = left
	]
\nextgroupplot[legend pos = north west,legend columns=3, ylabel = {uniaxial - $S_{ij}$}, ytick={-75, 0, 75, 150},yticklabels={-0.75, 0, 0.75, 1.5},
width=0.49*\textwidth, height=0.38*\textwidth,
xtick={0.5, 1, 1.5, 2},xticklabels={0.5, 1, 1.5, 2 },
ymin = -100, ymax = 155,
xlabel={$F_{11}$},
xmin = 0.45, xmax = 2.05,
ytick pos = left] % uniaxial 
\foreach \num in {11, 12, 13, 21, 22, 23, 31, 32, 33}{
 \addplot coordinates { (-2,-2) (-3,-3) };
\addlegendentryexpanded{$S_{\num}$}
 };
    \foreach \num in {0, 1, 2, 3, 4, 5, 6, 7, 8}{
    \addplot table [x =F11_training,y index=\num, only marks] 
    {plot_data/TI/P_uniaxial.txt};
    };
        \foreach \num in {0,1,2,3,4,5,6,7,8}{
    \addplot table [dashed, x =F11_training,y index=\num,  mark=none] {plot_data/TI/W_I/P_m_uniaxial.txt};
    };
        
   \nextgroupplot[ylabel = {shear - $S_{ij}$}, ytick={-25, 0, 25, 50},yticklabels={-0.25, 0, 0.25, 0.5},
width=0.49*\textwidth, height=0.38*\textwidth, xlabel={$F_{12}$},
xtick={0,0.25,0.5},xticklabels={0,0.25,0.5},
ymin = -30, ymax = 55,
xmin = -0.02, xmax = 0.52,
ytick pos = left] % shear 
\foreach \num in {11, 12, 13, 21, 22, 23, 31, 32, 33}{
 \addplot coordinates { (-2,-2) (-3,-3) };
 };
  \foreach \num in {0, 1, 2, 3, 4, 5, 6, 7, 8}{
   \addplot table [x =F12,y index=\num, only marks] 
    {plot_data/TI/P_pure_shear.txt};
    };
    
        \foreach \num in {0,1,2,3,4,5,6,7,8}{
    \addplot table [dashed, x =F12,y index=\num,  mark=none] {plot_data/TI/W_I/P_m_pure_shear.txt};
    }; 
    
    \nextgroupplot[legend pos = north west,legend columns=3, ylabel = {biaxial test - $S_{ij}$}, ytick={-75, 0, 75, 150},yticklabels={-0.75, 0, 0.75, 1.5},
width=0.49*\textwidth, height=0.38*\textwidth,
xtick={0.5, 1, 1.5, 2},xticklabels={0.5, 1, 1.5, 2},
ymin = -100, ymax = 160,
xlabel={$F_{11}$},
xmin = 0.45, xmax = 2.05,
ytick pos = left] % biaxial test 

\foreach \num in {11, 12, 13, 21, 22, 23, 31, 32, 33}{
 \addplot coordinates { (-2,-2) (-3,-3) };
 };
 
    \foreach \num in {0, 1, 2, 3, 4, 5, 6, 7, 8}{
    \addplot table [x =F11_biaxial_test,y index=\num, only marks] 
    {plot_data/TI/P_biax_test.txt};
    };

    \foreach \num in {0,1,2,3,4,5,6,7,8}{
    \addplot table [dashed, x =F11_biaxial_test,y index=\num] {plot_data/TI/W_I/P_m_biax_test.txt};
    };
    
    \nextgroupplot[legend pos = north west,legend columns=3, ylabel = {mixed test - $S_{ij}$}, ytick={-30, 0, 30, 60},yticklabels={-0.3, 0, 0.3, 0.6},
width=0.49*\textwidth, height=0.38*\textwidth,
xtick={0.8,1.0,1.2,1.4},xticklabels={0.8,1.0,1.2,1.4},
ymin = -40, ymax = 80,
xlabel={$F_{11}$},
xmin = 0.77, xmax = 1.53,
ytick pos = left] % mixed test 

\foreach \num in {11, 12, 13, 21, 22, 23, 31, 32, 33}{
 \addplot coordinates { (-2,-2) (-3,-3) };
 };
 
    \foreach \num in {0, 1, 2, 3, 4, 5, 6, 7, 8}{
    \addplot table [x =F11_mixed_test,y index=\num, only marks] 
    {plot_data/TI/P_mixed_test.txt};
    };

    \foreach \num in {0,1,2,3,4,5,6,7,8}{
    \addplot table [dashed, x =F11_mixed_test,y index=\num] {plot_data/TI/W_I/P_m_mixed_test.txt};
    };

\end{groupplot}
\end{tikzpicture}
}
\caption{Evaluation of $W^{\text{I}}$ for transverse isotropy. Points depict data from the analytical model $W^\text{SNE}_\text{ti}$, while lines depict the evaluation of the calibrated model $W^{\text{I}}$. Normal components of the stress tensor are shown in red colors, shear components in blue, less important components in gray; stress in $\left[\text{hPa}\right]$.}
\label{fig:TI_invariant_model}
\end{figure}
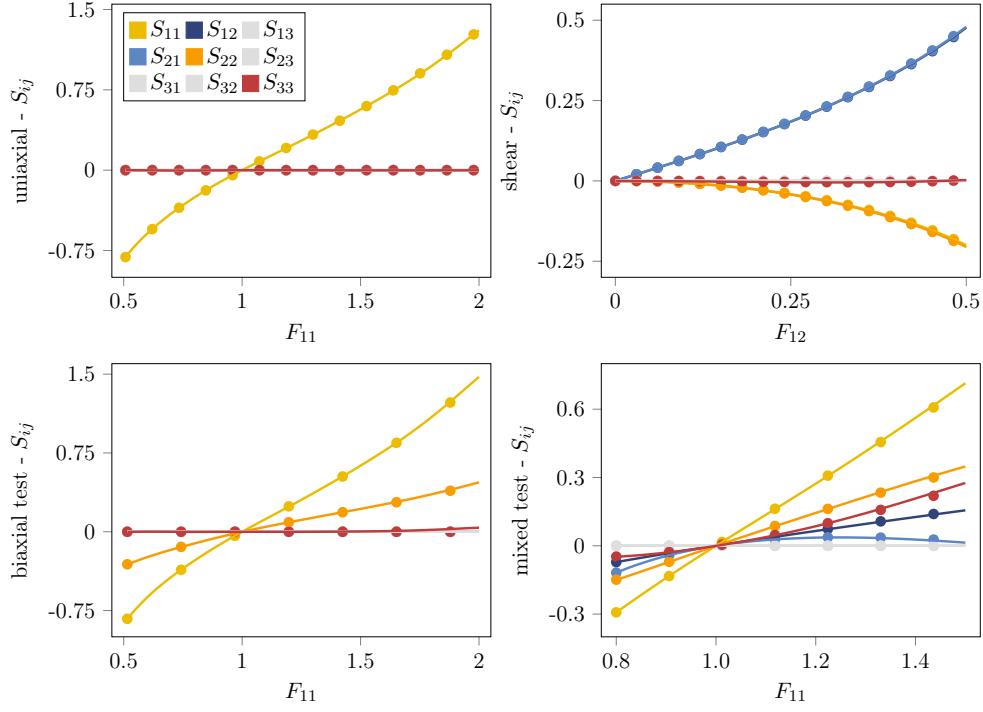

%% file: chapter/figures/TI_deformation_gradient_model.tex
\tikzsetnextfilename{TI_deformation_gradient_model}

\begin{figure}[tp]
\centering

\pgfplotstableset{
create on use/F11_training/.style={create col/copy column from table={plot_data/TI/uniaxial.txt}{0}}
}
\pgfplotstableset{
create on use/F12/.style={create col/copy column from table={plot_data/TI/pure_shear.txt}{1}}
}

\pgfplotstableset{
create on use/F11_mixed_test/.style={create col/copy column from table={plot_data/TI/mixed_test.txt}{0}}
}

\pgfplotstableset{
create on use/F11_biaxial_test/.style={create col/copy column from table={plot_data/TI/biax_test.txt}{0}}
}

\resizebox{0.8\textwidth}{!}{
\begin{tikzpicture}
\begin{groupplot}[
	group style = {group size = 2 by 2, vertical sep = 0.09*\textwidth,
								horizontal sep = 0.115*\textwidth},
	cycle list name=mycolorlist,
	xtick align = outside,
	ytick align = outside,
	xtick pos = left,
	%ytick pos = left
	]

\nextgroupplot[legend pos = north west,legend columns=3, ylabel = {uniaxial - $S_{ij}$}, ytick={-75, 0, 75, 150},yticklabels={-0.75, 0, 0.75, 1.5},
width=0.49*\textwidth, height=0.38*\textwidth,
xtick={0.5, 1, 1.5, 2},xticklabels={0.5, 1, 1.5, 2 },
ymin = -100, ymax = 155,
xlabel={$F_{11}$},
xmin = 0.45, xmax = 2.05,
ytick pos = left] % uniaxial 
\foreach \num in {11, 12, 13, 21, 22, 23, 31, 32, 33}{
 \addplot coordinates { (-2,-2) (-3,-3) };
\addlegendentryexpanded{$S_{\num}$}
 };

    \foreach \num in {0, 1, 2, 3, 4, 5, 6, 7, 8}{
    \addplot table [x =F11_training,y index=\num, only marks] 
    {plot_data/TI/P_uniaxial.txt};
    };

\pgfplotstableread{plot_data/TI/W_F/P_(1, 1)_uniaxial.txt}\data
\addplot [fill=color11,mark=none, draw=none, opacity=0.3] table {\data};
\pgfplotstableread{plot_data/TI/W_F/P_(1, 2)_uniaxial.txt}\data
\addplot [fill=color12,mark=none, draw=none, opacity=0.3] table {\data};
\pgfplotstableread{plot_data/TI/W_F/P_(1, 3)_uniaxial.txt}\data
\addplot [fill=color13,mark=none, draw=none, opacity=0.3] table {\data};
\pgfplotstableread{plot_data/TI/W_F/P_(2, 1)_uniaxial.txt}\data
\addplot [fill=color21,mark=none, draw=none, opacity=0.3] table {\data};
\pgfplotstableread{plot_data/TI/W_F/P_(2, 2)_uniaxial.txt}\data
\addplot [fill=color22,mark=none, draw=none, opacity=0.3] table {\data};
\pgfplotstableread{plot_data/TI/W_F/P_(2, 3)_uniaxial.txt}\data
\addplot [fill=color23,mark=none, draw=none, opacity=0.3] table {\data};
\pgfplotstableread{plot_data/TI/W_F/P_(3, 1)_uniaxial.txt}\data
\addplot [fill=color31,mark=none, draw=none, opacity=0.3] table {\data};
\pgfplotstableread{plot_data/TI/W_F/P_(3, 2)_uniaxial.txt}\data
\addplot [fill=color32,mark=none, draw=none, opacity=0.3] table {\data};
\pgfplotstableread{plot_data/TI/W_F/P_(3, 3)_uniaxial.txt}\data
\addplot [fill=color33,mark=none, draw=none, opacity=0.3] table {\data};

\pgfplotstableread{plot_data/TI/W_F/P_(1, 1)_uniaxial_min.txt}\data
\addplot [color11, fill=none,mark=none, very thick] table {\data};
\pgfplotstableread{plot_data/TI/W_F/P_(1, 1)_uniaxial_max.txt}\data
\addplot [color11, fill=none,mark=none, very thick] table {\data};
\pgfplotstableread{plot_data/TI/W_F/P_(1, 2)_uniaxial_min.txt}\data
\addplot [color12, fill=none,mark=none, very thick] table {\data};
\pgfplotstableread{plot_data/TI/W_F/P_(1, 2)_uniaxial_max.txt}\data
\addplot [color12, fill=none,mark=none, very thick] table {\data};
\pgfplotstableread{plot_data/TI/W_F/P_(1, 3)_uniaxial_min.txt}\data
\addplot [color13, fill=none,mark=none, very thick] table {\data};
\pgfplotstableread{plot_data/TI/W_F/P_(1, 3)_uniaxial_max.txt}\data
\addplot [color13, fill=none,mark=none, very thick] table {\data};
\pgfplotstableread{plot_data/TI/W_F/P_(2, 1)_uniaxial_min.txt}\data
\addplot [color21, fill=none,mark=none, very thick] table {\data};
\pgfplotstableread{plot_data/TI/W_F/P_(2, 1)_uniaxial_max.txt}\data
\addplot [color21, fill=none,mark=none, very thick] table {\data};
\pgfplotstableread{plot_data/TI/W_F/P_(2, 2)_uniaxial_min.txt}\data
\addplot [color22, fill=none,mark=none, very thick] table {\data};
\pgfplotstableread{plot_data/TI/W_F/P_(2, 2)_uniaxial_max.txt}\data
\addplot [color22, fill=none,mark=none, very thick] table {\data};
\pgfplotstableread{plot_data/TI/W_F/P_(2, 3)_uniaxial_min.txt}\data
\addplot [color23, fill=none,mark=none, very thick] table {\data};
\pgfplotstableread{plot_data/TI/W_F/P_(2, 3)_uniaxial_max.txt}\data
\addplot [color23, fill=none,mark=none, very thick] table {\data};
\pgfplotstableread{plot_data/TI/W_F/P_(3, 1)_uniaxial_min.txt}\data
\addplot [color31, fill=none,mark=none, very thick] table {\data};
\pgfplotstableread{plot_data/TI/W_F/P_(3, 1)_uniaxial_max.txt}\data
\addplot [color31, fill=none,mark=none, very thick] table {\data};
\pgfplotstableread{plot_data/TI/W_F/P_(3, 2)_uniaxial_min.txt}\data
\addplot [color32, fill=none,mark=none, very thick] table {\data};
\pgfplotstableread{plot_data/TI/W_F/P_(3, 2)_uniaxial_max.txt}\data
\addplot [color32, fill=none,mark=none, very thick] table {\data};
\pgfplotstableread{plot_data/TI/W_F/P_(3, 3)_uniaxial_min.txt}\data
\addplot [color33, fill=none,mark=none, very thick] table {\data};
\pgfplotstableread{plot_data/TI/W_F/P_(3, 3)_uniaxial_max.txt}\data
\addplot [color33, fill=none,mark=none, very thick] table {\data};

   \nextgroupplot[ylabel = {shear - $S_{ij}$}, ytick={-25, 0, 25, 50},yticklabels={-0.25, 0, 0.25, 0.5},
width=0.49*\textwidth, height=0.38*\textwidth, xlabel={$F_{12}$},
xtick={0,0.25,0.5},xticklabels={0,0.25,0.5},
ymin = -30, ymax = 55,
xmin = -0.02, xmax = 0.52,
ytick pos = left] % shear 
\foreach \num in {11, 12, 13, 21, 22, 23, 31, 32, 33}{
 \addplot coordinates { (-2,-2) (-3,-3) };
 };
  \foreach \num in {0, 1, 2, 3, 4, 5, 6, 7, 8}{
   \addplot table [x =F12,y index=\num, only marks] 
    {plot_data/TI/P_pure_shear.txt};
    };

\pgfplotstableread{plot_data/TI/W_F/P_(1, 1)_pure_shear.txt}\data
\addplot [fill=color11,mark=none, draw=none, opacity=0.3] table {\data};
\pgfplotstableread{plot_data/TI/W_F/P_(1, 2)_pure_shear.txt}\data
\addplot [fill=color12,mark=none, draw=none, opacity=0.3] table {\data};
\pgfplotstableread{plot_data/TI/W_F/P_(1, 3)_pure_shear.txt}\data
\addplot [fill=color13,mark=none, draw=none, opacity=0.3] table {\data};
\pgfplotstableread{plot_data/TI/W_F/P_(2, 1)_pure_shear.txt}\data
\addplot [fill=color21,mark=none, draw=none, opacity=0.3] table {\data};
\pgfplotstableread{plot_data/TI/W_F/P_(2, 2)_pure_shear.txt}\data
\addplot [fill=color22,mark=none, draw=none, opacity=0.3] table {\data};
\pgfplotstableread{plot_data/TI/W_F/P_(2, 3)_pure_shear.txt}\data
\addplot [fill=color23,mark=none, draw=none, opacity=0.3] table {\data};
\pgfplotstableread{plot_data/TI/W_F/P_(3, 1)_pure_shear.txt}\data
\addplot [fill=color31,mark=none, draw=none, opacity=0.3] table {\data};
\pgfplotstableread{plot_data/TI/W_F/P_(3, 2)_pure_shear.txt}\data
\addplot [fill=color32,mark=none, draw=none, opacity=0.3] table {\data};
\pgfplotstableread{plot_data/TI/W_F/P_(3, 3)_pure_shear.txt}\data
\addplot [fill=color33,mark=none, draw=none, opacity=0.3] table {\data};

\pgfplotstableread{plot_data/TI/W_F/P_(1, 1)_pure_shear_min.txt}\data
\addplot [color11, fill=none,mark=none, very thick] table {\data};
\pgfplotstableread{plot_data/TI/W_F/P_(1, 1)_pure_shear_max.txt}\data
\addplot [color11, fill=none,mark=none, very thick] table {\data};
\pgfplotstableread{plot_data/TI/W_F/P_(1, 2)_pure_shear_min.txt}\data
\addplot [color12, fill=none,mark=none, very thick] table {\data};
\pgfplotstableread{plot_data/TI/W_F/P_(1, 2)_pure_shear_max.txt}\data
\addplot [color12, fill=none,mark=none, very thick] table {\data};
\pgfplotstableread{plot_data/TI/W_F/P_(1, 3)_pure_shear_min.txt}\data
\addplot [color13, fill=none,mark=none, very thick] table {\data};
\pgfplotstableread{plot_data/TI/W_F/P_(1, 3)_pure_shear_max.txt}\data
\addplot [color13, fill=none,mark=none, very thick] table {\data};
\pgfplotstableread{plot_data/TI/W_F/P_(2, 1)_pure_shear_min.txt}\data
\addplot [color21, fill=none,mark=none, very thick] table {\data};
\pgfplotstableread{plot_data/TI/W_F/P_(2, 1)_pure_shear_max.txt}\data
\addplot [color21, fill=none,mark=none, very thick] table {\data};
\pgfplotstableread{plot_data/TI/W_F/P_(2, 2)_pure_shear_min.txt}\data
\addplot [color22, fill=none,mark=none, very thick] table {\data};
\pgfplotstableread{plot_data/TI/W_F/P_(2, 2)_pure_shear_max.txt}\data
\addplot [color22, fill=none,mark=none, very thick] table {\data};
\pgfplotstableread{plot_data/TI/W_F/P_(2, 3)_pure_shear_min.txt}\data
\addplot [color23, fill=none,mark=none, very thick] table {\data};
\pgfplotstableread{plot_data/TI/W_F/P_(2, 3)_pure_shear_max.txt}\data
\addplot [color23, fill=none,mark=none, very thick] table {\data};
\pgfplotstableread{plot_data/TI/W_F/P_(3, 1)_pure_shear_min.txt}\data
\addplot [color31, fill=none,mark=none, very thick] table {\data};
\pgfplotstableread{plot_data/TI/W_F/P_(3, 1)_pure_shear_max.txt}\data
\addplot [color31, fill=none,mark=none, very thick] table {\data};
\pgfplotstableread{plot_data/TI/W_F/P_(3, 2)_pure_shear_min.txt}\data
\addplot [color32, fill=none,mark=none, very thick] table {\data};
\pgfplotstableread{plot_data/TI/W_F/P_(3, 2)_pure_shear_max.txt}\data
\addplot [color32, fill=none,mark=none, very thick] table {\data};
\pgfplotstableread{plot_data/TI/W_F/P_(3, 3)_pure_shear_min.txt}\data
\addplot [color33, fill=none,mark=none, very thick] table {\data};
\pgfplotstableread{plot_data/TI/W_F/P_(3, 3)_pure_shear_max.txt}\data
\addplot [color33, fill=none,mark=none, very thick] table {\data};

    \nextgroupplot[legend pos = north west,legend columns=3, ylabel = {biaxial test - $S_{ij}$}, ytick={-75, 0, 75, 150},yticklabels={-0.75, 0, 0.75, 1.5},
width=0.49*\textwidth, height=0.38*\textwidth,
xtick={0.5, 1, 1.5, 2},xticklabels={0.5, 1, 1.5, 2},
ymin = -100, ymax = 160,
xlabel={$F_{11}$},
xmin = 0.45, xmax = 2.05,
ytick pos = left] % biaxial test 

\foreach \num in {11, 12, 13, 21, 22, 23, 31, 32, 33}{
 \addplot coordinates { (-2,-2) (-3,-3) };
 };
 
    \foreach \num in {0, 1, 2, 3, 4, 5, 6, 7, 8}{
    \addplot table [x =F11_biaxial_test,y index=\num, only marks] 
    {plot_data/TI/P_biax_test.txt};
    };

    \pgfplotstableread{plot_data/TI/W_F/P_(1, 1)_biax_test.txt}\data
\addplot [fill=color11,mark=none, draw=none, opacity=0.3] table {\data};
\pgfplotstableread{plot_data/TI/W_F/P_(1, 2)_biax_test.txt}\data
\addplot [fill=color12,mark=none, draw=none, opacity=0.3] table {\data};
\pgfplotstableread{plot_data/TI/W_F/P_(1, 3)_biax_test.txt}\data
\addplot [fill=color13,mark=none, draw=none, opacity=0.3] table {\data};
\pgfplotstableread{plot_data/TI/W_F/P_(2, 1)_biax_test.txt}\data
\addplot [fill=color21,mark=none, draw=none, opacity=0.3] table {\data};
\pgfplotstableread{plot_data/TI/W_F/P_(2, 2)_biax_test.txt}\data
\addplot [fill=color22,mark=none, draw=none, opacity=0.3] table {\data};
\pgfplotstableread{plot_data/TI/W_F/P_(2, 3)_biax_test.txt}\data
\addplot [fill=color23,mark=none, draw=none, opacity=0.3] table {\data};
\pgfplotstableread{plot_data/TI/W_F/P_(3, 1)_biax_test.txt}\data
\addplot [fill=color31,mark=none, draw=none, opacity=0.3] table {\data};
\pgfplotstableread{plot_data/TI/W_F/P_(3, 2)_biax_test.txt}\data
\addplot [fill=color32,mark=none, draw=none, opacity=0.3] table {\data};
\pgfplotstableread{plot_data/TI/W_F/P_(3, 3)_biax_test.txt}\data
\addplot [fill=color33,mark=none, draw=none, opacity=0.3] table {\data};

\pgfplotstableread{plot_data/TI/W_F/P_(1, 1)_biax_test_min.txt}\data
\addplot [color11, fill=none,mark=none, very thick] table {\data};
\pgfplotstableread{plot_data/TI/W_F/P_(1, 1)_biax_test_max.txt}\data
\addplot [color11, fill=none,mark=none, very thick] table {\data};
\pgfplotstableread{plot_data/TI/W_F/P_(1, 2)_biax_test_min.txt}\data
\addplot [color12, fill=none,mark=none, very thick] table {\data};
\pgfplotstableread{plot_data/TI/W_F/P_(1, 2)_biax_test_max.txt}\data
\addplot [color12, fill=none,mark=none, very thick] table {\data};
\pgfplotstableread{plot_data/TI/W_F/P_(1, 3)_biax_test_min.txt}\data
\addplot [color13, fill=none,mark=none, very thick] table {\data};
\pgfplotstableread{plot_data/TI/W_F/P_(1, 3)_biax_test_max.txt}\data
\addplot [color13, fill=none,mark=none, very thick] table {\data};
\pgfplotstableread{plot_data/TI/W_F/P_(2, 1)_biax_test_min.txt}\data
\addplot [color21, fill=none,mark=none, very thick] table {\data};
\pgfplotstableread{plot_data/TI/W_F/P_(2, 1)_biax_test_max.txt}\data
\addplot [color21, fill=none,mark=none, very thick] table {\data};
\pgfplotstableread{plot_data/TI/W_F/P_(2, 2)_biax_test_min.txt}\data
\addplot [color22, fill=none,mark=none, very thick] table {\data};
\pgfplotstableread{plot_data/TI/W_F/P_(2, 2)_biax_test_max.txt}\data
\addplot [color22, fill=none,mark=none, very thick] table {\data};
\pgfplotstableread{plot_data/TI/W_F/P_(2, 3)_biax_test_min.txt}\data
\addplot [color23, fill=none,mark=none, very thick] table {\data};
\pgfplotstableread{plot_data/TI/W_F/P_(2, 3)_biax_test_max.txt}\data
\addplot [color23, fill=none,mark=none, very thick] table {\data};
\pgfplotstableread{plot_data/TI/W_F/P_(3, 1)_biax_test_min.txt}\data
\addplot [color31, fill=none,mark=none, very thick] table {\data};
\pgfplotstableread{plot_data/TI/W_F/P_(3, 1)_biax_test_max.txt}\data
\addplot [color31, fill=none,mark=none, very thick] table {\data};
\pgfplotstableread{plot_data/TI/W_F/P_(3, 2)_biax_test_min.txt}\data
\addplot [color32, fill=none,mark=none, very thick] table {\data};
\pgfplotstableread{plot_data/TI/W_F/P_(3, 2)_biax_test_max.txt}\data
\addplot [color32, fill=none,mark=none, very thick] table {\data};
\pgfplotstableread{plot_data/TI/W_F/P_(3, 3)_biax_test_min.txt}\data
\addplot [color33, fill=none,mark=none, very thick] table {\data};
\pgfplotstableread{plot_data/TI/W_F/P_(3, 3)_biax_test_max.txt}\data
\addplot [color33, fill=none,mark=none, very thick] table {\data};

    \nextgroupplot[legend pos = north west,legend columns=3, ylabel = {mixed test - $S_{ij}$}, ytick={-30, 0, 30, 60},yticklabels={-0.3, 0, 0.3, 0.6},
width=0.49*\textwidth, height=0.38*\textwidth,
xtick={0.8,1.0,1.2,1.4},xticklabels={0.8,1.0,1.2,1.4},
ymin = -40, ymax = 80,
xlabel={$F_{11}$},
xmin = 0.77, xmax = 1.53,
ytick pos = left] % mixed test 

\foreach \num in {11, 12, 13, 21, 22, 23, 31, 32, 33}{
 \addplot coordinates { (-2,-2) (-3,-3) };
 };
 
    \foreach \num in {0, 1, 2, 3, 4, 5, 6, 7, 8}{
    \addplot table [x =F11_mixed_test,y index=\num, only marks] 
    {plot_data/TI/P_mixed_test.txt};
    };

        \pgfplotstableread{plot_data/TI/W_F/P_(1, 1)_mixed_test.txt}\data
\addplot [fill=color11,mark=none, draw=none, opacity=0.3] table {\data};
\pgfplotstableread{plot_data/TI/W_F/P_(1, 2)_mixed_test.txt}\data
\addplot [fill=color12,mark=none, draw=none, opacity=0.3] table {\data};
\pgfplotstableread{plot_data/TI/W_F/P_(1, 3)_mixed_test.txt}\data
\addplot [fill=color13,mark=none, draw=none, opacity=0.3] table {\data};
\pgfplotstableread{plot_data/TI/W_F/P_(2, 1)_mixed_test.txt}\data
\addplot [fill=color21,mark=none, draw=none, opacity=0.3] table {\data};
\pgfplotstableread{plot_data/TI/W_F/P_(2, 2)_mixed_test.txt}\data
\addplot [fill=color22,mark=none, draw=none, opacity=0.3] table {\data};
\pgfplotstableread{plot_data/TI/W_F/P_(2, 3)_mixed_test.txt}\data
\addplot [fill=color23,mark=none, draw=none, opacity=0.3] table {\data};
\pgfplotstableread{plot_data/TI/W_F/P_(3, 1)_mixed_test.txt}\data
\addplot [fill=color31,mark=none, draw=none, opacity=0.3] table {\data};
\pgfplotstableread{plot_data/TI/W_F/P_(3, 2)_mixed_test.txt}\data
\addplot [fill=color32,mark=none, draw=none, opacity=0.3] table {\data};
\pgfplotstableread{plot_data/TI/W_F/P_(3, 3)_mixed_test.txt}\data
\addplot [fill=color33,mark=none, draw=none, opacity=0.3] table {\data};

\pgfplotstableread{plot_data/TI/W_F/P_(1, 1)_mixed_test_min.txt}\data
\addplot [color11, fill=none,mark=none, very thick] table {\data};
\pgfplotstableread{plot_data/TI/W_F/P_(1, 1)_mixed_test_max.txt}\data
\addplot [color11, fill=none,mark=none, very thick] table {\data};
\pgfplotstableread{plot_data/TI/W_F/P_(1, 2)_mixed_test_min.txt}\data
\addplot [color12, fill=none,mark=none, very thick] table {\data};
\pgfplotstableread{plot_data/TI/W_F/P_(1, 2)_mixed_test_max.txt}\data
\addplot [color12, fill=none,mark=none, very thick] table {\data};
\pgfplotstableread{plot_data/TI/W_F/P_(1, 3)_mixed_test_min.txt}\data
\addplot [color13, fill=none,mark=none, very thick] table {\data};
\pgfplotstableread{plot_data/TI/W_F/P_(1, 3)_mixed_test_max.txt}\data
\addplot [color13, fill=none,mark=none, very thick] table {\data};
\pgfplotstableread{plot_data/TI/W_F/P_(2, 1)_mixed_test_min.txt}\data
\addplot [color21, fill=none,mark=none, very thick] table {\data};
\pgfplotstableread{plot_data/TI/W_F/P_(2, 1)_mixed_test_max.txt}\data
\addplot [color21, fill=none,mark=none, very thick] table {\data};
\pgfplotstableread{plot_data/TI/W_F/P_(2, 2)_mixed_test_min.txt}\data
\addplot [color22, fill=none,mark=none, very thick] table {\data};
\pgfplotstableread{plot_data/TI/W_F/P_(2, 2)_mixed_test_max.txt}\data
\addplot [color22, fill=none,mark=none, very thick] table {\data};
\pgfplotstableread{plot_data/TI/W_F/P_(2, 3)_mixed_test_min.txt}\data
\addplot [color23, fill=none,mark=none, very thick] table {\data};
\pgfplotstableread{plot_data/TI/W_F/P_(2, 3)_mixed_test_max.txt}\data
\addplot [color23, fill=none,mark=none, very thick] table {\data};
\pgfplotstableread{plot_data/TI/W_F/P_(3, 1)_mixed_test_min.txt}\data
\addplot [color31, fill=none,mark=none, very thick] table {\data};
\pgfplotstableread{plot_data/TI/W_F/P_(3, 1)_mixed_test_max.txt}\data
\addplot [color31, fill=none,mark=none, very thick] table {\data};
\pgfplotstableread{plot_data/TI/W_F/P_(3, 2)_mixed_test_min.txt}\data
\addplot [color32, fill=none,mark=none, very thick] table {\data};
\pgfplotstableread{plot_data/TI/W_F/P_(3, 2)_mixed_test_max.txt}\data
\addplot [color32, fill=none,mark=none, very thick] table {\data};
\pgfplotstableread{plot_data/TI/W_F/P_(3, 3)_mixed_test_min.txt}\data
\addplot [color33, fill=none,mark=none, very thick] table {\data};
\pgfplotstableread{plot_data/TI/W_F/P_(3, 3)_mixed_test_max.txt}\data
\addplot [color33, fill=none,mark=none, very thick] table {\data};

\end{groupplot}
\end{tikzpicture}
}
\caption{Evaluation of $W^{\text{F}}$ for transverse isotropy. Points depict data from the analytical model $W^\text{SNE}_\text{ti}$, while lines depict the evaluation of the  calibrated model $W^{\text{F}}$. Normal components of the stress tensor in red colors, shear components in blue, less important components in gray; stress in $\left[\text{hPa}\right]$.}
\label{fig:TI_deformation_gradient_model}
\end{figure}
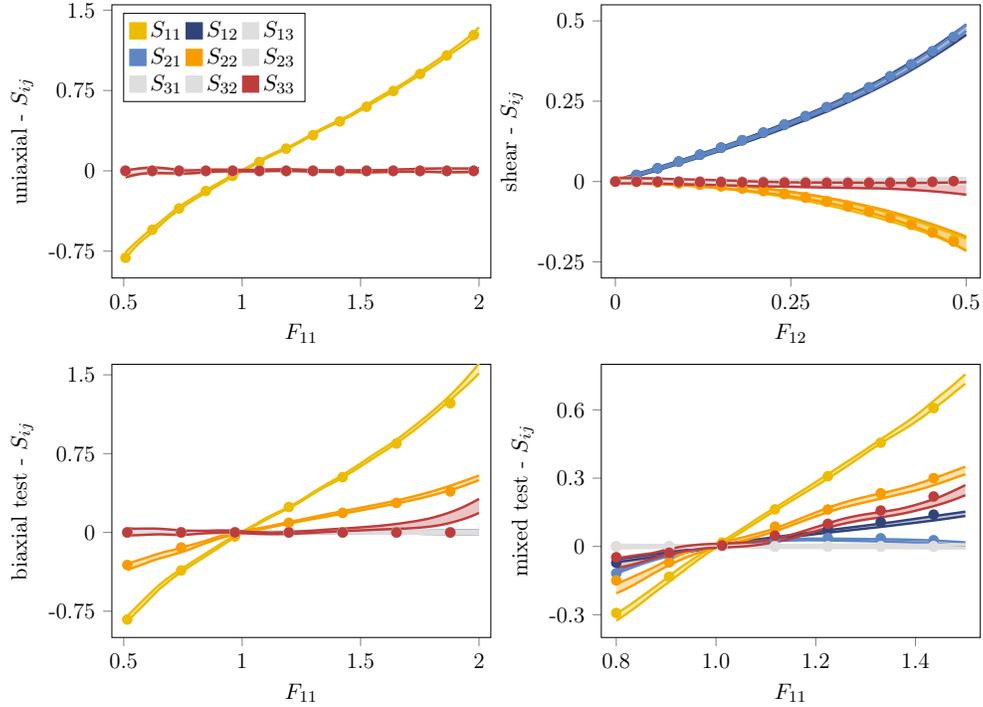

%% file: chapter/06_critique.tex
\section{A critique of machine learning in nonlinear elasticity theory}\label{sec:critique}

At this point, we would like to briefly discuss some general issues raised by the use of machine learning techniques in nonlinear elasticity theory. First and foremost, we would like to point out that these methods are not meant to serve as a \emph{replacement} for classical analytical models, but rather as an addition to the already existing extensive theoretical framework. More specifically, we want to address three interrelated shortcomings of the data-driven approach:
\begin{itemize}
	\item the lack of an intuitive interpretation of the model and its parameters;
	\item the unstable (and, in practice, even non-deterministic) dependence of the parameter values on the experimental data;
	\item the uncertainty of whether the resulting model is applicable to problems outside the range of prior experiments.
\end{itemize}
To a smaller extent, all three of these issues can be observed for a number of analytical models as well, especially some phenomenological models with a large number of parameters, which could be considered a precursor to the modern purely data-driven approaches.

\subsection{Analytical models}

For comparison, we first consider the the classical, isotropic \emph{Hencky strain energy}
\begin{equation}\label{eq:W_Hencky}
	\tDK{W^\text{H}}\colon\text{GL}^+(3)\to\mathbb{R}
	\,,\qquad
		\tDK{W^\text{H}}(F) = \mu\,\lVert\operatorname{dev}\log \sqrt{F^TF}\rVert^2+\frac{\kappa}{2}\,[\operatorname{tr}(\log \sqrt{F^TF})]^2
	\,,
\end{equation}
which depends solely on the two physical parameters $\mu$ (the shear modulus) and $\kappa$ (the bulk modulus).

While it is well known that the elasticity model induced by the Hencky energy does not provide an accurate description of very large deformations \cite{Anand79,ogden2004}, it is indeed highly accurate for up to moderate strains of about $20\%$ \cite{Anand79}. Moreover, the relation between the elastic behaviour predicted by the Hencky model and the experimental data used to determine the parameters is clearly accessible to direct \emph{interpretation}: The shear modulus and the bulk modulus are determined by the material's response to shear stresses and hydrostatic pressure, respectively, and in turn influence the stress response to certain modes of deformation, namely to simple shear and purely volumetric strain. In particular, this direct correspondence between the parameter values and the model's behaviour can be used to examine the \emph{plausibility} of a specific parameter set, even in the absence of additional test data.

Moreover, Hencky deduced his material model from a number of simple axiomatic assumptions \cite{Hencky1928,Hencky1929,neff2016,agn_neff2020axiomatic}. The applicability of his model to deformations not included in prior experiments can therefore be based on whether or not (or rather: to what degree) his postulates hold under the new circumstances. It is thereby possible to \emph{reasonably} assess the limitations of Hencky's model.

\medskip

This direct correspondence between the mechanical-geometrical interpretation of the model and its parameters can no longer be established for other hyperelastic material models, especially for so-called phenomenological (or \enquote{heuristic}) models.
For example, the \emph{Ogden energy}, which can be expressed in terms of the singular values $\lambda_1,\lambda_2,\lambda_3$ of $F$ via
\begin{equation}\label{eq:W_Ogden}
		\tDK{W^\text{OG}}\colon\text{GL}^+(3)\to\mathbb{R}
	\,,\qquad
	\tDK{W^\text{OG}}(F) = \sum_{i=1}^M \frac{\mu_i}{\alpha_i} (\lambda_1^{\alpha_i} + \lambda_2^{\alpha_i} + \lambda_3^{\alpha_i} - 3)
	\,,
\end{equation}
with $2M$ parameters $\mu_i > 0,\, \alpha_i\in\mathbb{R}$,
can provide a much better fit to empirical observations than the Hencky model for very large deformations \cite{ogden2004} if the number $2M$ of material parameters is sufficiently high. However, there is no longer any intuitive relation between these parameters and the predicted material behaviour.

Furthermore, the (globally) optimal choice of parameters for fitting the energy to a given dataset is difficult to determine due to the strongly nonlinear dependence of the induced stress-strain relation on the parameter values \cite{ogden2004}. Therefore, in practice, the result of the parameter optimization is not fully determined by the measured empirical data, but also affected by the random influence of, for example, the chosen starting points of the optimization algorithm. In particular, two possible outcomes of such an optimization procedure might yield a similar (or even an identical) quality of fit for the Ogden model to the limited experimental data, whereas the material behaviour predicted by those two distinct optimized models for \emph{other} deformations might differ significantly.
Therefore, a high degree of \emph{uncertainty} must remain about the prediction quality exhibited by material models such as Ogden's, particularly when applied to deformations which are not included in (or closely related to) the original experimental observations. This problem is closely related to the more general notion of \emph{overfitting}, i.e.\ optimizing too specifically to a given dataset, which tends to occur for model functions with a high number of parameters.

\subsection{Data-driven models}

Machine-learning based approaches share and even amplify these shortcomings of highly complex phenomenological models. Due to the general nature of data-driven methods, the number of parameters required for closely fitting such a model to a given specific dataset is necessarily rather high, even when compared to complex models such as the Ogden energy. In addition, for most machine learning methods used today (including neural networks), the resulting model cannot (easily) be stated explicitly in the form of a closed-form analytical expression. It is therefore extremely difficult, if not impossible, to develop an intuitive understanding of the relation between such a data-driven model and its parameters on the one hand and the predicted material behaviour on the other.

Of course, these problems have been recognized in many other fields of research where machine learning techniques have been applied, and a number of approaches have been suggested to determine not only the influence of different parameters on the prediction, but also the direct relation between the training data and the resulting output in a humanly comprehensible fashion \cite{lundberg2017,Shrikumar2017,strumbelj2014}. However, these techniques still lack the reliability and the direct intuition offered by more traditional models.

Similarly, the phenomenon of overfitting is a well known issue in the field of machine learning, and a number of precautions (such as the careful distinction between training and validation data) are taken in order to alleviate this problem. However, since any form of validation or testing is still based on available experimental data, any assumptions about the applicability of these models to circumstances outside the range of prior observations must remain unfounded even in the best of cases.

Thus, we consider it as extremely important to inform machine learning models with as much physical and mathematical structure as possible (as done here with the hyperelasticity, anisotropy, objectivity, and polyconvexity properties) and apply them only in cases where classical approaches cannot provide an acceptable accuracy (as is the case here due to the strong nonlinearity of the lattice microstructures).

\medskip

Finally, and perhaps most importantly, even if the resulting trained algorithm does indeed provide an accurate model in all practically relevant situations, it does not offer any insight as to \emph{why} its predictions are accurate. Again, this can be contrasted with the aforementioned Hencky elasticity model: Although not deduced \emph{ab initio}, the model has indeed been developed originally by Heinrich Hencky from simple geometrical and mechanical considerations, and a careful study of his deductions can without doubt further the reader's \emph{understanding} of continuum mechanics in a way that cannot be matched by inspecting the \enquote{black box} that results from training a machine learning algorithm. 
\medskip

Of course, the above considerations are not restricted to applications of data-driven models to nonlinear elasticity theory, but equally apply to many other domains of the natural sciences where machine learning has recently been demonstrated to yield promising results. Machine-learning based approaches can (and will) most certainly be employed to improve the accuracy of predictions and thus the quality of models and simulations in the years to come, most likely resulting in considerable technological advancements.
However, for the reasons outlined above, machine learning should not be considered as a full-fledged replacement of more traditional, analytical models, now or in the future, even if the outcomes seem to match or even surpass those resulting from more classical approaches.
In the past, humanity has found a major motivation for developing a further understanding of nature in the dependence of (practically applicable) \mbox{\emph{scientific techniques}} on the \emph{scientific method} \cite{russell1931}, and it would be most unfortunate if the success of machine learning in advancing the former would lead us to neglect the latter.

%% file: chapter/07_conclusion.tex
\section{Conclusion}
\label{sec:conclusion}
In the present work, two machine learning based constitutive models are proposed, which fulfill the polyconvexity condition by using input convex neural networks. This implies ellipticity of the constitutive models, which ensures material stability. The hyperelastic models are formulated for finite deformations and anisotropic material behavior. Furthermore, the neural networks yield highly flexible constitutive models, which are adaptable to a wide range of materials.

\medskip

The first model $W^{\text{I}}$ is based on a set of polyconvex, anisotropic invariants proposed by \textcite{Schroeder2010}, see eq.~\eqref{eq:invariants} and \eqref{pcnn_inv_based}, which fulfill the material objectivity and material symmetry conditions by construction. The neural network core is able to create highly nonlinear functions from the invariants, while conventional models are restricted to comparatively simple polynomials. Depending on the anisotropy class, various sets of invariants can be used.

The second model $W^{\text{F}}$ fully exploits the approximation capabilities of artificial neural networks. Formulated in the deformation gradient, its cofactor and determinant, the material objectivity condition is not fulfilled by construction, which would be a major drawback for conventional constitutive models. In the context of machine learning, however, the model is trained to approximate the objectivity condition, using data augmentation of the calibration dataset, cf.~\eqref{eq:data_augmentation}, which is possible due to the high flexibility of the models, and the high-performing optimization algorithms available in several machine learning libraries, e.g., TensorFlow. The present work not only uses the potential values, as done in \textcite{Ling2016}, but also the stress values. This is convenient as the stress is the quantity of interest for many applications of the constitutive model, and furthermore, the general approximation quality of the model can benefit from the additional information that the augmented stress values provide.
For incorporating the material symmetry, the group symmetrization introduced in \textcite{Fernandez2020} is used, see eq.~\eqref{eq:group_symmetrization}.

\medskip

The model capabilities are examined with synthetic homogenization data of cubic metamaterials used in \textcite{Fernandez2020}, and compared to the polyconvex model proposed by \textcite{Schroeder2010}. The simulation data offers a highly challenging benchmark case, with characteristics like lattice instabilities for several deformation modes. The analytical model from \cite{Schroeder2010} is not able to capture the behavior of the material, even for the uniaxial deformation case. The model $W^{\text{F}}$ shows excellent performance, not only for the calibration data, but also for several test scenarios which were not included in the training of the model. The evaluation of the model $W^{\text{I}}$ gave acceptable results for deformation gradients with dominating main diagonal elements, but failed to represent the stress response for shear deformations. 
\tOW{However, when fitted to data generated from the analytical, transversely isotropic model from \cite{Schroeder2008}, which represents a real-world material, the model $W^{\text{I}}$ also delivered excellent results. This shows that both models are applicable to a wide class of anisotropic materials, while $W^{\text{F}}$ is preferable for highly challenging metamaterials.}
Apparently, the polyconvexity conditions greatly improves the generalization capabilities of ANN-based constitutive models, such that they can be trained on fairly small training datasets. Here, deformation modes which are commonly applied in physical experiments were used.
Nevertheless, due to their high flexibility, the models can benefit from a wider range of calibration data and yield even better results. The data augmentation approach for the calibration of the model $W^{\text{F}}$ does not require additional simulation or experimental data, the extended dataset is created purely from mechanical considerations.

\medskip

The models are formulated as general as possible, and can be adapted to a wide range of anisotropic, hyperelastic materials. 
\tDK{It lies in the very nature of machine learning
that constitutive models based on neural networks are, to some extent, more complex than their conventional counterparts. As to what extent, the present work suggests that there are two major ways: For a wide range of materials with a moderately challenging behavior, very small ML models can be used, c.f.~the excellent performance of the small invariant based model in  Sect.~\ref{sec:application_to_ti}. The model complexity of this approach is close to the one of analytical potentials, without the need to manually construct a function for the specific material behavior at hand. However, for very complex material behavior, the full flexibility of neural networks can be utilized by using bigger networks, c.f.~the deformation gradient based model in Sect.~\ref{sec:application_to_cubic_metamaterials}. In both cases, the application to finite element simulations will be important for future research.
}
Considering the infinitely continuously differentiable neural network cores and the ellipticity of the proposed models, they offer a straightforward adaption for this, with favorable numerical behavior and trivial computation of stress and stiffness tensors, if automatic differentiation functionalities are considered. 
\tOW{For future work, it would be valuable to investigate the formulation of polyconvex FFNNs with a volumetric-deviatoric decomposition of the deformation gradient for (nearly) incompressible materials. Furthermore, the incorporation of parametric dependencies, such as the aspect ratio of a microstructure, into polyconvex FFNNs should be investigated. }

%% file: chapter/A_01_convex_neural_networks.tex
\section{Input convex feed-forward neural networks}
\label{app:aux_proofs}

FFNNs are a special class of artificial neural networks, which can be recursively defined as the composition of several vector-valued functions \cite{aggarwal2018, kollmannsberger2021}. The components of the vectors are referred to as nodes or neurons, the function in each neuron is referred to as activation function.
\begin{definition}[\textbf{Feed-forward neural networks (FFNNs)}]
\label{def:FFNNs}
The FFNN with vector-valued input $\uX$, $H$ hidden layers and scalar-valued output function $a$ is given by
\begin{equation}
\begin{aligned}\label{eq:FFNN}
\uX\in&\;\mathbb{R}^{n^{[0]}}
\\
\uA_1=\uA_1\left({{W}}^{[1]}\uX+\ub^{[1]}\right)\in&\;\mathbb{R}^{n^{[1]}},
\\
\uA_h=\uA_h\left({{W}}^{[h]}\uA_{h-1}+\ub^{[h]}\right)\in&\;\mathbb{R}^{n^{[h]}},\quad h=2,\dotsc,H
\\
a=a\left({{W}}^{[H+1]}\uA_H+\ub^{[H+1]}\right)\in&\;\mathbb{R}.
\end{aligned}
\end{equation}
Weights ${{W}}^{[h]}\in\mathbb{R}^{n^{[h]}\times n^{[h-1]}}$ and bias $\ub^{[h]}\in\mathbb{R}^{n^{[h]}}$ form the set of parameters, which is optimized when the model is calibrated. $\uX$ and $a$ are referred to as input and output layer, respectively, while the layers $\uA_h$ are referred to as hidden layers with component-wise applied activation functions according to
\begin{equation}\label{eq:weight_matrix}
\begin{aligned}
\left(\uA\left({{W}}\,\uX\right)\right)_i=A_i\left(\langle\uw^{[i]},\uX\rangle\right),
\quad\quad
{{W}}=\left(\uw^{[1]},\dotsc,\uw^{\left[n\right]}\right)^T.
\end{aligned}
\end{equation}
We apply the short notation for feed-forward neural networks
\begin{equation}
a\circ{\mathcal{A}}\circ\uX
\end{equation}
with the networks core ${\mathcal{A}}=\uA_H\circ\dotsc\circ\uA_1$. 
\end{definition}

\begin{definition}[\textbf{Input convex neural networks (ICNNs)}]
\label{def:convex_FFNNs}
The FFNN $\overline{a}=a\circ{\mathcal{A}}\circ\uX$ is called an ICNN, when the scalar-valued output $\overline{a}$ is convex w.r.t.~the vector-valued input $\uX$ \cite{Amos2016}. 
\end{definition}

Sufficiency conditions for the fulfillment of convexity in the case of function compositions are given in the following theorem.

\begin{theorem}%[\textbf{Convexity and function compositions}]
\label{theorem:convexity_and_compositions}
The function $\Bar{a}=a\circ\uB\circ\uX$ is convex in $\uX$ if the function $a$ is convex and non-decreasing in $\uB$, and $\uB$ is component-wise convex in $\uX$.
\end{theorem}
\begin{proof}
We have to show the positive semi-definiteness of the function's Hessian \cite{Hartmann2003,Silhavy2014}:
\begin{equation}\label{eq:convex_ANN_hessian}
   D^2_{\uX}\Bar{a}=\left(D_XB\right)^T\cdot D_B^2a \cdot D_XB+D_Ba \cdot D_X^2 B
\end{equation}
The derivatives $D_{\uX}\uB$ are mappings from the vector space $\uX$ into the vector space $\uB$. When $a$ is choosen as a convex function of $\uB$, the Hessian of $a$ w.r.t. $\uB$ is positive semi-definite and hence the first term in eq. \eqref{eq:convex_ANN_hessian} is positive semi-definite, see also observation 7.1.8 in \cite{horn2013}. The second term in eq. \eqref{eq:convex_ANN_hessian} is positive semi-definite when $a$ is non-decreasing in every component of $\uB$, and $\uB$ is component-wise convex in $\uX$.
\end{proof}

\begin{corollary}%[\textbf{Convex Neural Networks}]
\label{corollary:convex_ANN_compositions}
A FFNN is convex, when (i) the first hidden layer of the network's core is component-wise convex w.r.t.~the input, (ii) every following hidden layer is component-wise convex and non-decreasing w.r.t.~the previous layer, and (iii) the scalar-valued output function is convex and non-decreasing w.r.t.~the last hidden layer. 
\end{corollary}
\begin{proof}
This follows by recursively applying theorem \ref{theorem:convexity_and_compositions} to the components of the hidden layers.
\end{proof}

\begin{theorem}%[\textbf{Convexity and linear transformations}]
\label{theorem:convexity_and_linear_transformations}
Convexity is preserved under \tDK{affine} transformations.
\end{theorem}
The proof is clear but we provide it for the convenience of the reader.
\begin{proof}
The Hessian of the convex function ${a}$ applied on an \tDK{affine} transformation of its argument, i.e.,~${a}\big(\widetilde{X}\big)={a}\left(\uX\,\uuC+\uD\right)$ with arbitrary, but constant $\uuC$ and $\uD$, is positive semi-definite.
\begin{equation}\label{eq:affine_hessian}
   D^2_{\uX}{a}=\big(D_X\widetilde{X}\big)^T\cdot D_{\widetilde{X}}^2a \cdot D_X\widetilde{X}+D_{\widetilde{X}} a \cdot D_X^2 \widetilde{X}
\end{equation}
The positive semi-definiteness of the first summand follows equivalent to eq. \eqref{eq:convex_ANN_hessian}, while the second summand vanishes due to the linearity of $\widetilde{X}$ in $\uX$.
\end{proof}

\begin{corollary}%[\textbf{Convex Neural Networks}]
\label{corollary:convex_ANN_bias}
Due to its linearity, the bias of convex FFNNs can be choosen arbitrarily in every layer. The input of a convex FFNN can be multiplied by any constant matrix.
\end{corollary}

With corollaries \ref{corollary:convex_ANN_compositions} and \ref{corollary:convex_ANN_bias}, convex neural networks can be constructed. In the first step, this requires the choice of a convex and non-decreasing activation function. Furthermore, the activation function should be sufficiently smooth, as the calculation of gradients plays an important role in continuum mechanics. All of the former requirements can be fulfilled by the following functions.

\begin{theorem}[\textbf{Log-Sum-Exp function}]
\label{theorem:LSE}
The Log-Sum-Exp function is defined as 
\begin{equation}\label{eq:LSE0}
    f:\mathbb{R}^m\rightarrow\mathbb{R},\quad f \left(\ux\right)=\log \sum_{l=1}^me^{x_l}.
\end{equation}
We use the adaption $f_0\left(\ux\right)=f \left(0,\ux\right)$, since it is closely linked to the Softplus function, see corollary~\ref{cor:SP}.
Using the weight matrix ${{W}}\in\mathbb{R}^{m\times n}$ as defined in eq.~\eqref{eq:weight_matrix} and the bias vector $\ub\in\mathbb{R}^m$, we obtain the adapted Log-Sum-Exp function
\begin{equation}\label{eq:LSE}
    \LSE:\mathbb{R}^n\rightarrow\mathbb{R},\quad\LSE (\uX)=f_0\left({{W}}\,\uX+\ub\right)=\log \left[1+\sum_{l=1}^me^{\langle\uw^{[l]},\uX\rangle+b_l}\right]
\end{equation}
for neural networks.
The $\LSE$ function is convex for arbitrary weights and bias, and non-decreasing when all weights are non-negative. It is smooth for any choice of arguments or parameters, i.e., $\LSE\in C^\infty(\mathbb{R}^n)$.
\end{theorem}
\begin{proof}
\tRM{%
    We first show the convexity of the adapted softplus function $f_0$ by proving the positive semi-definiteness of the Hessian $D^2_{\ux}f_0\left(\ux\right)$. For this, we simply observe that the inequality
    \begin{align}
        0 \leq \uv\cdot D^2_{\ux}f_0\cdot\uv=
    	\frac{1}{\left(1+\sum_{l=1}^me^{x_l}\right)^2}\left(\left[\sum_{l=1}^m v_l^2 e^{x_l}\right]\left[1+\sum_{l=1}^me^{x_l}\right]-\left[\sum_{l=1}  ^mv_le^{x_l}\right]^2\right)
    \end{align}
    holds for any $\uv\in\mathbb{R}^m$ since, due to the Cauchy-Schwarz inequality,
    \begin{align}
    	\left[\sum_{l=1}^m v_le^{x_l}\right]^2
    	= \left[\sum_{l=1}^m v_le^{x_l/2}\cdot e^{x_l/2}\right]^2 
    	\leq \left[\sum_{l=1}^m v_l^2 e^{x_l}\right]\left[\sum_{l=1}^me^{x_l}\right]
    	\leq \left[\sum_{l=1}^m v_l^2 e^{x_l}\right]\left[1+\sum_{l=1}^me^{x_l}\right]
    	\,.
    \end{align}
}%
The $\LSE$ function is obtained through the linear transformation 
\begin{align}
\LSE (\uX)=f_0\left(\ux\right),\quad
    \ux={{W}}\,\uX+\ub
\end{align}
of the convex function $f_0$. Linear transformations preserve convexity, therefore the $\LSE$ function is also convex. 
The first derivative of the $\LSE$ function
\begin{equation}
      \left[D_{\uX}\LSE(\uX)\right]_i=\frac{\sum_{l=1}^mw^{[l]}_ie^{\langle\uw^{[l]},\uX\rangle}}{1+\sum_{l=1}^me^{\langle\uw^{[l]},\uX\rangle}}
\end{equation}
is non-negative for $w_i^{[l]}\geq 0\;\forall\, i,\,l$. The smoothness of the functions follows from the smoothness of the exponential function, and the smoothness of the logarithm on the positive domain.
\end{proof}

\begin{corollary}[\textbf{Softplus function}]
\label{cor:SP}
For $m=1$ in eq. \eqref{eq:LSE}, the $\LSE$ function is reduced to the Softplus ($\SP$) function
\begin{equation}\label{eq:SP}
    \SP:\mathbb{R}^n\rightarrow\mathbb{R},\quad\SP (\uX)=\log \left[1+e^{\langle\uw,\uX\rangle+b}\right]
\end{equation}
with weights $\uw\in\mathbb{R}^n$ and bias $b\in\mathbb{R}$. The $\SP$ function is convex for arbitrary weights and bias, and non-decreasing when all weights are non-negative. It is smooth for any choice of arguments or parameters.
\end{corollary}

\begin{prop}[\textbf{ICNNs using $\SP$ and $\LSE$ functions}]
\label{prop:LSE_SP_cores}
ICNNs based on $\SP$ functions are built from multiple layers, with several nodes using $\SP$ activation functions in each layer. For the first hidden layer, the weights of the softplus functions can take arbitrary values, while the other layers must be non-decreasing functions and, therefore, use non-negative weights. The networks output is given as the non-negative weighted sum of the last $\SP$ layer. The bias in every layer can be chosen arbitrarily.

ICNNs based on $\LSE$ functions can be composed of either a single $\LSE$ function, or of several $\LSE$ functions.
In the first case, the $\LSE$ function can be understood as a composition of one layer with several nodes using exponential activation functions, which are summed up and logarithmized in the next layer. The function achieves its approximation properties by increasing the amount of nodes in the exponential layer. As the exponential layer is the first hidden layer, the weights can be chosen arbitrarily.
Alternatively, the convex neural network can be constructed using several $\LSE$ functions. We obtain this approach similar to the one introduced for $\SP$ functions, just by replacing the $\SP$ functions by $\LSE$ functions. In both cases, the bias in every layer can be chosen arbitrarily.
\end{prop}

\begin{remark}
\label{rem:invariant_layer}
Constitutive models are often formulated in sets of invariants. In doing so, several important properties are fulfilled already by the choice of the input quantity. When a set of invariants is used for a ML model, the networks core must not only be convex, but also non-decreasing in its input, which follows directly from eq.~\eqref{eq:convex_ANN_hessian}. 
Since the invariants are created by nonlinear functions, e.g., $I_1=\tr(\uuF^T\uuF)$, the subsequent function processing the invariants must be convex and non-decreasing in order to be convex in $\left(\uuF,\,\cof \uuF,\,\det \uuF\right)$. For cores based on $\SP$ or $\LSE$ functions, this can easily be achieved by using non-negative weights in the first layer.
\end{remark}

\begin{remark}
The $\LSE$ activation function is not included in the current TensorFlow version, and was manually implemented. While $\LSE$ based cores should benefit from the highly flexible activation function and yield excellent results, the convergence behavior during training was very slow, and no satisfying results could be obtained. However, this should be seen as a numerical drawback of the implementation and optimization approaches used in this work rather than a general drawback of the function itself. Thus, only $\SP$ based cores were used for the numerical investigations shown in Sect.~\ref{sec:application_to_cubic_metamaterials}. 
\end{remark}

%% file: bibliography.bib
@phdthesis{Ebbing2010,
    title    = {Design of Polyconvex Energy Functions for All Anisotropy Classes},
    school   = {Universität Duisburg-Essen},
    author   = {Ebbing, V.},
    year     = {2010}, 
}

@book{Silhavy2014,
  title={The Mechanics and Thermodynamics of Continuous Media},
  author={Silhavy, M.},
  isbn={9783662033906},
  series={Theoretical and Mathematical Physics},
  year={2014},
  publisher={Springer Berlin Heidelberg},
  edition={1}
}

@article{Zee1983,
  title={Ordinary and strong ellipticity in the equilibrium theory of incompressible hyperelastic solids},
  author={L. Zee and E. R. Sternberg},
  journal={Archive for Rational Mechanics and Analysis},
  year={1983},
  volume={83},
  pages={53-90},
  doi={10.1007/BF00281087}
}

@article{Ball1976,
title = {Convexity conditions and existence theorems in nonlinear elasticity},
author = {Ball, J. M.},
year = {1976},
doi = {10.1007/BF00279992},
volume = {63},
pages = {337--403},
journal = {Archive for Rational Mechanics and Analysis},
publisher = {Springer},
number = {4},
}

@InProceedings{Ball1977,
author="Ball, J. M.",
editor="Knops, R. J.",
title="Constitutive inequalities and existence theorems in nonlinear elasto-statics",
booktitle="Herriot Watt Symposion: Nonlinear Analysis and Mechanics",
year="1977",
volume = "1",
publisher="Pitman",
address="London",
pages="187--241",
isbn="9780273011286"
}

@book{Holzapfel2000,
  title={Nonlinear Solid Mechanics: A Continuum Approach for Engineering},
  author={Holzapfel, G. A.},
  isbn={9780471823193},
  year={2000},
  publisher={Wiley},
  edition={2}
}

@book{TruesdellNoll,
  title={The Non-Linear Field Theories of Mechanics},
  author={Truesdell, C. and Noll, W.},
  isbn={9783662103883},
  year={2004},
  publisher={Springer Berlin Heidelberg},
  edition={3}
}

@book{Haupt2002,
  title={Continuum Mechanics and Theory of Materials},
  author={Haupt, P.},
  isbn={9783540431114},
  year={2002},
  publisher={Springer Berlin Heidelberg},
  edition={2}
}

@article{Fernandez2020,
    author       = {Fernández, M. and Jamshidian, M. and Böhlke, T. and Kersting, K. and Weeger, O.},
    year         = {2021},
    title        = {Anisotropic hyperelastic constitutive models for finite deformations combining material theory and data-driven approaches with application to cubic lattice metamaterials},
    volume       = {67},
    number       = {2},
    pages        = {653–677},
    journal      = {Computational Mechanics},
    doi          = {10.1007/s00466-020-01954-7},
    publisher    = {{Springer}},
}

@article{Schroeder2003,
author = {Schröder, J. and Neff, P.},
year = {2003},
pages = {401-445},
title = {Invariant formulation of hyperelastic transverse isotropy based on polyconvex free energy functions},
volume = {40},
journal = {International Journal of Solids and Structures},
doi = {10.1016/S0020-7683(02)00458-4}
}

@InProceedings{Schroeder2010,
author="Schröder, J.
and Neff, P.
and Ebbing, V.",
editor="Hackl, K.",
title="Polyconvex energies for trigonal, tetragonal and cubic symmetry groups",
booktitle="IUTAM Symposium on Variational Concepts with Applications to the Mechanics of Materials",
year="2010",
publisher="Springer Netherlands",
address="Dordrecht",
pages="221--232",
isbn="978-90-481-9195-6"
}

@article{Schroeder2008,
author = {Schröder, J. and Neff, P. and Ebbing, V.},
year = {2008},
pages = {3486-3506},
title = {Anisotropic polyconvex energies on the basis of crystallographic motivated structural tensors},
volume = {56},
journal = {Journal of the Mechanics and Physics of Solids},
doi = {10.1016/j.jmps.2008.08.008}
}

@article{Zheng1993,
title = "Tensors which characterize anisotropies",
journal = "International Journal of Engineering Science",
volume = "31",
number = "5",
pages = "679 - 693",
year = "1993",
doi = "10.1016/0020-7225(93)90118-E",
author = "Zheng, Q.-S. and Spencer, A. J."
}

@article{Calafiore2020,
  author={Calafiore, G. C. and Gaubert, S. and Possieri, C.},
  journal={IEEE Transactions on Neural Networks and Learning Systems}, 
  title={Log-sum-exp neural networks and posynomial models for convex and log-log-convex data}, 
  year={2020},
  volume={31},
  number={3},
  pages={827-838},
  doi={10.1109/TNNLS.2019.2910417}
  }

@article{Calafiore2020a,
author = {Calafiore, G. C. and Gaubert, S. and Possieri, C.},
year = {2020},
pages = {1-10},
title = {A universal approximation result for difference of log-sum-exp neural networks},
volume = {PP},
journal = {IEEE Transactions on Neural Networks and Learning Systems},
doi = {10.1109/TNNLS.2020.2975051}
}

@article{Jamshidian2020,
title = "Multiscale modelling of soft lattice metamaterials: Micromechanical nonlinear buckling analysis, experimental verification, and macroscale constitutive behaviour",
journal = "International Journal of Mechanical Sciences",
volume = "188",
pages = "105956",
year = "2020",
doi = "10.1016/j.ijmecsci.2020.105956",
author = "M. Jamshidian and N. Boddeti and D. W. Rosen and O. Weeger",
}

@article{Martin2018a,
   title={A non-ellipticity result, or the impossible taming of the logarithmic strain measure},
   volume={102},
   DOI={10.1016/j.ijnonlinmec.2018.02.011},
   journal={International Journal of Non-Linear Mechanics},
   publisher={Elsevier BV},
   author={Martin, R. J. and Ghiba, I.-D. and Neff, P.},
   year={2018},
   pages={147–158}
}

@article{Martin2017,
author = {Martin, R. J. and Ghiba, I.-D. and Neff, P.},
title = {A polyconvex extension of the logarithmic Hencky strain energy},
journal = {Analysis and Applications},
volume = {17},
number = {03},
pages = {349-361},
year = {2019},
doi = {10.1142/S0219530518500173},
}

@article{Martin2018b,
author = {Martin, R. J. and Voss, J. and Ghiba, I.-D. and Sander, O. and Neff, P.},
year = {2020},
title = {The quasiconvex envelope of conformally invariant planar energy functions in isotropic hyperelasticity},
journal = {Journal of Nonlinear Science},
volume = {30},
pages = {2885–2923},
doi = {10.1007/s00332-020-09639-4}
}

@article{Martin2020,
author = {Martin, R. J. and Voss, J. and Ghiba, I.-D. and Neff, P.},
year = {2020},
title = {Rank-one convexity vs. ellipticity for isotropic functions},
	journal = {Pre-print under review},
	      eprint={2008.11631},
      archivePrefix={arXiv},
}

@article{Kambouchev2007,
author = {Kambouchev, N. and Fernandez, J. and Radovitzky, R.},
year = {2007},
pages = {451},
title = {A polyconvex model for materials with cubic symmetry},
volume = {15},
journal = {Modelling and Simulation in Materials Science and Engineering},
doi = {10.1088/0965-0393/15/5/006}
}

@article{Linka2020,
author = {Linka, K. and Hillgärtner, M. and Abdolazizi, K. and Aydin, R. and Itskov, M. and Cyron, C.},
year = {2020},
pages = {110010},
title = {Constitutive artificial neural networks: A fast and general approach to predictive data-driven constitutive modeling by deep learning},
journal = {Journal of Computational Physics},
doi = {10.1016/j.jcp.2020.110010}
}

@article{Hornik1991,
  author = {Hornik, K.},
  doi = {10.1016/0893-6080(91)90009-T},
  journal = {Neural Networks},
  number = 2,
  pages = {251--257},
  title = {{Approximation capabilities of multilayer feedforward networks}},
  volume = 4,
  year = 1991
}

@article{Itskov2004,
author = {Itskov, M. and Aksel, N.},
year = {2004},
pages = {3833–3848},
title = {A class of orthotropic and transversely isotropic hyperelastic constitutive models based on a polyconvex strain energy function},
volume = {41},
journal = {International Journal of Solids and Structures},
doi = {10.1016/j.ijsolstr.2004.02.027}
}

@article{Hartmann2003,
author = {Hartmann, S. and Neff, P.},
year = {2003},
pages = {2767-2791},
title = {Polyconvexity of generalized polynomial-type hyperelastic strain energy funtions for near-incompressibility},
volume = {40},
journal = {International Journal of Solids and Structures},
doi = {10.1016/S0020-7683(03)00086-6}
}

@article{Ghiba2018,
author = {Ghiba, I.-D. and Martin, R. J. and Neff, P.},
year = {2018},
title = {Rank-one convexity implies polyconvexity in isotropic planar incompressible elasticity},
journal = {Journal de Mathématiques Pures et Appliqués},
volume = {116},
doi = {10.1016/j.matpur.2018.06.009}
}

@article{Itskov2001,
author = {Itskov, M.},
year = {2001},
pages = {1777 - 1799},
title = {A generalized orthotropic hyperelastic material model with application to incompressible shells},
volume = {50},
journal = {International Journal for Numerical Methods in Engineering},
doi = {10.1002/nme.86}
}

@article{Jiang2016,
author = {Jiang, Y.},
year = {2016},
pages = {34147},
title = {Highly-stretchable 3D-architected mechanical metamaterials},
volume = {6},
journal = {Scientific Reports},
doi = {10.1038/srep34147}
}

@article {Lee2012,
	Title = {Micro-/nanostructured mechanical metamaterials},
	Author = {Lee, J.-H. and Singer, J. P. and Thomas, E. L.},
	DOI = {10.1002/adma.201201644},
	Number = {36},
	Volume = {24},
	Year = {2012},
	Journal = {Advanced materials (Deerfield Beach, Fla.)},
	Pages = {4782—4810}
}

@article{Liu2016,
author = {Liu, J. and Gu, T. and Shan, S. and Kang, S. and Weaver, J. and Bertoldi, K.},
year = {2016},
pages = {},
title = {Harnessing buckling to design architected materials that exhibit effective negative swelling},
volume = {28},
journal = {Advanced Materials},
doi = {10.1002/adma.201600812}
}

@article{Matous2016,
author = {Matous, K. and Geers, M. and Kouznetsova, V. and Gillman, A.},
year = {2016},
pages = {},
title = {A review of predictive nonlinear theories for multiscale modeling of heterogeneous materials},
volume = {330},
journal = {Journal of Computational Physics},
doi = {10.1016/j.jcp.2016.10.070}
}

@article{Ehret2007,
author = {Ehret, A. and Itskov, M.},
year = {2007},
pages = {8853-8863},
title = {A polyconvex hyperelastic model for fiber-reinforced materials in application to soft tissues},
volume = {42},
journal = {Journal of Materials Science},
doi = {10.1007/s10853-007-1812-6}
}

@InProceedings{Amos2016, 
 title = {Input convex neural networks}, 
 author = {Amos, B. and
               Xu, L. and
               Kolter, J. Z.}, 
 booktitle = {Proceedings of the 34th International Conference on Machine Learning}, 
 pages = {146--155}, 
 year = {2017}, 
 editor = {Precup, D. and Teh, Y. W.}, volume = {70}, 
 series = {Proceedings of Machine Learning Research}, 
 publisher = {{PMLR}}, 
   archivePrefix = {arXiv},
  eprint    = {1609.07152}
 }

@article{Ghaderi2020,
author = {Ghaderi, A. and Morovati, V. and Dargazany, R.},
year = {2020},
title = {A physics-informed assembly of feed-forward neural network engines to predict inelasticity in cross-linked polymers},
volume = {12},
journal = {Polymers},
doi = {10.3390/polym12112628}
}

@article{Ling2016,
author = {Ling, J. and Jones, R. and Templeton, J.},
year = {2016},
title = {Machine learning strategies for systems with invariance properties},
volume = {318},
journal = {Journal of Computational Physics},
doi = {10.1016/j.jcp.2016.05.003}
}

@article{Xu2020,
  author = {Xu, K. and Huang, D. and Darve, E.},
year = {2021},
title = {Learning constitutive relations using symmetric positive definite neural networks},
volume = {428},
journal = {Journal of Computational Physics},
doi = {10.1016/j.jcp.2020.110072}
}

@article{Gonzalez2019,
author = {González, D. and Chinesta, F. and Cueto, E.},
year = {2019},
pages = {14},
title = {Learning corrections for hyperelastic models from data},
volume = {6},
journal = {Frontiers in Materials},
doi = {10.3389/fmats.2019.00014}
}

@article{Gonzalez2020,
author = {González, D. and García, A. and Chinesta, F. and Cueto, E.},
year = {2020},
pages = {2319},
title = {A data-driven learning method for constitutive modeling: Application to vascular hyperelastic soft tissues},
volume = {13},
journal = {Materials},
doi = {10.3390/ma13102319}
}

@book{horn2013, 
author= {Horn, R. A. and Johnson, C. R.}, 
title = {Matrix Analysis}, 
year = {2013}, 
isbn = {0521548233}, 
publisher = {Cambridge University Press}, 
address = {USA}, 
edition = {2}, 
 }

@article{fernandez2021,
	title = {Material modeling for parametric finite hyperelasticity based on machine learning with application in optimization of metamaterials},
	author = {Fernández, M. and Fritzen, F. and Weeger, O.},
	journal = {Pre-print under review},
	year = {2021},
	doi = {10.13140/RG.2.2.21536.10242}
}

@article{surjadi19,
	author = {Surjadi, J. U. and Gao, L. and Du, H. and Li, X. and Xiong, X. and Fang, N. Xuanlai and Lu, Y.},
	title = {Mechanical metamaterials and their engineering applications},
	journal = {Advanced Engineering Materials},
	volume = 21,
	number = 3,
	pages = 1800864,
	doi = {10.1002/adem.201800864},
	year = 2019,
}

@article{bertoldi2017,
	title = {Flexible mechanical metamaterials},
	volume = {2},
	copyright = {2017 Macmillan Publishers Limited},
	doi = {10.1038/natrevmats.2017.66},
	number = {11},
	journal = {Nature Reviews Materials},
	author = {Bertoldi, K. and Vitelli, V. and Christensen, J. and van Hecke, M.},
	year = {2017},
	note = {Number: 11
Publisher: Nature Publishing Group},
	pages = {1--11},
}

@article{xiang2020a,
	title = {A {review} of {physically} {based} and {thermodynamically} {based} {constitutive} {models} for {soft} {materials}},
	volume = {87},
	doi = {10.1115/1.4047776},
	number = {11},
	journal = {Journal of Applied Mechanics},
	author = {Xiang, Y. and Zhong, D. and Rudykh, S. and Zhou, H. and Qu, S. and Yang, W.},
	year = {2020}
}

@article{chagnon2015,
	title = {Hyperelastic {energy} {densities} for {soft} {biological} {tissues}: {A} {review}},
	volume = {120},
	shorttitle = {Hyperelastic {Energy} {Densities} for {Soft} {Biological} {Tissues}},
	doi = {10.1007/s10659-014-9508-z},
	number = {2},
	journal = {Journal of Elasticity},
	author = {Chagnon, G. and Rebouah, M. and Favier, D.},
	year = {2015},
	pages = {129--160},
}

@article{Balzani2006,
title = {A polyconvex framework for soft biological tissues. Adjustment to experimental data},
journal = {International Journal of Solids and Structures},
volume = {43},
number = {20},
pages = {6052-6070},
year = {2006},
doi = {10.1016/j.ijsolstr.2005.07.048},
author = {D. Balzani and P. Neff and J. Schröder and G. A. Holzapfel},
}

@article{Schroeder2005,
title = {A variational approach for materially stable anisotropic hyperelasticity},
journal = {International Journal of Solids and Structures},
volume = {42},
number = {15},
pages = {4352-4371},
year = {2005},
doi = {10.1016/j.ijsolstr.2004.11.021},
author = {J. Schröder and P. Neff and D. Balzani},
}

@article{kirchdoerfer2016,
	title = {Data-driven computational mechanics},
	volume = {304},
	doi = {10.1016/j.cma.2016.02.001},
	journal = {Computer Methods in Applied Mechanics and Engineering},
	author = {Kirchdoerfer, T. and Ortiz, M.},
	year = {2016},
	pages = {81--101},
}

@article{nguyen2018,
	title = {A data-driven approach to nonlinear elasticity},
	volume = {194},
	doi = {10.1016/j.compstruc.2017.07.031},
	journal = {Computers and Structures},
	author = {Nguyen, L. T. K. and Keip, M.-A.},
	year = {2018},
	pages = {97--115},
}

@article{carrara2020,
	title = {Data-driven fracture mechanics},
	volume = {372},
	doi = {10.1016/j.cma.2020.113390},
	journal = {Computer Methods in Applied Mechanics and Engineering},
	author = {Carrara, P. and De Lorenzis, L. and Stainier, L. and Ortiz, M.},
	year = {2020},
	pages = {113390},
}

@article{yvonnet2007,
	title = {The reduced model multiscale method ({R3M}) for the non-linear homogenization of hyperelastic media at finite strains},
	volume = {223},
	doi = {10.1016/j.jcp.2006.09.019},
	number = {1},
	journal = {Journal of Computational Physics},
	author = {Yvonnet, J. and He, Q.-C.},
	year = {2007},
	pages = {341--368},
}

@article{YVONNET2009,
	title = {Numerically explicit potentials for the homogenization of nonlinear elastic heterogeneous materials},
	volume = {198},
	doi = {10.1016/j.cma.2009.03.017},
	number = {33},
	journal = {Computer Methods in Applied Mechanics and Engineering},
	author = {Yvonnet, J. and Gonzalez, D. and He, Q.-C.},
	year = {2009},
	pages = {2723 -- 2737}
}

@article{le2015,
	title = {Computational homogenization of nonlinear elastic materials using neural networks},
	volume = {104},
	copyright = {Copyright © 2015 John Wiley \& Sons, Ltd.},
	doi = {10.1002/nme.4953},
	number = {12},
	journal = {International Journal for Numerical Methods in Engineering},
	author = {Le, B. A. and Yvonnet, J. and He, Q.-C.},
	year = {2015},
	pages = {1061--1084},
}

@article{yang2019a,
	title = {Derivation of heterogeneous material laws via data-driven principal component expansions},
	volume = {64},
	doi = {10.1007/s00466-019-01728-w},
	number = {2},
	journal = {Computational Mechanics},
	author = {Yang, H. and Guo, X. and Tang, S. and Liu, W. K.},
	year = {2019},
	pages = {365--379},
}

@article{liu2020,
	title = {A generic physics-informed neural network-based constitutive model for soft biological tissues},
	volume = {372},
	doi = {10.1016/j.cma.2020.113402},
	journal = {Computer Methods in Applied Mechanics and Engineering},
	author = {Liu, M. and Liang, L. and Sun, W.},
	year = {2020},
	pages = {113402},
}

@article{flaschel2021,
	title = {Unsupervised discovery of interpretable hyperelastic constitutive laws},
	journal = {Computer Methods in Applied Mechanics and Engineering},
	author = {Flaschel, M. and Kumar, S. and De Lorenzis, L.},
	year = {2021},
	doi = {10.1016/j.cma.2021.113852},
volume = {381},
pages = {113852},
}

@article{fritzen2018,
	title = {Two-stage data-driven homogenization for nonlinear solids using a reduced order model},
	volume = {69},
	doi = {10.1016/j.euromechsol.2017.11.007},
	journal = {European Journal of Mechanics - A/Solids},
	author = {Fritzen, F. and Kunc, O.},
	year = {2018},
	pages = {201--220},
}

@article{fritzen2019a,
	title = {On-the-fly adaptivity for nonlinear twoscale simulations using artificial neural networks and reduced order modeling},
	volume = {6},
	doi = {10.3389/fmats.2019.00075},
	journal = {Frontiers in Materials},
	author = {Fritzen, F. and Fernández, M. and Larsson, F.},
	year = {2019},
	pages = {75},
}

@article{glaesener2020b,
	title = {Continuum representation of nonlinear three-dimensional periodic truss networks by on-the-fly homogenization},
	doi = {10.1016/j.ijsolstr.2020.08.013},
	journal = {International Journal of Solids and Structures},
	author = {Glaesener, R. N. and Träff, E. A. and Telgen, B. and Canonica, R. M. and Kochmann, D. M.},
	year = {2020},
}

@article{khajehtourian2021,
	title = {Soft {adaptive} {mechanical} {metamaterials}},
	volume = {8},
	doi = {10.3389/frobt.2021.673478},
	journal = {Frontiers in Robotics and AI},
	author = {Khajehtourian, R. and Kochmann, D.},
	year = {2021},
	pages = {673478},
}

@article{khajehtourian2020,
	title = {A continuum description of substrate-free dissipative reconfigurable metamaterials},
	volume = {147},
	doi = {10.1016/j.jmps.2020.104217},
	journal = {Journal of the Mechanics and Physics of Solids},
	author = {Khajehtourian, R. and Kochmann, D.},
	year = {2020},
}

@incollection{schroder2010,
	address = {Vienna},
	series = {{CISM} {International} {Centre} for {Mechanical} {Sciences}},
	title = {Anisotropic polyconvex energies},
	isbn = {978-3-7091-0174-2},
	booktitle = {Poly-, {Quasi}- and {Rank}-{One} {Convexity} in {Applied} {Mechanics}},
	publisher = {Springer},
	author = {Schröder, J.},
	editor = {Schröder, J. and Neff, P.},
	year = {2010},
	doi = {10.1007/978-3-7091-0174-2_3},
	pages = {53--105},
}

@article{bonet2015,
	title = {A computational framework for polyconvex large strain elasticity},
	volume = {283},
	doi = {10.1016/j.cma.2014.10.002},
	journal = {Computer Methods in Applied Mechanics and Engineering},
	author = {Bonet, J. and Gil, A. J. and Ortigosa, R.},
	year = {2015},
	pages = {1061--1094},
}

@article{pfefferkorn2019,
	title = {Extension of the {enhanced} {assumed} {strain} {method} {based} on the {structure} of {polyconvex} {strain}‐{energy} {functions}},
	volume = {121},
	doi = {10.1002/nme.6284},
	journal = {International Journal for Numerical Methods in Engineering},
	author = {Pfefferkorn, R. and Betsch, P.},
	year = {2019},
}

@article{heider2020,
	title = {{SO}(3)-invariance of informed-graph-based deep neural network for anisotropic elastoplastic materials},
	volume = {363},
	doi = {10.1016/j.cma.2020.112875},
	journal = {Computer Methods in Applied Mechanics and Engineering},
	author = {Heider, Y. and Wang, K. and Sun, W.},
	year = {2020},
}

@article{vlassis2020a,
	title = {Geometric deep learning for computational mechanics {part} {I}: {Anisotropic} hyperelasticity},
	volume = {371},
	shorttitle = {Geometric deep learning for computational mechanics {Part} {I}},
	doi = {10.1016/j.cma.2020.113299},
	journal = {Computer Methods in Applied Mechanics and Engineering},
	author = {Vlassis, N. and Ma, R. and Sun, W.},
	year = {2020},
}

@article{neff2015,
title={The exponentiated Hencky-logarithmic strain energy. Part I: Constitutive issues and rank-one convexity},
author={Neff, P. and Ghiba, I.-D. and Lankeit, J.},
volume={121},
journal={Journal of Elasticity},
doi={10.1007/s10659-015-9524-7},
year={2015},
pages={143-234}
}

@article{neff2016,
	title =		{Geometry of logarithmic strain measures in solid mechanics},
	author =	{Neff, P. and Eidel, B. and Martin, R. J.},
	journal =	{Archive for Rational Mechanics and Analysis},
	volume =	{222},
	number =	{2},
	pages =		{507-572},
	year =		{2016},
	doi =		{10.1007/s00205-016-1007-x},
	class =	{mathematics}
}

@book{bertram2021,
  title={Elasticity and Plasticity of Large Deformations},
  author={Bertram, A.},
  isbn={978-3-030-72328-6},
  year={2021},
  publisher={Springer International Publishing},
  edition={4}
}

@article{Cai2021,
       author = {{Cai}, R. and {Holweck}, F. and {Feng}, Z.-Q. and {Peyraut}, F.},
        title = {Integrity basis of polyconvex invariants for modeling hyperelastic orthotropic materials {\textemdash} Application to the mechanical response of passive ventricular myocardium},
      journal = {International Journal of Non-Linear Mechanics},
         year = 2021,
       volume = {133},
        pages = {103713},
          doi = {10.1016/j.ijnonlinmec.2021.103713},
}

@article{e2020,
      title={Integrating machine learning with physics-based modeling}, 
      author={W. E and J. Han and L. Zhang},
      year={2020},
      eprint={2006.02619},
      archivePrefix={arXiv},
      journal={Pre-print under review}
}

@article{willard2020,
      title={Integrating physics-based modeling with machine learning: A survey}, 
      author={J. Willard and X. Jia and S. Xu and M. Steinbach and V. Kumar},
      year={2020},
      eprint={2003.04919},
      archivePrefix={arXiv},
      journal={Pre-print under review}
}

@article{karniadakis2021,
title={Physics-informed machine learning},
author={G. E. Karniadakis and I. G. Kevrekidis and L. Lu and P. Perdikaris and S. Wang and L. Yang},
year={2021},
doi={10.1038/s42254-021-00314-5},
journal={Nature Reviews Physics},
SN = {2522-5820}
}

@article{wang2020,
  author    = {S. Wang and
               X. Yu and
               P. Perdikaris},
  title     = {When and why PINNs fail to train: {A} neural tangent kernel perspective},
  journal   = {Pre-print under review},
  year      = {2020},
  archivePrefix = {arXiv},
  eprint    = {2007.14527},
}

@book{kruzik2019,
author = {Kru\v{z}\'{i}k, M. and Roub\'{i}\v{c}ek, T.},
year = {2019},
title = {Mathematical Methods in Continuum Mechanics of Solids},
isbn = {978-3-030-02064-4},
edition={1},
publisher={Springer International Publishing}
}

@article{schroeder2018,
author = {Schröder, J. and von Hoegen, M. and Neff, P.},
year = {2018},
pages = {657-685},
title = {The exponentiated Hencky energy: Anisotropic extension and case studies},
volume = {61},
journal = {Computational Mechanics},
doi = {10.1007/s00466-017-1466-4}
}

@InProceedings{Shrikumar2017,
  title = {Learning important features through propagating activation differences},
  author = {A. Shrikumar and P. Greenside and A. Kundaje},
  booktitle = {Proceedings of the 34th International Conference on Machine Learning},
  pages = {3145-3153},
  year = {2017},
  editor = {Precup, D. and Teh, Y. W.},
  volume = {70},
  series = {Proceedings of Machine Learning Research},
  publisher = {PMLR},
    archivePrefix = {arXiv},
  eprint    = {1704.02685},
}

@InProceedings{lundberg2017,
  title={A unified approach to interpreting model predictions},
  author={Lundberg, S. M. and Lee, S.-I.},
  booktitle={Advances in Neural Information Processing Systems},
  pages={4765-4774},
  year={2017},
    archivePrefix = {arXiv},
  eprint    = {1705.07874}
}

@article{strumbelj2014,
  title={Explaining prediction models and individual predictions with feature contributions},
  author={{\v{S}}trumbelj, E. and Kononenko, I.},
  journal={Knowledge and information systems},
  volume={41},
  number={3},
  pages={647-665},
  year={2014},
  publisher={Springer},
  doi={10.1007/s10115-013-0679-x}
}

@article{Anand79,
  author =	{L. Anand},
  title =	{On {H.~H}encky's approximate strain energy function for moderate deformations},
  journal =	{Journal of Applied Mechanics},
  volume =	{46},
  pages =	{78-82},
  year =	{1979},
  doi={10.1115/1.3424532}
}

@article{ogden2004,
  title =		{Fitting hyperelastic models to experimental data},
  author =		{Ogden, R. W. and Saccomandi, G. and Sgura, I.},
  journal =		{Computational Mechanics},
  volume =		{34},
  number =		{6},
  pages =		{484-502},
  year =		{2004},
  publisher =	{Springer},
  doi={10.1007/s00466-004-0593-y}
}

@article{Hencky1928,
  title = 	{{\"U}ber die {F}orm des {E}lastizit{\"a}tsgesetzes bei ideal elastischen {S}toffen},
  author =	{H. Hencky},
  journal =	{Zeitschrift f\"ur technische {P}hysik},
  volume =	{9},
  pages =	{215-220},
  year =	{1928}
}

@article{Hencky1929,
  title = 	{{W}elche {U}mst{\"a}nde bedingen die {V}erfestigung bei der bildsamen {V}erformung von festen isotropen {K\"o}rpern?},
  author =	{H. Hencky},
  journal =	{Zeitschrift f{\"u}r {P}hysik},
  volume =	{55},
  pages =	{145-155},
  year =	{1929}
}

@book{russell1931,
  title={The Scientific Outlook},
  author={Russell, B.},
  year={1931},
  publisher={George Allen \& Unwin}
}

@article{agn_neff2020axiomatic,
  title={The axiomatic introduction of arbitrary strain tensors by Hans Richter -- A commented translation of ‘Strain tensor, strain deviator and stress tensor for finite deformations’},
  author={Neff, P. and Graban, K. and Schweickert, E. and Martin, R. J.},
  journal={Mathematics and Mechanics of Solids},
  volume={25},
  number={5},
  pages={1060-1080},
  year={2020},
  publisher={Sage Publications},
        eprint={1909.05998},
      archivePrefix={arXiv},
}

@article{tac2021,
      title={Data-driven modeling of the mechanical behavior of anisotropic soft biological tissue}, 
      author={V. Tac and V. D. Sree and M. K. Rausch and A. B. Tepole},
      year={2021},
      eprint={2107.05388},
      archivePrefix={arXiv},
      journal={Pre-print under review}
}

@article{kunc2019,
   title={Finite strain homogenization using a reduced basis and efficient sampling},
   volume={24},
   DOI={10.3390/mca24020056},
   number={2},
   journal={Mathematical and Computational Applications},
   author={Kunc, O. and Fritzen, F.},
   year={2019},
   pages={56}
}

@article{scipy,
  author  = {Virtanen, P. and Gommers, R. and Oliphant, T. E. and
            Haberland, M. and Reddy, T. and Cournapeau, D. and
            Burovski, E. and Peterson, P. and Weckesser, W. and
            Bright, J. and {van der Walt}, S. J. and
            Brett, M. and Wilson, J. and Millman, K. J. and
            Mayorov, N. and Nelson, A. R. J. and Jones, E. and
            Kern, R. and Larson, E. and Carey, C J and
            Polat, {\.I}. and Feng, Y. and Moore, E. W. and
            {VanderPlas}, J. and Laxalde, D. and Perktold, J. and
            Cimrman, R. and Henriksen, I. and Quintero, E. A. and
            Harris, C. R. and Archibald, A. M. and
            Ribeiro, A. H. and Pedregosa, F. and
            {van Mulbregt}, P. and {SciPy 1.0 Contributors}},
  title   = {{{SciPy} 1.0: Fundamental algorithms for scientific
            computing in python}},
  journal = {Nature Methods},
  year    = {2020},
  volume  = {17},
  pages   = {261--272},
  doi     = {10.1038/s41592-019-0686-2},
}

@article{Baydin2018,
author = {Baydin, A. and Pearlmutter, B. and Radul, A. and Siskind, J.},
year = {2018},
pages = {1-43},
title = {Automatic differentiation in machine learning: A survey},
volume = {18},
journal = {Journal of Machine Learning Research},
      eprint={1502.05767},
      archivePrefix={arXiv},
}

@book{kollmannsberger2021,
  author    = {S. Kollmannsberger and
               D. D'Angella and
               M. Jokeit and
               L. Herrmann},
  title     = {Deep Learning in Computational Mechanics},
  series    = {Studies in Computational Intelligence},
  volume    = {977},
  publisher = {Springer},
  year      = {2021},
  doi       = {10.1007/978-3-030-76587-3},
  isbn      = {978-3-030-76586-6},
}

@book{aggarwal2018,
author = {Aggarwal, C. C.},
title = {Neural Networks and Deep Learning},
year = {2018},
isbn = {3319944622},
publisher = {Springer International Publishing},
edition = {1},
}
